\documentclass[%
 article,
superscriptaddress,
 amsmath,amssymb,
 aps, 
]{revtex4-2}
\usepackage[colorlinks=true,linkcolor=blue]{hyperref}

\usepackage{xspace}
\usepackage{graphicx}
\usepackage{float}
\usepackage{amsmath}
\usepackage{tikz}
\usepackage{tikz-feynman}

\usepackage{amssymb}
\usepackage{comment}
\usepackage{xcolor}

\newcommand{\ave}[1]{\left\langle #1 \right\rangle}
\newcommand{\spave}[1]{\overline{#1}}

\newcommand{\imag}{\mathring{\imath}}

\newcommand{\plaind}{\mathrm{d}}
\newcommand{\dint}[1]{\mathchoice{\!\plaind#1\,}{\!\plaind#1\,}{\!\plaind#1\,}{\!\plaind#1\,}}
\newcommand{\ddint}[1]{\ddintx{#1}{d}}
\newcommand{\ddintx}[2]{\mathchoice{\!\plaind^{#2}#1\,}{\!\plaind^{#2}#1\,}{\!\plaind^{#2}#1\,}{\!\plaind^{#2}#1\,}}
\newcommand{\dTWOint}[1]{\ddintx{#1}{2}}
\newcommand{\dTHREEint}[1]{\ddintx{#1}{3}}

\newcommand{\dbar}{\plaind\mkern-6mu\mathchar'26}
\newcommand{\deltabar}{\delta\mkern-6mu\mathchar'26}
\newcommand{\dintbar}[1]{\mathchoice{\!\dbar#1\,}{\!\dbar#1\,}{\!\dbar#1\,}{\!\dbar#1\,}}

\newcommand{\dTWOintbar}[1]{\mathchoice{\!\dbar^{2}#1\,}{\!\dbar^{2}#1\,}{\!\dbar^{2}#1\,}{\!\dbar^{2}#1\,}}
\newcommand{\dTHREEintbar}[1]{\mathchoice{\!\dbar^{3}#1\,}{\!\dbar^{3}#1\,}{\!\dbar^{3}#1\,}{\!\dbar^{3}#1\,}}
\newcommand{\Dint}[1]{\mathcal{D}\!#1\,}

\newcommand{\MCLineVsMathEnv}[2]{\mathchoice{#1}{#2}{#2}{#2}}

\newcommand{\ddX}[1]{\MCLineVsMathEnv{\frac{\plaind}{\plaind #1}}{\plaind/\plaind #1}}

\newcommand{\ddx}{\ddX{x}}

\usepackage{dsfont}

\newcommand{\gpset}[1]{\mathds{#1}}

\newcommand{\canetset}[1]{{\mathchoice {\hbox{$\sf\textstyle #1\kern-0.4em #1$}}
{\hbox{$\sf\textstyle #1\kern-0.4em #1$}}
{\hbox{$\sf\scriptstyle #1\kern-0.3em #1$}}
{\hbox{$\sf\scriptscriptstyle #1\kern-0.2em #1$}}}}

\newcommand{\Nset}{\gpset{N}}
\newcommand{\Zset}{\gpset{Z}}

\def\nbZ{{\mathchoice {\hbox{$\sf\textstyle Z\kern-0.4em Z$}}
{\hbox{$\sf\textstyle Z\kern-0.4em Z$}}
{\hbox{$\sf\scriptstyle Z\kern-0.3em Z$}}
{\hbox{$\sf\scriptscriptstyle Z\kern-0.2em Z$}}}}

\newcommand{\gpvec}[1]{\mathbf{#1}}

\newcommand{\zerovec}{\gpvec{0}}
\newcommand{\nullvec}{\zerovec}

\newcommand{\evec}{\gpvec{e}}

\newcommand{\kvec}{\gpvec{k}}

\newcommand{\nvec}{\gpvec{n}}
\newcommand{\pvec}{\gpvec{p}}
\newcommand{\qvec}{\gpvec{q}}
\newcommand{\rvec}{\gpvec{r}}

\newcommand{\xvec}{\gpvec{x}}

\usepackage{bm}
\newcommand{\xivec}{\bm{\xi}}
\newcommand{\zetavec}{\bm{\zeta}}

\newcommand{\transpose}{\mathsf{T}}

\newcommand{\ident}{\mathds{1}}

\newcommand{\AC}{\mathcal{A}}

\newcommand{\LC}{\mathcal{L}}
\newcommand{\NC}{\mathcal{N}}
\newcommand{\OC}{\mathcal{O}}

\newcommand{\utilde}{\tilde{u}}

\newcommand{\chitilde}{\tilde{\chi}}

\newcommand{\fourth}{\mathchoice{\frac{1}{4}}{(1/4)}{\frac{1}{4}}{(1/4)}}

\newcommand{\quarter}{\fourth}

\newcommand{\Exp}[1]{\operatorname{exp}\left(#1\right)}
\renewcommand{\exp}[1]{\mathchoice{\mathrm{e}^{#1}}{\operatorname{exp}\left(#1\right)}{\operatorname{exp}\left(#1\right)}{\operatorname{exp}\left(#1\right)}}

\newcommand{\elabel}[1]{\label{eq:#1}}
\newcommand{\eref}[1]{(\ref{eq:#1})}
\newcommand{\Eref}[1]{Eq.~(\ref{eq:#1})}
\newcommand{\Erefs}[1]{Eqs.~(\ref{eq:#1})}

\newcommand{\Sref}[1]{Section~\ref{sec:#1}}

\newcommand{\latin}[1]{{\it #1}}
\newcommand{\ie}{\latin{i.e.}\@\xspace}
\newcommand{\eg}{\latin{e.g.}\@\xspace}

\newcommand{\gcomment}[1]{{\textcolor{orange}{\bf GP: #1}}}

\newlength \standardfigwidth
\newcounter{exercise}
{\addtocounter{exercise}{1}\begin{center}\begin{minipage}{0.8\linewidth}\textbf{Exercise
\arabic{exercise}:}\begin{itshape}}
{\end{itshape}\end{minipage}\end{center}}

\makeatletter
\newcommand{\creat}[3][]{\@ifempty{#1}{#2^{\dagger}}{\left(#2^{\dagger}\right)^{#1}}\@ifempty{#3}{}{\!(#3)}}

\newcommand{\creatDoi}[3][]{\@ifempty{#1}{\tilde{#2}}{\left(\tilde{#2}\right)^{#1}}\@ifempty{#3}{}{(#3)}}

\newcommand{\annih}[3][]{#2\@ifempty{#1}{}{^{#1}}\@ifempty{#3}{}{(#3)}}

\makeatother

% ...

\newlength{\bibmarkkeyAleft}

\newlength{\bibmarkkeyBleft}

\newlength{\bibmarkkeyCleft}

\newlength{\bibmarkkeyDleft}

\newcommand{\action}{\AC}

\tikzset{
xxtsubstrate/.style={decorate, 
line width=1pt,
draw=olive, 
decoration=snake, 
segment amplitude=0.75mm, 
line after snake=0.25mm,
line before snake=0.25mm
},
tsubstrate/.style={decorate, 
line width=1pt,
draw=olive, 
decoration=snake, 
segment amplitude=0.5mm, 
segment length=5pt,
segment amplitude=0.2mm, 
line after snake=1mm,
line before snake=1mm
},
Bsubstrate/.style={decorate, 
line width=1pt,
draw=olive, 
decoration=snake,
segment length=5pt,
segment aspect=0,
segment amplitude=0.5mm, 
line after snake=0mm,
line before snake=0mm
},
substrate/.style={decorate, 
line width=1pt,
draw=olive, 
decoration=snake, 
segment length=5pt,
segment amplitude=0.5mm, 
line after snake=0.5mm,
line before snake=0.5mm
},
activity/.style={very thick,draw=red,postaction={decorate},
decoration={markings,mark=at position .5 with
{\arrow[draw=red]{>}}}},
tactivity/.style={thick,draw=red,postaction={decorate},
decoration={markings,mark=at position .5 with
{\arrow[draw=red]{>}}}},
tEPSactivity/.style={thick,draw=red,postaction={decorate},
decoration={markings,mark=at position .55 with
{\arrow[draw=red]{>}}}},
tAactivity/.style={thick,draw=red},
Aactivity/.style={very thick,draw=red},
Bactivity/.style={very thick,draw=blue,dashed},
Baractivity/.style={very thick,draw=black},
Cactivity/.style={very thick,draw=cyan,snake},
tSactivity/.style={thick,draw=red,postaction={decorate},
decoration={markings,mark=at position .7 with
{\arrow[draw=red]{>}}}},
Sactivity/.style={very thick,draw=red,postaction={decorate},
decoration={markings,mark=at position .7 with
{\arrow[draw=red]{>}}}},
ABPactivity/.style={very thick,draw=red,decorate,decoration={coil,segment length=6pt}},
Arrowactivity/.style={decoration={markings,mark=at position 1 with
    {\arrow[scale=1.6,>=stealth]{>}}},postaction={decorate}},
triangle/.style = {fill=blue!10, regular polygon, regular polygon sides=3,minimum size=3mm},
    node rotated/.style = {rotate=180},
    border rotated/.style = {shape border rotate=180},}

\newcommand{\bareprop}[2]{\tikz[baseline=-2.5pt]{
\draw[Aactivity] (180:0.6) -- (0,0)  node[at start ,above]{$#1$} ;
\draw[Aactivity] (0:0.6) -- (0,0) node[at start,above] {$#2$};}}

\newcommand{\pertprop}[2]{\tikz[baseline=-2.5pt]{
\draw[Aactivity] (180:0.1) -- (0,0)  node[at start ,above]{$#1$} ;
\draw[Bactivity] (270:0.5) -- (0,0) ;
\draw[Aactivity] (0:0.1) -- (0,0) node[at start,above] {$#2$};
\draw[Baractivity](-0.2,0.1)--(-0.2,-0.1);
\draw[Baractivity](-0.1,-0.2)--(0.1,-0.2);
\draw [very thick,draw=blue] (0,-0.6) circle (0.1);}}

\newcommand{\pertpropZHIJIAO}[2]{\tikz[baseline=-2.5pt]{
\draw[Aactivity] (180:0.1) -- (0,0)  node[at start ,above]{$#1$} ;
\draw[Bactivity] (270:0.5) -- (0,0) ;
\draw[Aactivity] (0:0) -- (0,0) node[at start,above] {$#2$};
\draw[Baractivity](-0.2,0.1)--(-0.2,-0.1);
\draw[Baractivity](-0.1,-0.2)--(0.1,-0.2);
\draw [very thick,draw=blue] (0,-0.6) circle (0.1);}}

\newcommand{\singlepertpropPot}[3]{\tikz[baseline=-2.5pt]{
\draw[Aactivity] (180:0.4) -- (0,0)  node[at start ,above]{$#1$} ;
\draw[Bactivity] (270:0.5) -- (0,0) ;
\draw[Aactivity] (0:0.4) -- (0,0) node[at start,above] {$#2$};
\draw[Baractivity](-0.2,0.1)--(-0.2,-0.1);
\draw[Baractivity](-0.1,-0.2)--(0.1,-0.2);
\draw [very thick,draw=blue] (0,-0.6) circle (0.1) node[right] {$#3$}; }}

\newcommand{\barepropX}[4]{\tikz[baseline=-2.5pt]{
\draw[Aactivity]   (-0.8,0) --(0,0)   node[at start, above] {$#1$}  node[at start, below] {$#2$};
\draw[Aactivity] (0.8,0) -- (0,0) node[at start,above] {$#3$}
node[at start, below] {$#4$};
}}

\newcommand{\barepropS}[4]{\tikz[baseline=-2.5pt]{
\draw[Aactivity]   (-0.3,0) --(0,0)   node[at start, above] {$#1$}  node[at start, below] {$#2$};
\draw[Aactivity] (0.3,0) -- (0,0) node[at start,above] {$#3$}
node[at start, below] {$#4$};
}}

\newcommand{\ABPpropX}[4]{\tikz[baseline=-2.5pt]{
\draw[ABPactivity]   (-0.8,0) --(0,0)   node[at start, above] {$#1$}  node[at start, below] {$#2$};
\draw[ABPactivity] (0.8,0) -- (0,0) node[at start,above] {$#3$}
node[at start, below] {$#4$};
\draw [Arrowactivity] (0,0) -- (-0.1,0); 
}}

\newcommand{\ABPpropS}[4]{\tikz[baseline=-2.5pt]{
\draw[ABPactivity]   (-0.3,0) --(0,0)   node[at start, above] {$#1$}  node[at start, below] {$#2$};
\draw[ABPactivity] (0.1,0) -- (0,0) node[at start,above] {$#3$}
node[at start, below] {$#4$};
}}

\newcommand{\velpert}{\tikz[baseline=-2.5pt]{
\draw[Aactivity]  (0,0)--(-0.3,0);
\draw[Aactivity]  (0,0)--(0.3,0) ;
\draw[red,fill=red] (0,0) circle (1mm);
\draw[Aactivity]  (-0.2,-0.1)--(-0.2,0.1) ;
}}

\newcommand{\velpertX}[2]{
\tikz[baseline=-2.5pt]{
\draw[Aactivity]  (0,0)--(180:0.6)  node[at end ,below]{$#1$};
\draw[Aactivity]  (0,0)--(0:0.6)  node[at end ,below]{$#2$};
\draw[Aactivity]  (-0.2,-0.1)--(-0.2,0.1);
\draw[red,fill=red] (0,0) circle (1mm);
}}

\newcommand{\Liouville}{\LC}
\newcommand{\driftVec}{\gpvec{w}}
\newcommand{\probab}{P}
\newcommand{\extPot}{\Upsilon}
\newcommand{\mass}{r}
\newcommand{\barepropG}[2]{G\left(#1,#2\right)}
\newcommand{\barepropGwPert}[4]{H_{#3}^{#4}\left(#1,#2\right)}
\newcommand{\barepropGMathieu}[4]{G_{#3}^{#4}\left(#1,#2\right)}
\newcommand{\MathieuX}[5]{#1_{#2}^{#3}\left(#4,#5\right)}
\newcommand{\MMprojectionX}[4]{#1_{#2}^{#3}(#4)}
\makeatletter
\newcommand{\MathieuA}[3]{A_{#1}^{#2}\@ifempty{#3}{}{(#3)}}
\newcommand{\MathieuANought}[3]{\MathieuA{#1,0}{#2}{#3}}
\newcommand{\SphericalHarmonics}[3]{Y_{#1}^{#2}\@ifempty{#3}{}{(#3)}}
\newcommand{\Legendre}[3]{P_{#1}^{#2}\@ifempty{#3}{}{(#3)}}
\newcommand{\MathieuLambda}[3]{\lambda_{#1}^{#2}\@ifempty{#3}{}{(#3)}}
\makeatother
\newcommand{\Mathieu}[5]{\MathieuX{u}{#2}{#3}{#4}{#5}}
\newcommand{\Mathieutilde}[5]{\MathieuX{\widetilde{u}}{#2}{#3}{#4}{#5}}
\newcommand{\MMprojection}[4]{\MMprojectionX{\Delta}{#2}{#3}{#4}}

\newcommand{\Appref}[1]{App.~\ref{sec:#1}}

\newcommand{\order}[1]{\OC\!\left(#1\right)}

\newcommand{\drift}{w}
\newcommand{\diffusion}{\mathsf{D}}
\newcommand{\rotDiffusion}{\diffusion_r}
\newcommand{\transDiffusion}{\diffusion_t}
\newcommand{\effDiffusion}[1]{\diffusion^{\text{#1}}_{\text{eff}}}
\newcommand{\density}{\rho}
\newcommand{\sysL}{L}

\begin{document}

\preprint{}

\title{Field Theory of Active Brownian Particles in Potentials}

\author{Ziluo Zhang}
\affiliation{%
Department of Mathematics and Centre of Complexity Science, 180 Queen's Gate, Imperial College London, London SW7 2AZ, UK
}%
\author{Lili Fehértói-Nagy}
\affiliation{%
Department of Physics, Exhibition Road, Imperial College London, London SW7 2AZ, UK
}%
\author{Maria Polackova}
\affiliation{%
Department of Physics, Exhibition Road, Imperial College London, London SW7 2AZ, UK
}%
\author{Gunnar Pruessner}
 \email{g.pruessner@imperial.ac.uk}
 \homepage{https://www.imperial.ac.uk/non-equilibrium-systems/}
\affiliation{%
Department of Mathematics and Centre of Complexity Science, 180 Queen's Gate, Imperial College London, London SW7 2AZ, UK
}%


\date{\today}

\begin{abstract}
The Active Brownian Particle (ABP) model exemplifies a wide class of active matter particles. In this work, we demonstrate how this model can be cast into a field theory in both two and three dimensions. Our aim is manifold: we wish both to extract useful features of the system, as well as to build a framework which can be used to study more complex systems involving ABPs, such as those involving interaction. Using the two-dimensional model as a template, we calculate the mean squared displacement exactly, and the one-point density in an external potential perturbatively. We show how the effective diffusion constant appears in the barometric density formula to leading order, and determine the corrections to it. We repeat the calculation in three dimensions, clearly a more challenging setup. Comparing different ways to capture the self-propulsion, we find that its perturbative treatment results in more tractable derivations without loss of exactness, where this is accessible.





\end{abstract}

\maketitle


\section{Introduction}
Active Brownian particles (ABPs) \cite{Howse:2007} 
are one of the paradigmatic models of active matter \cite{Bechinger:2016}. 
Such particles move with a constant self-propulsion \emph{speed} $\drift_0$, in a time-varying direction, represented by a vector subject to rotational diffusion with constant $\rotDiffusion$. At the same time, the particles are subject to thermal diffusion with constant $\transDiffusion$ \cite{Cates:2013, Cates:2015}.
As opposed to Run-and-Tumble (RnT) motion \cite{Cates:2013}, both degrees of freedom, the vectorial velocity $\driftVec(t)$ as well as the position $\rvec(t)$, are continuous functions of time $t$.
ABPs approximate so-called Janus particles particularly well \cite{Howse:2007,Bechinger:2016}. Similarly, \emph{monotrichous} bacteria vary their orientation more smoothly and are much more reliant on rotational diffusion \cite{LiTamTang:2008}. In contrast, the motion of the much-studied \emph{E.~coli} is closer to RnT.


%
%
The motion of ABPs is most directly described by a pair of Langevin equations for their position $\rvec(t)$ as a function of time $t$,
\begin{subequations}
\begin{align}
    \partial_t\rvec(t)&=\driftVec(t) + \xivec(t)
    &&\quad\text{with}\quad&
    \ave{\xivec(t)\cdot\xivec(t')} &= 2d\transDiffusion\delta(t'-t)\\
    \elabel{Langevin_rot}
    \partial_t\driftVec(t)&=\zetavec_{\perp\driftVec(t)}(t)
    &&\quad\text{with}\quad&
    \ave{\zetavec_{\perp\driftVec(t)}(t)\cdot\zetavec_{\perp\driftVec(t')}(t')} &= 2(d-1)\drift_0^2\rotDiffusion\delta(t'-t)
\end{align}
\end{subequations}
where the angular brackets $\ave{\bullet}$ indicate an ensemble average. Both $\xivec(t)$ and $\zetavec_{\perp\driftVec(t)}(t)$ represent white noise with vanishing mean and correlators as stated. 
The translational noise $\xivec(t)$ is the commonly used Gaussian displacement in $d$ dimensions.
The rotational noise $\zetavec_{\perp\driftVec(t)}(t)$ is confined to the $(d-1)$-dimensional tangential plane orthogonal to $\driftVec(t)$ at any time
\cite{Carlsson:2010}, so that $\driftVec(t)$ stays on the surface of a sphere of radius $\drift_0$ at any time. This is guaranteed by $\ave{\driftVec(t)\cdot\zetavec_{\perp\driftVec(t)}(t)}=0$,
which requires an Ito interpretation of \Eref{Langevin_rot}. In \Sref{ABPintro} we cast the above in Fokker-Planck form, avoiding any such issues of stochastic calculus.

The motion of ABPs in two dimensions has been analysed in various ways using classical methods \cite{FrankeGruler:1990,Cates:2013,Cates:2015,Bechinger:2016}, while ABPs in three dimensions have received much less attention. This might be related to the complications arising from the curvature of the surface the rotational diffusion takes place on. The aim of the present work is to develop the mathematical toolkit that will allow us to study ABPs using field theory. Some of us have studied ABPs in two dimensions before \cite{ZhangPruessner:2022}. We will now study ABPs in two dimensions from a different perspective, primarily to provide us with a template for the treatment in three dimensions. 

Studying ABPs in two and three dimensions through field theory will not only give direct insight about ABPs --- developing and verifying this framework will also open the door to add interaction. We take a first step in this work by considering external potentials.

There are, in principle, always two ways of describing single-particle dynamics using field theory. Either, one can use a suitable eigensystem in which to expand the fields, or proceed by considering the motion as a perturbation about pure diffusion. The former approach is, in theory, more powerful, as the particle dynamics is fully contained in a set of eigenfunctions, whose eigenvalues determine the ``decay" rates. However, a combination of the two approaches is sometimes called for, particularly when there are multiple intertwined degrees of freedom, not all of which can be included in an appropriate eigensystem. The motion of Run-and-Tumble particles in a harmonic potential is an example of such a system \cite{Garcia-MillanPruessner:2021}. In the present case of ABPs, the choice seems obvious in two dimensions, as, after suitable simplifications, we encounter the Mathieu equation. However, in three dimensions, no such simplifications are available. This necessitates exploring new methods, in particular determining the eigenfunctions themselves perturbatively, or treating self-propulsion as a perturbation about pure diffusion. This latter method is first tried out in the easier case of two dimensions.

The complexity of the methods increases further when we allow for an external potential, which must be treated perturbatively. As a test bed for the methods, we return to ABPs in two dimensions~\cite{ZhangPruessner:2022}, before moving on to three dimensions. In both cases, we determine the mean squared displacement (MSD) in closed form, as well as the stationary density in the presence of an external potential. These two observables are related: the MSD naturally gives rise to the notion of an \emph{effective diffusion constant}, which may or may not feature as the effective temperature in a Boltzmann-like distribution of particles in the stationary state. 

In the following, we introduce the ABP model, first through a Fokker-Planck description, and subsequently in the form of an action. In \Sref{ABP2D}, we proceed to characterise the MSD and stationary distribution in two dimensions. We do this first using Mathieu functions, and then perturbatively. These methods and results serve as the template for the treatment of ABPs in three dimensions in \Sref{ABP3D}.

\subsection{Active Brownian Particles}
\label{sec:ABPintro}
Active Brownian Particles (ABP) are particles that move by self-propulsion by a constant speed $\drift_0$ as as well as due to thermal noise characterised by translational diffusion constant $\transDiffusion$. The direction of the self-propulsion is itself subject to diffusion. In the following, we will generally denote this direction by $\Omega$, and more specifically in the plane the polar angle by $\varphi\in[0,2\pi)$ and in spherical coordinates the azimuthal angle by $\varphi\in[0,2\pi)$ and the polar angle by $\theta\in[0,\pi)$. The direction of the self-propulsion is then 
\begin{equation}\elabel{def_driftVecOmega}
	\driftVec(\Omega) = 
	\left(
	\begin{array}{c}
	\drift_x\\
	\drift_y
	\end{array}
	\right)
	=\drift_0
	\left(
	\begin{array}{c}
	\cos\varphi\\
	\sin\varphi
	\end{array}
	\right)
	\quad\text{ and }\quad
	\driftVec(\Omega) = 
	\left(
	\begin{array}{c}
	\drift_x\\
	\drift_y\\
	\drift_z
	\end{array}
	\right)
	=\drift_0
	\left(
	\begin{array}{c}
	\sin\theta\cos\varphi\\
	\sin\theta\sin\varphi\\
	\cos\theta
	\end{array}
	\right)
\end{equation}
in two and three dimensions respectively. The rotational diffusion is parameterised by the constant $\rotDiffusion$, specifically by a Fokker-Planck equation of the form $\partial_t \probab(\Omega,t)=\rotDiffusion\nabla^2_\Omega \probab(\Omega,t)$ with 
\begin{equation}\elabel{def_nabla2Omega}
\nabla^2_\Omega=\partial_\varphi^2
	\quad\text{ and }\quad
\nabla^2_\Omega=
\frac{1}{\sin \theta}\partial_\theta\sin\!\theta\ \partial_\theta + 
\frac{1}{\sin^2\! \theta}\partial_\varphi^2
\end{equation}
in two and three dimensions respectively. 

The Fokker-Planck equation of active Brownian particles is 
\begin{equation}\elabel{FPE}
	\partial_t \probab(\rvec,\Omega,t) = \Liouville\probab(\rvec,\Omega,t)
	\quad\text{with}\quad
	\Liouville=
	\transDiffusion\nabla_\rvec^2 
	- \driftVec(\Omega)\cdot\nabla_\rvec
	+ \nabla_\rvec\cdot\Big(\nabla_\rvec\extPot(\rvec)\Big)
	+ \rotDiffusion\nabla^2_\Omega
	\end{equation}
where $\probab(\rvec,\Omega,t)$ is the probability to find a particle after some initialization at time $t$ at position $\rvec$ with the director pointing to $\Omega$ and $\nabla_\rvec$ the usual spatial gradient as opposed to $\nabla_\Omega$ acting on the director, \Eref{def_nabla2Omega}. The self-propulsion is implemented by the term $\driftVec\cdot\nabla_\rvec$, which is the only term mixing the degrees of freedom $\Omega$ and $\rvec$. The effect of an additional external potential $\extPot(\rvec)$ which acts equally on particles without orientation is implemented by $\nabla_\rvec\cdot(\nabla_\rvec\extPot)$, which is structurally identical to that of a drift $\driftVec$ and reduces to that term when $\nabla_\rvec\extPot$ is constant in space. We write $\extPot'$ for $\nabla_\rvec\extPot$ in the following. When the operator $\nabla_\rvec\cdot\extPot'$ acts on $\probab$ in \Eref{FPE}, only the leftmost spatial gradient $\nabla_\rvec$ acts on the product of $\extPot'$ and $\probab$, whereas $\nabla_\rvec\extPot$ results solely in $\extPot'$.
For ease of notation below, we further introduce 
\begin{subequations}
\elabel{def_Liouvilles}
\begin{align}\elabel{def_Liouville1}
	\Liouville_1&=\transDiffusion\nabla_\rvec^2- \driftVec(\Omega)\cdot\nabla_\rvec + \rotDiffusion\nabla^2_\Omega\\
	\elabel{def_Liouville2}
	\Liouville_2&=\transDiffusion\nabla_\rvec^2 + \rotDiffusion \nabla^2_\Omega \ .
\end{align}
\end{subequations}

From the single-particle Fokker-Planck \Eref{FPE} the multiple, non-interacting particle action 
\begin{equation}\elabel{def_action_general}
	\action[\chitilde,\chi]=\int\ddint{r}\int\dint{\Omega}\int\dint{t}
    \chitilde(\rvec,\Omega,t)
    \big(
    -\partial_t 
	+ \Liouville 
	- \mass
    \big) 
    \chi(\rvec,\Omega,t)
\end{equation}
follows immediately \cite{PruessnerGarcia-Millan:2022}, with an additional mass $\mass\downarrow0$ as a regulator. At this stage, we leave the dimension $d\in\{2,3\}$ and the integral $\int\dint{\Omega}$ unspecified. Expectations  of the annihilator field $\chi(\rvec,\Omega,t)$ and the creator field $\chitilde(\rvec,\Omega,t)$ can be calculated in a path integral\cite{PruessnerGarcia-Millan:2022}
\begin{equation}
	\ave{\bullet} = \int\Dint{\chi}\Dint{\chitilde} \exp{\action[\chitilde,\chi]}\bullet \ =\ave{\bullet\exp{-\action_P[\chitilde,\chi]}}_0 \quad\text{with}\quad \ave{\bullet}_0=\int\Dint{\chitilde}\Dint{\chi}\bullet\exp{-\action_0[\chitilde,\chi]}\ ,
\end{equation}
which allows for a perturbative treatment by splitting the action into a harmonic part $\action_0$, whose path-integral is readily taken, and a perturbative part $\action_P$, whose exponential is expanded,
\begin{equation}\elabel{perturbative_action_use}
    \ave{\bullet} = \int\Dint{\chi}\Dint{\chitilde} \exp{\action_0[\chitilde,\chi]}\bullet\sum_{n=0}^\infty\frac{\action_P^n}{n!} 
    = \ave{\bullet\sum_{n=0}^\infty\frac{\action_P^n}{n!}}_0 \ ,
\end{equation}
where we have introduced $\ave{\bullet}_0$ as an average on the basis of the harmonic part of the action only. 
Which part of the action \Eref{def_action_general} is treated perturbatively depends crucially on the eigensystem the fields $\chi$ and $\chitilde$ are written in, as it determines which part of the operator becomes diagonal. Because the external potential is arbitrary, it cannot be diagonalised in its present generality. 

We may thus consider two schemes: In the first, the fields are written in eigenfunctions of $\Liouville_1$, \Eref{def_Liouvilles}, with the harmonic and perturbative part of the action
\begin{subequations}
\elabel{def_action1}
\begin{align}\elabel{def_action01}
	\action_{01}&=\int\ddint{r}\int\dint{\Omega}\int\dint{t}
    \chitilde(\rvec,\Omega,t)
    \big(
    -\partial_t 
	+ \Liouville_1
	- \mass
    \big) 
    \chi(\rvec,\Omega,t)\\
    \elabel{def_actionP1}
	\action_{P1}&=\int\ddint{r}\int\dint{\Omega}\int\dint{t}
    \chitilde(\rvec,\Omega,t)
	\nabla_\rvec\cdot\big(\nabla_\rvec\extPot(\rvec)\big)
    \chi(\rvec,\Omega,t) \ ,
\end{align} 
\end{subequations}
which is more efficiently dealt with using Gauss' theorem in $\action_{P1}$ to make $\nabla_\rvec$ act on $\chitilde$ rather than $\extPot'\chi$.

The benefit of 
including the self-propulsion in the harmonic part of the action $\action_{01}$, \Eref{def_action01}, via $\Liouville_1$, \Eref{def_Liouville1}, at the expense of having to find suitable eigenfunctions,
is to have only one perturbative vertex to worry about, namely that representing the external potential. The major disadvantage is that the diagonalising effect of the eigenfunctions on $\Liouville_1$ relies on momentum conservation, which is notably broken in the presence of an external potential, so that the external potential becomes significantly more difficult to handle. A similar problem would be encountered if we were to allow for pair interaction, in which case the action contains terms that are not bilinear, which result in difficult integrals involving products of more than two eigenfunctions. 

The eigenfunctions of the angular part of $\Liouville_1$ in two spatial dimensions, after some manipulation, are the $\pi$-periodic Mathieu functions \cite{ZIENER2012}, as discussed further below. Compared to other orthonormal systems, they are far less well studied \cite{abramowitzStegun}. In three dimensions, we shall call them "three-dimensional Mathieu functions", but as far as we are aware, they have not been introduced in the literature of orthonormal functions in their own right. We have characterised them to some extent in \Appref{app_3DMathieu} using similar tools as outlined in \cite{ZIENER2012}. Of course, the complete framework of Sturm-Liouville theory provides broad theoretical backing. Once their existence is established, the eigenfunctions need to be determined only to the extent that the observables require it.

In the second scheme, the fields are written in terms of eigenfunctions of the simpler $\Liouville_2$ in \Eref{def_Liouvilles}, with the harmonic and perturbative part of the action
\begin{subequations}
\begin{align}\elabel{def_action02}
	\action_{02}&=\int\ddint{r}\int\dint{\Omega}\int\dint{t}
    \chitilde(\rvec,\Omega,t)
    \big(
    -\partial_t 
	+ \Liouville_2
	- \mass
    \big) 
    \chi(\rvec,\Omega,t)\\
    \elabel{def_actionP2}
	\action_{P2}&=\int\ddint{r}\int\dint{\Omega}\int\dint{t}
    \chitilde(\rvec,\Omega,t)
	\nabla_\rvec\cdot\big(\nabla_\rvec\extPot(\rvec)-\driftVec(\Omega)\big)
    \chi(\rvec,\Omega,t).
	\end{align}     
\end{subequations}
This results in simpler expressions overall, but suffers from the significant disadvantage of an additional vertex mediating the self-propulsion. One may expect a significantly enlarged number of diagrams to be considered in any interesting observable, but this is not the case for the observables considered here.

We proceed by determining the MSD and the stationary density for ABPs in two dimensions. In the course, we will discuss the details of the action, the diagrammatics and the eigenfunctions.

\section{Active Brownian Particles in $2$ dimensions}\label{sec:ABP2D}
\subsection{Mathieu functions}\label{sec:ABP2D_mathieu}
The starting point of the present derivation is 
the operator $\Liouville_1$ in \Eref{def_Liouville1}
and
the action \Eref{def_action01}, restated here with the explicit parameterisation of the director
\begin{equation}\elabel{action_2DABP}
    \action_{01}[\chitilde,\chi]=\int\dTWOint{r}\int_0^{2\pi}\dint{\varphi}\int\dint{t}
    \chitilde(\rvec,\varphi,t)
    \big(
    -\partial_t + \Liouville_1 - \mass
    \big) 
    \chi(\rvec,\varphi,t)
\end{equation}
for convenience. The key difficulty here is to find eigenfunctions of the operator $-\driftVec\cdot\nabla_\rvec+\nabla_\Omega^2$. This is greatly helped by measuring the azimuthal angle relative to the reciprocal vector $\kvec$, as we will discuss in the following. We first introduce
\begin{subequations}
\elabel{orthoSystem_2DABP}
\begin{align}
    \chi(\rvec,\varphi,t) &= \int\dintbar{\omega}\dTWOintbar{k}\sum_{\ell=0}^\infty
    \exp{-\imag\omega t}\exp{\imag\kvec\cdot\rvec}
    \Mathieu{2}{\ell}{}{\frac{\varphi-\sigma(\kvec)}{2}}{q(|\kvec|)}
    \chi_\ell(\kvec,\omega)\\
    \chitilde(\rvec,\varphi,t) &= \int\dintbar{\omega}\dTWOintbar{k}\sum_{\ell=0}^\infty
    \exp{-\imag\omega t}\exp{\imag\kvec\cdot\rvec}
    \Mathieutilde{2}{\ell}{}{\frac{\varphi-\sigma(-\kvec)}{2}}{q(|-\kvec|)}
    \chitilde_\ell(\kvec,\omega)
\end{align}
\end{subequations}
where 
$\Mathieu{2}{\ell}{}{\gamma}{q(k)}$
and
$\Mathieutilde{2}{\ell}{}{\gamma}{q(k)}$
%
%
%
are suitably normalised $\pi$-periodic Mathieu functions
\cite{ZIENER2012,abramowitzStegun}, depending on the dimensionless, purely imaginary parameter 
\begin{equation}\elabel{def_q}
q(k)=\frac{2\imag\drift_0k}{\rotDiffusion}
\end{equation}
which is a function of the (absolute) magnitude $k=|\kvec|=|-\kvec|$ of the $\kvec$ vector.
The function $\sigma(\kvec)$, hitherto undetermined, allows a crucial simplification of the action further below. For the time being it suffices to state orthogonality even in the presence of the shift by $\sigma$,
\begin{equation}\elabel{orthonormality_Mathieu2D}
    \deltabar(\kvec+\kvec')
    \int_0^{2\pi}\dint{\varphi}
    \Mathieu{2}{\ell}{}{\frac{\varphi-\sigma(\kvec)}{2}}{q(|\kvec|)}
    \Mathieutilde{2}{\ell'}{}{\frac{\varphi-\sigma(-\kvec')}{2}}{q(|-\kvec'|)}
    =
    \deltabar(\kvec+\kvec')\delta_{\ell,\ell'}
\end{equation}
as the $\deltabar(\kvec+\kvec')$ forces this shift to be identical, $\sigma(\kvec)=\sigma(-\kvec')$, which can therefore be absorbed into the dummy variable and subsequently into a change of the integration limits, from where it disappears, because of the periodicity of the integrand. 

Using \Eref{orthoSystem_2DABP} in the action \Eref{action_2DABP} produces
\begin{multline}\elabel{action_2DABP_rewritten}
    \action_{01}[\chitilde,\chi]=
    -\int\dTWOintbar{k}
    \sum_{\ell,\ell'}
    \int\dintbar{\omega}
    \chitilde_{\ell'}(-\kvec,-\omega)
    \chi_\ell(\kvec,\omega)\\
    \times
    \int_0^{2\pi}\dint{\varphi}
    \utilde_{\ell'}\left(\frac{\varphi-\sigma(\kvec)}{2},q(k)\right)
    \big(
    -\imag\omega + \transDiffusion \kvec^2 + \imag \drift_0 k
    \cos(\varphi-\alpha)
    - \rotDiffusion \partial_\varphi^2
    \big) 
    u_\ell\left(\frac{\varphi-\sigma(\kvec)}{2},q(k)\right) \ ,
\end{multline}
where $\alpha$ is the polar angle of $\kvec=k(\cos\alpha,\sin\alpha)^\transpose$, so that $\kvec\cdot\driftVec(\varphi)=k\drift(\cos\alpha\cos\varphi+\sin\alpha\sin\varphi)=k\drift\cos(\varphi-\alpha)$.
Choosing $\sigma(\kvec)=\alpha$ allows a change of variables and a change of integration limits, so that the last integral in \Eref{action_2DABP_rewritten} reads
\begin{subequations}
\elabel{alpha_simplification}
\begin{align}
&\phantom{=}
\int_0^{2\pi}\dint{\varphi}
    \utilde_{\ell'}\left(\frac{\varphi-\sigma(\kvec)}{2},q(k)\right)
    \bigg(
    -\imag\omega + \transDiffusion \kvec^2 + \imag \drift_0 k
    \cos(\varphi-\alpha)
    - \rotDiffusion \partial_\varphi^2
    + \mass
    \bigg) 
    u_\ell\left(\frac{\varphi-\sigma(\kvec)}{2},q(k)\right)\\
    &=2\int_0^{\pi}\dint{\gamma}
    \utilde_{\ell'}\left(\gamma,q\right)
    \bigg(
    -\imag\omega + \transDiffusion \kvec^2 + \quarter \rotDiffusion(2 q
    \cos(2\gamma)
    - \partial_\gamma^2)
    + \mass
    \bigg) 
    u_\ell\left(\gamma,q\right)\\
    &=
    \delta_{\ell',\ell}
    \bigg(
    -\imag\omega + \transDiffusion \kvec^2 + \quarter \rotDiffusion\lambda_\ell(q)
    +\mass
    \bigg) 
\end{align}
\end{subequations}
where in the last equality we have used 
that the Mathieu functions obey 
\begin{equation}\elabel{Mathieu_eqn}
    \left(\partial_\gamma^2 - 2 q \cos(2\gamma)
    \right)
    u_\ell(\gamma,q)=-\lambda_\ell(q) u_\ell(\gamma,q)
\end{equation}
and the orthonormality  \Eref{orthonormality_Mathieu2D}.

With this simplification the action \Eref{action_2DABP}  becomes diagonal
\begin{equation}\elabel{action_2DABP_final}
    \action_{01}[\chitilde,\chi]=
    - \int\dTWOintbar{k}
    \sum_{\ell}
    \int\dintbar{\omega}
    \chitilde_{\ell}(-\kvec,-\omega)
    \chi_\ell(\kvec,\omega)
    \bigg(
    -\imag \omega + \transDiffusion\kvec^2 + \rotDiffusion \lambda_\ell(q(k))/4 + \mass
    \bigg)
    \ ,
\end{equation}
and the (bare) propagator can immediately be read off
\begin{subequations}
\elabel{2DABP_propagator}
\begin{align}
    \ave{
    \chi_{\ell}(\kvec,\omega)
    \chitilde_{\ell_0}(\kvec_0,\omega_0)
    }_0
    =
    \frac{\delta_{\ell,\ell_0}\deltabar(\kvec + \kvec_0)\deltabar(\omega + \omega_0) }{-\imag \omega + \transDiffusion\kvec^2 + \rotDiffusion \lambda_\ell(q(k))/4 + \mass}
    = 
    \delta_{\ell,\ell_0} \deltabar(\kvec + \kvec_0)\deltabar(\omega + \omega_0)
    \barepropGMathieu{\kvec}{\omega}{\ell}{}\\
    \corresponds
    \ABPpropX{\kvec,\omega}{\ell}{\kvec_0,\omega_0}{\ell_0}
    \ ,
\end{align}
\end{subequations}
where we have introduced $\barepropGMathieu{\kvec}{\omega}{\ell}{}$ as a shorthand, as well as the diagrammatic representation to be used in \Sref{2DABP_Mathieu_extPot}.


\subsubsection{Mean squared displacement}\label{sec:2DABP_Mathieu_MSD}
The mean-squared displacement (MSD)  $\spave{\rvec^2}(t)$ of a two-dimensional ABP (2DABP) can be calculated on the basis of a few algebraic properties of the $u_\ell(\gamma,q)$. The probability of finding a 2DABP with director $\varphi$ at position $\rvec$ at time $t$, after being placed at $\rvec_0$ with director $\varphi_0$ at time $t_0$ is $\ave{\chi(\rvec,\varphi,t)\chitilde(\rvec_0,\varphi_0,t_0)}$, which is easily determined from \Eref{2DABP_propagator} as
\begin{equation}
\ave{\chi(\kvec,\varphi,t)\chitilde(\kvec_0,\varphi_0,t_0)}
= \deltabar(\kvec+\kvec_0)
    \sum_{\ell=0}^\infty 
    u_\ell\left(\frac{\varphi-\sigma(\kvec)}{2},q(k)\right)
    \utilde_\ell\left(\frac{\varphi_0-\sigma(\kvec)}{2},q(k)\right)
    \exp{-(\transDiffusion k^2 + \quarter \rotDiffusion \lambda_\ell(q(k))t}
\end{equation}
up to an inverse of Fourier-transform of $\kvec$.
Making use of the Fourier transform, the MSD of a 2DABP irrespective of the final director is
\begin{equation}\elabel{MSDv1}
    \spave{\rvec^2}(t) = -\left.\nabla_{\kvec}^2\right|_{\kvec=\nullvec} \int_0^{2\pi}\dint{\varphi}
    \sum_{\ell=0}^\infty 
    u_\ell\left(\frac{\varphi-\sigma(\kvec)}{2},q(k)\right)
    \utilde_\ell\left(\frac{\varphi_0-\sigma(\kvec)}{2},q(k)\right)
    \exp{-(\transDiffusion k^2 + \quarter \rotDiffusion \lambda_\ell(q(k))t}
\end{equation}
Having integrated over $\varphi$, it is obvious that the MSD cannot depend on the initial orientation of the director, $\varphi_0$, as changing it, amounts to a rotation of the isotropic plane. This is indeed trivial to show when $\drift_0=0$, in which case $q(k)=0$ and the Mathieu functions degenerate into trigonometric functions. However, demonstrating this when $\drift_0\ne0$ is far from trivial.
We proceed by rewriting the MSD as an average 
$\int\dint{\varphi_0}(2\pi)^{-1}$
of \Eref{MSDv1} over the initial angle $\varphi_0$,
so that both Mathieu functions, $u_\ell$ and $\utilde_\ell$ reduce to a single coefficient via 
\begin{equation}\elabel{MathieuIntegral}
    \int_0^{2\pi} \dint{\varphi} 
    u_\ell(\varphi/2,q(k))=
    2 \pi 
    \MathieuANought{\ell}{}{q(k)}
    \ ,
\end{equation}
which vanishes in fact for all odd indices $\ell$. We thus arrive at
\begin{align}
    \spave{\rvec^2}(t) 
\elabel{MSDv3}
    &= -\left.\nabla_{\kvec}^2\right|_{\kvec=\nullvec} 
    2 \exp{-\transDiffusion k^2t} \sum_{\ell=0}^\infty
    \left(
    \MathieuANought{\ell}{}{q(k)}
    \right)^2 \exp{-\quarter \rotDiffusion \lambda_\ell(q(k)t}
\end{align}

The $u_\ell(\gamma,q)$ are in fact the even $\pi$-periodic Mathieu functions for even $\ell$ and odd $\pi$-periodic Mathieu functions for odd $\ell$ \cite{ZIENER2012}.
The $\Mathieutilde{2}{\ell}{}{\gamma}{q}$ differ from the $\Mathieu{2}{\ell}{}{\gamma}{q}$ only by a factor $1/\pi$ \cite{ZhangPruessner:2022},
\begin{equation}\elabel{def_Mathieu2Tilde}
    \Mathieutilde{2}{\ell}{}{\gamma}{q}=\pi^{-1} \Mathieu{2}{\ell}{}{\gamma}{q} \ ,
\end{equation}
which is a convenient way to meet the orthogonality relation \Eref{orthonormality_Mathieu2D} while maintaining that the non-tilde $\Mathieu{2}{\ell}{}{\gamma}{q}$ are standard Mathieu functions.
At this stage, all we need to know about the Mathieu functions is \cite{ZIENER2012}
\begin{subequations}
\elabel{Mathieu_properties}
\begin{align}
    \MathieuANought{0}{}{q} & = \frac{1}{\sqrt{2}}  - \frac{q^2}{16\sqrt{2}} + \order{q^3}\\
    \MathieuANought{2}{}{q}& = \frac{q}{4} + \order{q^2}\\
    \MathieuANought{2j}{}{q} & = \order{q^j} \\
	\MathieuANought{2j+1}{}{q} &= 0\\
    \lambda_0 & = \order{q^2}\\
    \lambda_2 & = 4 + \order{q^2}
\end{align}
\end{subequations}
for $j\in\Nset_0$, 
which means that only two of the $\MathieuANought{\ell}{}{}$ in the sum \Eref{MSDv3} contribute, namely $\ell=0$ and $\ell=2$, as all other $\MathieuANought{\ell}{}{}$ vanish at $q(k=0)=0$ even when differentiated twice. With \Eref{Mathieu_properties} the MSD becomes
\cite{ZhangPruessner:2022} 
\begin{equation}
    \elabel{MSD_2d}
	\overline{\rvec^2}(t) = 4\transDiffusion t + 2 \frac{\drift_0^2}{\rotDiffusion^2}\left(e^{-\rotDiffusion t} - 1 + \rotDiffusion t\right) \ ,
\end{equation}
identical to that of Run-and-Tumble particles, Eq.~(49) of \cite{ZhangPruessner:2022}, if the tumble rate $\alpha$ is $\rotDiffusion$.
In asymptotically large $t$, the MSD of a 2DABP is equivalent to that of conventional diffusion with effective diffusion constant 
\begin{equation}\elabel{def_effDiffusion}
\effDiffusion{2D} = \transDiffusion + \frac{\drift_0^2}{2\rotDiffusion}
\ ,
\end{equation}
and RnT motion with tumbling rate 
$\alpha = \rotDiffusion$
\cite{Cates:2013, Cates:2015,ZhangPruessner:2022}.

We will use the above as a template for the derivation of the MSD of 3DABP. 
There, we will make use again of the isotropy of the initial condition $\varphi_0$ and further of the isotropy of the displacement in any of the three spatial directions. In fact, picking a "preferred direction" can greatly reduce the difficulty of the present calculation.

\subsubsection{External Potential}\label{sec:2DABP_Mathieu_extPot}
Next, we place the 2DABP in a (two-dimensional) potential $\extPot(\rvec)$. Without self-propulsion, the density $\density_0(\rvec)$ at stationarity is given by the barometric formula $\density_0(\rvec)\propto\exp{-\extPot(\rvec)/\transDiffusion}$. As a sanity check for the present field-theoretic framework, we derive it in \Appref{app-field-theory}. In the present section we want to determine the effect of the self-propulsion, which we expect is partially covered by $\transDiffusion$ in the Boltzmann distribution being replaced by $\effDiffusion{2D}$, \Eref{def_effDiffusion}.

As the external potential is treated perturbatively, stationarity even at vanishing potential requires a finite volume, which we assume is a periodic square, \ie a torus, with linear extent $\sysL$. This requires a small adjustment of the Fourier representation introduced in \Eref{orthoSystem_2DABP}, which now reads
\begin{subequations}
\elabel{orthoSystem_2DABP_finite}
\begin{align}
    \chi(\rvec,\varphi,t) &= \int\dintbar{\omega}\frac{1}{\sysL^2}\sum_{\nvec\in\Zset^2}\sum_{\ell=0}^\infty
    \exp{-\imag\omega t}\exp{\imag\kvec_\nvec\cdot\rvec}
    \Mathieu{2}{\ell}{}{\frac{\varphi-\sigma(\kvec_\nvec)}{2}}{q(|\kvec_\nvec|)}
    \chi_\ell(\kvec_\nvec,\omega)\\
    \chitilde(\rvec,\varphi,t) &= \int\dintbar{\omega}\frac{1}{\sysL^2}\sum_{\nvec\in\Zset^2}\sum_{\ell=0}^\infty
    \exp{-\imag\omega t}\exp{\imag\kvec_\nvec\cdot\rvec}
    \Mathieutilde{2}{\ell}{}{\frac{\varphi-\sigma(-\kvec_\nvec)}{2}}{q(|-\kvec_\nvec|)}
    \chitilde_\ell(\kvec_\nvec,\omega)
\end{align}
\end{subequations}
where $\sysL^{-2}\sum_{\nvec\in\Zset^2}$ has replaced the integration over $\kvec$ in \Eref{orthoSystem_2DABP}, with the couple $\nvec=(n_x,n_y)^\transpose$ running over the entire $\Zset^2$ and $\kvec_{\nvec}=\nvec \sysL/(2\pi)$. The bare propagator \Eref{2DABP_propagator} is adjusted simply by replacing $\deltabar(\kvec+\kvec_0)$ by, say, $\sysL^2\delta_{\nvec+\pvec,\nullvec}$.
With the introduction of Fourier sums, we implement periodic boundary conditions, effectively implementing periodic observables and, below, a periodic potential.


As far as the Mathieu functions are concerned this adjustment to a finite domain has no significant consequences. 
Due to the Mathieu functions, however, the perturbative part of the action $\action_{P1}$ \Eref{def_actionP1} gets significantly more difficult, as the functions are orthogonal only if the $\kvec$-modes match,
\begin{subequations}
\begin{align}\elabel{action_2DABP_pert_X}
	\action_{P1}&=\int\ddint{r}\int\dint{\Omega}\int\dint{t}
    \chitilde(\rvec,\Omega,t)
	\nabla_\rvec\cdot\big(\nabla_\rvec\extPot(\rvec)\big)
    \chi(\rvec,\Omega,t)\\
    \elabel{action_2DABP_pert_k}
    &=\int\dintbar{\omega}
    \frac{1}{\sysL^{6}}\sum_{\nvec_1,\nvec_2,\nvec_3}
    \sysL^2\delta_{\nvec_1+\nvec_2+\nvec_3,\nullvec}
    \sum_{\ell_1,\ell_3}
    (\kvec_{\nvec_1}\cdot\kvec_{\nvec_2}) 
    \chitilde_{\ell_1}(\kvec_{\nvec_1},-\omega)
    \extPot_{\nvec_2}
    \chi_{\ell_3}(\kvec_{\nvec_3},\omega)\\
    &\quad\times\int_0^{2\pi}\dint{\varphi}
    \Mathieu{2}{\ell_3}{}{\frac{\varphi-\sigma(\kvec_{\nvec_3})}{2}}{q(|\kvec_{\nvec_3}|)}
    \Mathieutilde{2}{\ell_1}{}{\frac{\varphi-\sigma(-\kvec_{\nvec_1})}{2}}{q(|\kvec_{\nvec_1}|)}
\end{align}
\end{subequations}
where each $\nvec_1$, $\nvec_2$ and $\nvec_3$ runs over all $\Zset^2$ in the sum and the Kronecker $\delta$-function $\delta_{\nvec_1+\nvec_2+\nvec_3,\nullvec}$ enforces $\nvec_1+\nvec_2+\nvec_3=\nullvec$. Similarly, $\ell_1$ and $\ell_3$ run over all non-negative integers indexing the Mathieu functions. The potential $\extPot(\rvec)$ now enters via its modes $\extPot_{\nvec}$, more explictly
\begin{subequations}\elabel{def_extPot_nvec}
\begin{align}
    \extPot(\rvec)&=\frac{1}{\sysL^2}\sum_{\nvec\in\Zset^2} \extPot_\nvec\exp{\imag\kvec_\nvec\cdot\rvec}\\
    \extPot_\nvec&=\iint 
    \limits_0^{\quad\sysL}
    \dTWOint{r}\extPot(\rvec)\exp{-\imag\kvec_\nvec\cdot\rvec}
    \ ,
\end{align}
\end{subequations}
rendering the potential effectively periodic.

Of course, a non-trivial $\extPot(\rvec)$ spoils translational invariance and correspondingly $\extPot_\nvec$ provides a \emph{source} of momentum in \Eref{action_2DABP_pert_k}, so that $\kvec_{\nvec_3}$ is generally not equal to $-\kvec_{\nvec_1}$, with the sole exception of $\nvec_2=\nullvec$. As a result, generally $\sigma(\kvec_{\nvec_3})\ne\sigma(-\kvec_{\nvec_1})$ 
and 
$q(|\kvec_{\nvec_3}|)\neq q(|\kvec_{\nvec_1}|)$, \Eref{def_q}. The integral 
%
\begin{equation}\elabel{def_MM2}
    \MMprojection{2}{\ell_1,\ell_3}{}{-\kvec_{\nvec_1},-\kvec_{\nvec_3}}
= \int_0^{2\pi}\dint{\varphi}
    \Mathieu{2}{\ell_3}{}{\frac{\varphi-\sigma(\kvec_{\nvec_3})}{2}}{q(|\kvec_{\nvec_3}|)}
    \Mathieutilde{2}{\ell_1}{}{\frac{\varphi-\sigma(-\kvec_{\nvec_1})}{2}}{q(|\kvec_{\nvec_1}|)}
\end{equation}
however, is generally diagonal in $\ell_1,\ell_3$ only when $\kvec_{\nvec_1}=-\kvec_{\nvec_3}$. It is this projection, \Eref{def_MM2}, that is at the heart of the complications to come. It is notably absent in the case of purely diffusive particles, \Appref{barometric_formula}. With the help of \Eref{def_MM2}, the perturbative action \Eref{action_2DABP_pert_k} can be written as 
\begin{equation}\elabel{action_2DABP_pert_k2}
\action_{P1}=\int\dintbar{\omega}
    \frac{1}{\sysL^{6}}\sum_{\nvec_1,\nvec_2,\nvec_3}
    \sysL^2\delta_{\nvec_1+\nvec_2+\nvec_3,\nullvec}
    \sum_{\ell_1,\ell_3}
    (\kvec_{\nvec_1}\cdot\kvec_{\nvec_2}) 
    \chitilde_{\ell_1}(\kvec_{\nvec_1},-\omega)
    \extPot_{\nvec_2}
    \chi_{\ell_3}(\kvec_{\nvec_3},\omega)
	\MMprojection{2}{\ell_1,\ell_3}{}{-\kvec_{\nvec_1},-\kvec_{\nvec_3}} \ .
\end{equation}




The perturbative part of the action, \Eref{action_2DABP_pert_k}, may diagrammatically be written as
\begin{equation}\elabel{extPot_vertex}
\ABPpropS{\kvec,\omega}{\ell}{}{}\!\!\!\!\pertprop{}{}\!\!\!\!\ABPpropS{}{}{\kvec_0,\omega_0}{\ell_0}
\ ,
\end{equation}
where a dash across any line serves as a reminder that the vertex carries a factor of $\kvec$ carried by the leg. Unless there are further singularities in $\kvec$, any such line vanishes at $\kvec=\nullvec$. The two  factors of $\kvec$ are multiplied in an inner product. The dangling bauble in \Eref{extPot_vertex} represents the external potential, which is a source of momentum. 

We will determine in the following the stationary density 
\begin{equation}\elabel{def_density2DMathieu}
    \density_0(\rvec)=\lim_{t_0\to-\infty} 
    \int_0^{2\pi}\dint{\varphi}
    \ave{\chi(\rvec,\varphi,t)\chitilde(\rvec_0,\varphi_0,t_0)}
\end{equation} 
at position $\rvec$, starting with a single particle placed at $\rvec_0$ with orientation $\varphi_0$ at time $t_0$. We will determine this density
to first order in the potential and second order in $q=2\imag\drift_0k/\rotDiffusion$, \Eref{def_q}, starting from the diagrammatic representation in inverse space,
\begin{equation}\elabel{density2DMathieu_diagrams}
    \ave{\chi(\kvec_\nvec,\ell,\omega)\chitilde(\kvec_\pvec,\ell_0,\omega_0)} \corresponds
        \ABPpropX{\kvec_\nvec,\omega}{\ell}{\kvec_\pvec,\omega_0}{\ell_0}
        +
        \ABPpropX{\kvec_\nvec,\omega}{\ell}{}{}\!\!\!\!\pertprop{}{}\!\!\!\!\ABPpropX{}{}{\kvec_\pvec,\omega_0}{\ell_0}
        +
        \ldots \ ,
\end{equation}
where the first diagram represents the bare propagator as introduced in \Eref{2DABP_propagator} and the second diagram the contribution to first order due to the external potential \Eref{extPot_vertex}. The limit $t_0\to-\infty$ needs to be taken after an inverse Fourier transform in $\omega$ and $\omega_0$. The structure of the propagator \Eref{2DABP_propagator}, essentially of the form $(-\imag\omega+\lambda+\mass)^{-1}$ with some non-negative $\lambda\ge0$ dependent on the spatial and angular modes, means that every such Fourier transform results in an expression proportional to $\exp{-(\lambda+\mass)(t-t_0)}$. Removing the mass by $\mass\downarrow0$ and taking $t_0\to-\infty$ results in the expression vanishing, except when $\Re(\lambda)\le0$ \cite{Zhen:2022}, which is the case, by inspection of \Eref{2DABP_propagator}, only when $\kvec=\nullvec$ and $\Re{\lambda_\ell(q(\kvec))}\le0$, which requires $\ell=0$, as 
\begin{subequations}
\elabel{eigenvalue_Mathieu_even}
\begin{align}
    \lambda_{2\ell}(q)&=4\ell^2+\order{q}\ ,\\
     \lambda_{2\ell+1}(q)&=4(\ell+1)^2+\order{q}\ .
\end{align}
\end{subequations}
\Eref{eigenvalue_Mathieu_even} is easily verified as the periodic solutions of the Mathieu \Eref{Mathieu_eqn} are trigonometric functions. Any propagator carrying a factor of $\kvec$ therefore vanishes under this operation. The only pole of the propagator to be considered is the one at $\omega=-\imag\mass$. Because only the rightmost, incoming leg in any of the diagrams is undashed, taking the inverse Fourier transform followed by the limit $t_0\to-\infty$ thus amounts to replacing the incoming legs by $\delta_{\nvec,0}$ and every occurance of $\omega$ in the diagram by $0$. This is equivalent to an amputation of the incoming leg \cite{Zhen:2022}. At this point, the inverse Fourier transforms from $\kvec$ to $\rvec$ and the transforms of azimuthal degree of freedom from $\ell$ to $\varphi$ are still to be taken.

We may summarise by writing
\begin{equation}\elabel{limit_chichitilde_2DMathieu}
    \lim_{t_0\to-\infty}\ave{\chi(\kvec_\nvec,\ell,t)\chitilde(\kvec_\pvec,\ell_0,t_0)} \corresponds
        \delta_{\ell,\ell_0}\delta_{\ell_0,0}\sysL^2\delta_{\nvec+\pvec,0}\delta_{\pvec,\nullvec}
        +
        \ABPpropX{\kvec_\nvec,0}{\ell}{}{}\!\!\!\!\pertprop{}{}\times\Big\{\delta_{\ell_0,0}\delta_{\pvec,\nullvec}\Big\}
        +
        \ldots \ .
\end{equation}
Using \Erefs{2DABP_propagator} and \eref{action_2DABP_pert_k2}, the last diagram evaluates to
\begin{equation}\elabel{Pot2Dk_MathieuAmputated}
        \ABPpropX{\kvec_\nvec,0}{\ell}{}{}\!\!\!\!\pertprop{}{}\times\Big\{\delta_{\ell_0,0}\delta_{\pvec,\nullvec}\Big\}
		=
		(-\kvec_\nvec\cdot\kvec_{\nvec+\pvec})\extPot_{\nvec+\pvec}\barepropGMathieu{\kvec_\nvec}{0}{\ell}{}\delta_{\ell_0,0}\delta_{\pvec,\nullvec}
		\MMprojection{2}{\ell,\ell_0}{}{\kvec_\nvec,\kvec_\pvec}
\end{equation}
The projection $\MMprojection{2}{\ell,\ell_0}{}{\kvec_\nvec,\kvec_\pvec}$ needs to be known only for $\ell_0=0$ and $\kvec_\pvec=\nullvec$, and because $\Mathieu{2}{\ell=0}{}{\gamma}{q=0}=1/\sqrt{2}$, \Erefs{MathieuIntegral} and \eref{def_MM2} give
\begin{equation}\elabel{Pot2Dk_MathieuAmputated_MM2term}
	\MMprojection{2}{\ell,0}{}{\kvec_\nvec,\nullvec}=\sqrt{2}
	\MathieuANought{\ell}{}{q(|\kvec_\nvec|)}
\end{equation}
To carry out the integral over $\varphi$ in \Eref{def_density2DMathieu}, we have to transform \Eref{limit_chichitilde_2DMathieu} from $\ell,\ell_0$ to $\varphi,\varphi_0$ using \Eref{orthoSystem_2DABP} first,
\begin{multline}\elabel{limit_chichitilde_2DMathieu_step2}
    \lim_{t_0\to-\infty}\ave{\chi(\kvec_\nvec,\varphi,t)\chitilde(\kvec_\pvec,\varphi_0,t_0)} 
	=        
	\Mathieu{2}{0}{}{\frac{\varphi-\sigma(\kvec_\nvec)}{2}}{q(|\kvec_\nvec|)}
	\Mathieutilde{2}{0}{}{\frac{\varphi_0-\sigma(-\kvec_\pvec)}{2}}{q(|\kvec_\pvec|)}
	\sysL^2\delta_{\nvec+\pvec,0}\delta_{\pvec,\nullvec}\\
        +
		\sum_{\ell=0}^\infty
		(-\kvec_\nvec\cdot\kvec_{\nvec+\pvec})\extPot_{\nvec+\pvec}\barepropGMathieu{\kvec_\nvec}{0}{\ell}{}\delta_{\pvec,\nullvec}
		\sqrt{2}
		\MathieuANought{\ell}{}{q(|\kvec_\nvec|)}
	\Mathieu{2}{\ell}{}{\frac{\varphi-\sigma(\kvec_\nvec)}{2}}{q(|\kvec_\nvec|)}
	\Mathieutilde{2}{0}{}{\frac{\varphi_0-\sigma(-\kvec_\pvec)}{2}}{q(|\kvec_\pvec|)}
	+
        \ldots \ .
\end{multline}
In the first term, $\kvec_\nvec=\kvec_\pvec=\nullvec$, so that the product of the two Mathieu functions degenerates to $1/(2\pi)$, cancelling with the integration over $\varphi$ still to be performed. In the second term, integration over $\varphi$ can again be done with the help of \Eref{MathieuIntegral}, while the final Mathieu function $\Mathieutilde{2}{0}{}{\frac{\varphi_0-\sigma(-\kvec_\pvec)}{2}}{q(|\kvec_\pvec|)}$ is in fact $1/(\sqrt{2}\pi)$, once $\kvec_\pvec=\nullvec$ is taken into account,
\begin{equation}\elabel{limit_chichitilde_2DMathieu_step3}
    \int_0^{2\pi}\dint{\varphi}\lim_{t_0\to-\infty}\ave{\chi(\kvec_\nvec,\varphi,t)\chitilde(\kvec_\pvec,\varphi_0,t_0)} 
	=        
	\sysL^2\delta_{\nvec+\pvec,0}\delta_{\pvec,\nullvec}
        +
		\sum_{\ell=0}^\infty
		(-\kvec_\nvec\cdot\kvec_{\nvec+\pvec})\extPot_{\nvec+\pvec}\barepropGMathieu{\kvec_\nvec}{0}{\ell}{}\delta_{\pvec,\nullvec}
		2 \big(
		\MathieuANought{\ell}{}{q(|\kvec_\nvec|)}
		\big)^2
+
        \ldots \ .
\end{equation}
Since $\MathieuANought{\ell}{}{q}$, \Eref{Mathieu_properties}, vanish for odd $\ell$ and otherwise behave to leading order like $q^{\ell/2}$, \Eref{Mathieu_properties}, 
expanding these coefficients to second order means to retain only $\ell=0,2$ in the sum, so that with explicit $\barepropGMathieu{\kvec_\nvec}{0}{\ell}{}$, \Eref{2DABP_propagator}, we arrive at 
\begin{multline}\elabel{lim_of_prop}
    \int_0^{2\pi}\dint{\varphi}\lim_{t_0\to-\infty}\ave{\chi(\kvec_\nvec,\varphi,t)\chitilde(\kvec_\pvec,\varphi_0,t_0)} 
	=        
	\sysL^2\delta_{\nvec+\pvec,0}\delta_{\pvec,\nullvec}
        +
		(-\kvec_\nvec\cdot\kvec_{\nvec+\pvec})\extPot_{\nvec+\pvec}\delta_{\ell_0,0}\delta_{\pvec,\nullvec}\\
		\times
		\Big\{
		\big(\transDiffusion k_\nvec^2 + \rotDiffusion\lambda_0(q)/4 + \mass\big)^{-1} \bigg(1-\frac{q^2}{16}\bigg)^2
		+
		\big(\transDiffusion k_\nvec^2 + \rotDiffusion\lambda_2(q)/4 + \mass\big)^{-1} \frac{q^2}{8}
		\Big\}+
		\order{q^4}+
        \ldots \ ,
\end{multline}
where $q^2=-4\drift_0^2k_\nvec^2\rotDiffusion^{-2}$ and $k_\nvec=|\kvec_\nvec|$. The limit $\mass\downarrow0$ is yet to be taken and $\ldots$ on the right of \Eref{lim_of_prop} refers to the higher orders in the external potential.
Expanding the eigenvalues $\MathieuLambda{\ell}{}{}$ beyond \Eref{Mathieu_properties},
\begin{subequations}
\elabel{Mathieu_properties2}
\begin{align}
	\MathieuLambda{0}{}{} & = -q^2/2 + \order{q^4} = 2\drift_0^2k_\nvec^2\rotDiffusion^{-2} + \order{q^4}\\
    \MathieuLambda{2}{}{} & = 4 + \order{q^2} \ ,
\end{align}
\end{subequations}
suggests that $\kvec_\nvec\cdot\kvec_{\nvec+\pvec}\delta_{\pvec,\nullvec}$ can be cancelled with $k_\nvec^2$ in the denominator when $\ell=0$. However, for $\nvec=\nullvec$ the denominator does not vanish as $\mass>0$. Performing the inverse Fourier transform to return real space gives after $\mass\downarrow0$,
\begin{multline}\elabel{2DABPdensity_final}
\density_0(\rvec)
=\frac{1}{\sysL^4}\sum_{\nvec,\pvec}
\exp{\imag(\kvec_\nvec\cdot\rvec+\kvec_\pvec\cdot\rvec_0)}
\int_0^{2\pi}\dint{\varphi}\lim_{t_0\to-\infty}\ave{\chi(\kvec_\nvec,\varphi,t)\chitilde(\kvec_\pvec,\varphi_0,t_0)} 
\\=
\sysL^{-2} - 
\frac{1}{\sysL^4}\sum_{\nvec\ne\nullvec}\exp{\imag\kvec_\nvec\cdot\rvec}
\extPot_{\nvec}
\Bigg\{
	\big(\transDiffusion + \drift_0^2/(2\rotDiffusion) + \order{q^4}/k_\nvec^2 \big)^{-1} 
	+
	\frac{k_\nvec^4\drift_0^2 }{2\rotDiffusion^{2}}
	\Big[
	\big(\transDiffusion k_\nvec^2 + \drift_0^2k_\nvec^2/(2\rotDiffusion) + \order{q^4} \big)^{-1}\\
	-
	\big(\transDiffusion k_\nvec^2 + \rotDiffusion + \rotDiffusion \order{q^2}\big)^{-1}
	\Big]
\Bigg\}
		+ \order{q^4}+
        \ldots \ ,
\end{multline}
where $\nvec=\nullvec$ is explicitly omitted from the sum and the first term in the curly brackets shows the expected $\effDiffusion{2D} = \transDiffusion + \drift_0^2/(2\rotDiffusion)
$, \Eref{def_effDiffusion}, amended, however, by a correction of order $q^4/k_\nvec^2$. To identify the lowest order correction in $k_\nvec$ and weak potentials, we expand further than \Eref{Mathieu_properties2},
\begin{equation}
\elabel{Mathieu_properties3}
	\lambda_0(q)=-\frac{q^2}{2}+\frac{7q^4}{128}+\order{q^6}=
	2\drift_0^2k_\nvec^2\rotDiffusion^{-2} 
	+
	\frac{7}{8}\drift_0^4k_\nvec^4\rotDiffusion^{-4} 
	+\order{q^6}
\end{equation}
in the first (of three terms, in the curly bracket) correction term in the sum in \Eref{2DABPdensity_final},
which gives
\begin{equation}\elabel{2DABPdensity_final_in_k}
\density_0(\rvec)
=
\sysL^{-2} - 
\frac{1}{\sysL^4}\sum_{\nvec\ne\nullvec}\exp{\imag\kvec_\nvec\cdot\rvec}
\frac{\extPot_{\nvec}}{\effDiffusion{2D}}
\left\{
1+\frac{\drift_0^2k_\nvec^2}{2\rotDiffusion^2}
\bigg(
1-\frac{7}{16}\frac{\drift_0^2}{\rotDiffusion\effDiffusion{2D}}
\bigg)
+\order{k_\nvec^4}
\right\}
+\order{\extPot^2}
\end{equation}
by dropping the last term in \Eref{2DABPdensity_final}, as it is $\order{k_\nvec^4}$, and not expanding the second term any further, as it is $\order{k_\nvec^2}$ as it stands. 
The expression above it to be compared to the first order term in the barometric formula, \Eref{density_expanded_in_E_easy}, which produces only $-\extPot_\nvec/\transDiffusion$ as the first-order summand  when setting $\driftVec_0=\nullvec$. 
The corrections beyond that are generically due to the stochastic self-propulsion and amount to more than replacing $\transDiffusion$ by $\effDiffusion{2D}$.
\Eref{2DABPdensity_final_in_k} completes the derivation in the present section. In the next section, we will re-derive MSD $\overline{\rvec^2}(t)$ and density $\density_0(\rvec)$ of active Brownian particles in two dimensions at stationarity, in an attempt to recover \Erefs{MSD_2d} and \eref{2DABPdensity_final} or \eref{2DABPdensity_final_in_k} in a perturbation theory in small $\drift_0$ and small potential.

\subsection{Perturbation in small $\drift_0$}\label{sec:ABP2D_pert}
The starting point of the present derivation is 
the operator $\Liouville_2$ in \Eref{def_Liouville2}
and
the harmonic part of the action \Eref{def_action02}, 
\begin{equation}\elabel{action_2DABP_wPert}
    \action_{02}[\chitilde,\chi]=\int\dTWOint{r}\int_0^{2\pi}\dint{\varphi}\int\dint{t}
    \chitilde(\rvec,\varphi,t)
    \big(
    -\partial_t + \transDiffusion\nabla_\rvec^2 + \nabla^2_\varphi - \mass
    \big) 
    \chi(\rvec,\varphi,t)
\end{equation}
which becomes diagonal with 
\begin{subequations}
\elabel{orthoSystem_2DABP_wPert}
\begin{align}
    \chi(\rvec,\varphi,t) &= \int\dintbar{\omega}\dTWOintbar{k}\sum_{\ell=-\infty}^\infty
    \exp{-\imag\omega t}\exp{\imag\kvec\cdot\rvec}
	\exp{\imag\ell\varphi}
    \chi_\ell(\kvec,\omega)\\
    \chitilde(\rvec,\varphi,t) &= \int\dintbar{\omega}\dTWOintbar{k}\sum_{\ell=-\infty}^\infty
    \exp{-\imag\omega t}\exp{\imag\kvec\cdot\rvec}
\frac{\exp{-\imag\ell\varphi}}{2\pi} 
    \chitilde_\ell(\kvec,\omega) \ .
\end{align}
\end{subequations}
Compared to \Eref{orthoSystem_2DABP}, in \Eref{orthoSystem_2DABP_wPert} the Mathieu functions have been replaced by exponentials, 
\begin{subequations}
\begin{align}
    \Mathieu{2}{\ell}{}{\frac{\varphi-\sigma(\kvec)}{2}}{q(|\kvec|)}
&\mapsto
	\exp{\imag\ell\varphi}\\
    \Mathieutilde{2}{\ell}{}{\frac{\varphi-\sigma(\kvec)}{2}}{q(|\kvec|)}
&\mapsto
	\frac{\exp{-\imag\ell\varphi}}{2\pi}
\end{align}
\end{subequations}
which are orthonormal irrespective of $\kvec$. The bare propagator of \Eref{action_2DABP_wPert} can immediately be read off,
\begin{equation}\elabel{def_barepropGwPert}
    \ave{
    \chi_{\ell}(\kvec,\omega)
    \chitilde_{\ell_0}(\kvec_0,\omega_0)
    }_0
    =
    \frac{\delta_{\ell,\ell_0} \deltabar(\kvec + \kvec_0)\deltabar(\omega + \omega_0)}{-\imag \omega + \transDiffusion\kvec^2 + \rotDiffusion \ell^2 + \mass}
    = 
    \delta_{\ell,\ell_0} \deltabar(\kvec + \kvec_0)\deltabar(\omega + \omega_0)
    \barepropGwPert{\kvec}{\omega}{\ell}{}\\
    \corresponds\barepropX{\kvec,\omega}{\ell}{\kvec_0,\omega_0}{\ell_0}
    \ .
\end{equation}
The perturbative part of the action \Eref{def_actionP2} includes the self-propulsion, but its precise form depends on whether the system is finite or infinite and thus left to be stated in the sections ahead.

\subsubsection{Mean squared displacement}\label{sec:2DABP_wPert_MSD}
It is instructive to derive the MSD with harmonic action \Eref{action_2DABP_wPert}, orthogonal system \Eref{orthoSystem_2DABP_wPert} and perturbative action \Eref{def_actionP2}, rewritten as
\begin{equation}
	    \elabel{def_actionP2_2DABP}
	\action_{P2}=
	\int\dTWOintbar{k}\dTWOintbar{k'}\dintbar{\omega}
	\sum_{\ell,\ell'=-\infty}^\infty
	\chi_\ell(\kvec,\omega)
    \chitilde_{\ell'}(\kvec',-\omega)
	\frac{\drift_0}{2}\deltabar(\kvec+\kvec')
	\big[
	-\imag k_x (\delta_{\ell',\ell+1}+\delta_{\ell',\ell-1})
	+ k_y (\delta_{\ell',\ell-1}-\delta_{\ell',\ell+1})
	\big]
\end{equation}
in the absence of an external potential. To ease notation, we may introduce the tumbling vertex 
\newcommand{\tumbleW}[3]{W_{#1}^{#2}(#3)}
\begin{subequations}
\begin{align}\elabel{def_tumbleW}
\tumbleW{\ell',\ell}{}{\kvec}&=
- \drift_0
\int\dint{\varphi}
	\frac{\exp{-\imag\ell'\varphi}}{2\pi}
(\imag k_x\cos\varphi+\imag k_y\sin\varphi)
	\exp{\imag\ell\varphi}
\\&=
	\frac{\drift_0}{2}
	\big[
	-\imag k_x 
	(\delta_{\ell',\ell+1}+\delta_{\ell',\ell-1})
	+ k_y 
	(\delta_{\ell',\ell-1}-\delta_{\ell',\ell+1})
	\big]
\end{align}
\end{subequations}
with $\kvec=(k_x,k_y)^\transpose$ and $\rvec=(x,y)^\transpose$. 
The expression $\imag k_x\cos\varphi+\imag k_y\sin\varphi$ is due to $\driftVec\cdot\nabla_\rvec$ of the self-propulsion hitting $\exp{\imag\kvec\rvec}$, \Eref{def_driftVecOmega}.
For the diagrammatics
we introduce the self-propulsion vertex
\begin{equation}
\tumbleW{\ell',\ell}{}{\kvec}\corresponds\velpertX{\ell}{\ell'} \ ,
\end{equation}
so that the full propagator reads
\begin{multline}\elabel{2D_propagator_perturbative}
        \ave{
    \chi_{\ell}(\kvec,\omega)
    \chitilde_{\ell_0}(\kvec_0,\omega_0)
    }
\corresponds
\barepropX{\kvec,\omega}{\ell}{\kvec_0,\omega_0}{\ell_0}+\barepropS{\kvec,\omega}{\ell}{}{}\!\!\!\!\velpert\!\!\!\!\barepropS{}{}{\kvec_0,\omega_0}{\ell_0}+\barepropS{\kvec,\omega}{\ell}{}{}\!\!\!\!\velpert\!\!\!\!\barepropS{}{}{}{}\!\!\!\!\velpert\!\!\!\!\barepropS{}{}{\kvec_0,\omega_0}{\ell_0}+\dots
\\
\corresponds
\deltabar(\kvec + \kvec_0)\deltabar(\omega + \omega_0) 
\Big(
\barepropGwPert{\kvec}{\omega}{\ell}{}
\delta_{\ell,\ell_0}
+
\barepropGwPert{\kvec}{\omega}{\ell}{}
\tumbleW{\ell,\ell_0}{}{\kvec}
\barepropGwPert{\kvec}{\omega}{\ell_0}{}\\
+
\sum_{\ell_1}
\barepropGwPert{\kvec}{\omega}{\ell}{}
\tumbleW{\ell,\ell_1}{}{\kvec}
\barepropGwPert{\kvec}{\omega}{\ell_1}{}
\tumbleW{\ell_1,\ell_0}{}{\kvec}
\barepropGwPert{\kvec}{\omega}{\ell_0}{}
+ \ldots
\Big)
\end{multline}
using the notation of \Eref{def_barepropGwPert}. 
A marginalisation over $\varphi$ amounts to evaluating \Eref{2D_propagator_perturbative} at $\ell=0$, \Eref{orthoSystem_2DABP_wPert}.
The MSD can be determined by double differentiation with respect to $\kvec$ as done in \Erefs{MSDv1} and \eref{MSD_2d}. This calculation is simplified by calculating the mean squared displacement in $x$ only, \ie taking $\partial_{k_x}^2$ rather than $\partial_{k_x}^2+\partial_{k_y}^2$. This ``trick'', however also requires uniform initialisation in $\varphi_0$ to avoid any bias. In summary
\begin{equation}\elabel{MSDx2D}
\overline{\rvec^2}(t-t_0) = 2 \overline{x^2}(t-t_0) = - 2 \left.\partial_{k_x}^2\right|_{\substack{k_x=0\\ k_y=0}} \int_0^{2\pi}\dint{\varphi}
\frac{1}{2\pi}\int_0^{2\pi}\dint{\varphi_0}
\int\dTWOintbar{k_0}
\ave{
    \chi(\kvec,\varphi,t)
    \chitilde(\kvec_0,\varphi_0,t_0)
    }
\end{equation}
where the integral over $\kvec_0$ is the inverse Fourier transform with $\xvec_0=\nullvec$ and is trivially taken by using $\deltabar(\kvec + \kvec_0)$ due to translational invariance. 

The two integrals over $\varphi$ and $\varphi_0$ in \Eref{MSDx2D} amount to setting $\ell=0$ and $\ell_0=0$, with all factors of $2\pi$ cancelling. Of the three terms in \Eref{2D_propagator_perturbative}, the second one, which is the first order correction in the self-propulsion carrying a single factor of the tumbling vertex $\tumbleW{\ell,\ell_0}{}{\kvec}$, \Eref{def_tumbleW}, is odd in $k_x$. Differentiating twice with respect to $k_x$ and evaluating at $k_x=k_y=0$ will thus make it vanish. The third term is quadratic in $k_x$ in the numerator and not singular in $k_x$ in the denominator. The only way for the third term not to vanish at $k_x=k_y=0$ is when $\tumbleW{\ell,\ell_1}{}{\kvec}\tumbleW{\ell_1,\ell_0}{}{\kvec}$ is differentiated twice with respect to $k_x$,
\begin{equation}\elabel{tumble_derivative}
\left.\partial_{k_x}^2\right|_{\substack{k_x=0\\ k_y=0}} 
 \tumbleW{0,\ell_1}{}{\kvec}\tumbleW{\ell_1,0}{}{\kvec}
 =
	\frac{-\drift_0^2}{2}
	\big[
	\delta_{1,\ell_1}+\delta_{-1,\ell_1}
	\big] \ .
\end{equation}
No higher order terms in $w_0$ can contribute, as they all carry higher powers of $k_x$ as a pre-factor. We thus arrive at
\begin{equation}\elabel{MSDx2D_omega}
\overline{\rvec^2}(t-t_0) = 
\int\dintbar{\omega}\exp{-\imag\omega(t-t_0)}
\Big\{
\frac{4\transDiffusion}{(-\imag\omega+\mass)^2}
+
\frac{\drift_0^2}{(-\imag\omega+\mass)^2}
\frac{2}{-\imag\omega+\rotDiffusion+\mass}
\Big\}
\end{equation}
where the term $2/(-\imag\omega+\rotDiffusion+\mass)$ originates from 
$\barepropGwPert{\kvec}{\omega}{\ell_1}{}$ at $\ell_1=-1,1$ according to \Eref{tumble_derivative}. Taking the inverse Fourier transform from $\omega$ to $t-t_0$ then indeed reproduces \Eref{MSD_2d} in the limit $\mass\downarrow0$ with the repeated poles producing the desired factors of $t-t_0$. This confirms that even when the self-propulsion is dealt with perturbatively, observables can be derived in closed form. This is not a triviality, as $\drift_0^2t/\transDiffusion$ is a dimensionless quantity and thus might enter to all orders.
We proceed in the present framework by allowing for an external potential.

\subsubsection{External Potential}\label{sec:2DABP_wPert_extPot}
In this section, we aim to re-derive the density $\density_0(\rvec)$ of active Brownian Particles in an external potential, as initially determined in \Sref{2DABP_Mathieu_extPot}. As the self-propulsion is a perturbation, the full perturbative action \Erefs{def_actionP2}, \eref{def_actionP2_2DABP} reads 
\begin{multline}
 	    \elabel{def_actionP2_2DABP_wPert_k}
 	\action_{P2}=
 	\int\dintbar{\omega}
 	\frac{1}{\sysL^{4}}\sum_{\nvec_1,\nvec_2}
 	\sum_{\ell_1,\ell_2=-\infty}^\infty
 	\chi_{\ell_1}(\kvec_{\nvec_1},\omega)
     \chitilde_{\ell_2}(\kvec_{\nvec_2},-\omega)
 	\sysL^2\delta_{\nvec_1+\nvec_2,\nullvec}
 	\tumbleW{\ell_2,\ell_1}{}{\kvec_{\nvec_1}}
 	\\
 	+
 	\frac{1}{\sysL^{6}}\sum_{\nvec_1,\nvec_2,\nvec_3}
     \sysL^2\delta_{\nvec_1+\nvec_2+\nvec_3,\nullvec}
     \sum_{\ell_1,\ell_3}
     (\kvec_{\nvec_1}\cdot\kvec_{\nvec_2}) 
     \chitilde_{\ell_1}(\kvec_{\nvec_1},-\omega)
     \extPot_{\nvec_2}
     \chi_{\ell_3}(\kvec_{\nvec_3},\omega)
\end{multline}
using the notation $\kvec_{\nvec_1}=(k_{n_{1x}},k_{n_{1y}})^\transpose$.
The action \Eref{def_actionP2_2DABP_wPert_k} is based on 
discrete Fourier modes similar to \Eref{orthoSystem_2DABP_finite}
as the system's volume is finite. It deviates from 
\Eref{orthoSystem_2DABP_finite}
but using Fourier modes for the director as in \Eref{orthoSystem_2DABP_wPert}
\begin{subequations}
\elabel{orthoSystem_2DABP_wPert_finite}
\begin{align}
    \chi(\rvec,\varphi,t) &= \int\dintbar{\omega}\frac{1}{\sysL^2}\sum_{\nvec\in\Zset^2}\sum_{\ell=-\infty}^\infty
    \exp{-\imag\omega t}\exp{\imag\kvec_\nvec\cdot\rvec}
	\exp{\imag\ell\varphi}
	\chi_\ell(\kvec_\nvec,\omega)\\
    \chitilde(\rvec,\varphi,t) &= \int\dintbar{\omega}\frac{1}{\sysL^2}\sum_{\nvec\in\Zset^2}\sum_{\ell=-\infty}^\infty
    \exp{-\imag\omega t}\exp{\imag\kvec_\nvec\cdot\rvec}
	\frac{\exp{-\imag\ell\varphi}}{2\pi} 
    \chitilde_\ell(\kvec_\nvec,\omega)\ .
\end{align}
\end{subequations}
Compared to \Eref{action_2DABP_pert_k}, \Eref{def_actionP2_2DABP_wPert_k} has the great benefit of the diagonality of the angular modes in the potential term.

To calculate the particle density $\density_0(\rvec)$ in the stationary state, \Eref{def_density2DMathieu}, we expand the full propagator diagrammatically,
\begin{subequations}
\elabel{chichitilde_diagrams_withPot_wPert_2D}
\begin{align}
\ave{\chi(\kvec,\varphi,\omega)\chitilde(\kvec_0,\varphi_0,\omega_0)}
\elabel{chichitilde_diagrams_withPot_wPert_2D_line1}
&= \barepropX{\kvec,\omega}{\ell}{\kvec_0,\omega_0}{\ell_0}
+
\barepropX{\kvec,\omega}{\ell}{}{}\!\!\!\!\!\!\!\!\velpert\!\!\!\!\barepropS{}{}{\kvec_0,\omega_0}{\ell_0}
+
\barepropS{\kvec,\omega}{\ell}{}{}\!\!\!\!\velpert\!\!\!\!\barepropS{}{}{}{}\!\!\!\!\velpert\!\!\!\!\barepropS{}{}{\kvec_0,\omega_0}{\ell_0}+\ldots\\
\elabel{chichitilde_diagrams_withPot_wPert_2D_line2}
&\ +\barepropX{\kvec,\omega}{\ell}{}{}\!\!\!\!\!\!\!\!\pertprop{}{}\!\!\!\!\barepropS{}{}{\kvec_0,\omega_0}{\ell_0}\\
\elabel{chichitilde_diagrams_withPot_wPert_2D_line3}
&\ +
\barepropS{\kvec,\omega}{\ell}{}{}\!\!\!\!\velpert\!\!\!\!\barepropS{}{}{}{}\!\!\!\!\pertprop{}{}\!\!\!\!\barepropS{}{}{\kvec_0,\omega_0}{\ell_0}
+
\barepropS{\kvec,\omega}{\ell}{}{}\!\!\!\!\pertprop{}{}\!\!\!\!\barepropS{}{}{}{}\!\!\!\!\velpert\!\!\!\!\barepropS{}{}{\kvec_0,\omega_0}{\ell_0}
\\
\elabel{chichitilde_diagrams_withPot_wPert_2D_line4}
&\ +\barepropS{\kvec,\omega}{\ell}{}{}\!\!\!\!\velpert\!\!\!\!\barepropS{}{}{}{}\!\!\!\!\velpert\!\!\!\!\barepropS{}{}{}{}\!\!\!\!\pertprop{}{}\!\!\!\!\barepropS{}{}{\kvec_0,\omega_0}{\ell_0}
+
\barepropS{\kvec,\omega}{\ell}{}{}\!\!\!\!\velpert\!\!\!\!\barepropS{}{}{}{}\!\!\!\!\pertprop{}{}\!\!\!\!\barepropS{}{}{}{}\!\!\!\!\velpert\!\!\!\!\barepropS{}{}{\kvec_0,\omega_0}{\ell_0}
+
\barepropS{\kvec,\omega}{\ell}{}{}\!\!\!\!\pertprop{}{}\!\!\!\!\barepropS{}{}{}{}\!\!\!\!\velpert\!\!\!\!\barepropS{}{}{}{}\!\!\!\!\velpert\!\!\!\!\barepropS{}{}{\kvec_0,\omega_0}{\ell_0}
\\
\elabel{chichitilde_diagrams_withPot_wPert_2D_line5}
&+\barepropS{\kvec,\omega}{\ell}{}{}\!\!\!\!\!\!\!\!\pertprop{}{}\!\!\!\!\barepropS{}{}{}{}\!\!\!\!\pertprop{}{}\!\!\!\!\barepropS{}{}{\kvec_0,\omega_0}{\ell_0}
+
\barepropS{\kvec,\omega}{\ell}{}{}\!\!\!\!\!\!\!\!\pertprop{}{}\!\!\!\!\barepropS{}{}{}{}\!\!\!\!\pertprop{}{}\!\!\!\!\barepropS{}{}{}{}\!\!\!\!\velpert\!\!\!\!\barepropS{}{}{\kvec_0,\omega_0}{\ell_0}
+
\barepropS{\kvec,\omega}{\ell}{}{}\!\!\!\!\!\!\!\!\pertprop{}{}\!\!\!\!\barepropS{}{}{}{}\!\!\!\!\velpert\!\!\!\!\barepropS{}{}{}{}\!\!\!\!\pertprop{}{}\!\!\!\!\barepropS{}{}{}{}\!\!\!\!\velpert\!\!\!\!\barepropS{}{}{\kvec_0,\omega_0}{\ell_0}+
\ldots
\end{align}
\end{subequations}
As in \Sref{2DABP_Mathieu_extPot}, taking the limit $t_0\to-\infty$ amounts to an amputation upon which any diagram vanishes that carries a dashed propagator before (\ie to the right of) a source of momentum. As a result, only the five leftmost diagrams in \Eref{chichitilde_diagrams_withPot_wPert_2D} contribute at stationarity. In the following, we will restrict ourselves to diagrams with at most one bauble, \ie to first order in the potential. Further, we will incorporate the self-propulsion to second order, leaving us with the four leftmost diagrams in \Erefs{chichitilde_diagrams_withPot_wPert_2D_line1}-\eref{chichitilde_diagrams_withPot_wPert_2D_line4}.

As seen in \Sref{2DABP_Mathieu_extPot}, \Eref{chichitilde_diagrams_withPot_wPert_2D_line1} contributes to the density merely with $1/\sysL^2$. The other three diagrams require more detailed attention. 
Firstly, \Eref{chichitilde_diagrams_withPot_wPert_2D_line2} gives
\begin{equation}\elabel{chichitilde_diagrams_withPot_wPert_2D_line2_v2}
D_b(\kvec_\nvec, \ell, \omega;\kvec_\pvec, \ell_0, \omega_0) \corresponds
\barepropS{\kvec_\nvec,\omega}{\ell}{}{}\!\!\!\!\!\!\!\!\pertprop{}{}\!\!\!\!\barepropS{}{}{\kvec_\pvec,\omega_0}{\ell_0}
\corresponds\delta_{\ell,\ell_0}\deltabar(\omega+\omega_0)
\barepropGwPert{\kvec_\nvec}{\omega}{\ell}{}
(-\kvec_\nvec\cdot\kvec_{\nvec+\pvec})
\extPot_{\nvec+\pvec}
\barepropGwPert{-\kvec_\pvec}{\omega}{\ell_0}{}\ ,
\end{equation}
using \Eref{def_barepropGwPert} for $\barepropGwPert{\kvec_\nvec}{\omega}{\ell}{}$. Upon taking the inverse Fourier transform in $\omega$ and $\omega_0$, the limit $t_0\to-\infty$ can be taken, replacing the incoming leg 
$\barepropGwPert{\kvec_\pvec}{\omega_0}{\ell_0}{}$ effectively by $\delta_{\pvec,\nullvec}\delta_{\ell_0,0}$. After transforming from $\ell,\ell_0$ to $\varphi,\varphi_0$ and taking the integral over $\varphi$ all factors of $2\pi$ have cancelled, leaving of \Eref{chichitilde_diagrams_withPot_wPert_2D_line2_v2}
\begin{equation}\elabel{chichitilde_diagrams_withPot_wPert_2D_line2_v3}
\lim_{t_0\to-\infty} \int_0^{2\pi}\dint{\varphi} 
D_b(\rvec, \varphi, t;\rvec_0, \varphi_0, t_0)
=
\frac{1}{\sysL^4}\sum_{\nvec}\exp{\imag\kvec_\nvec\cdot\rvec}
\barepropGwPert{\kvec_\nvec}{0}{0}{}
(-\kvec_\nvec\cdot\kvec_{\nvec})
\extPot_{\nvec}
\stackrel{\mass\downarrow0}{=}
-\frac{1}{\sysL^4}\sum_{\nvec\ne\nullvec}\exp{\imag\kvec_\nvec\cdot\rvec}
\frac{\extPot_{\nvec}}{\transDiffusion}
\end{equation}
as $\barepropGwPert{\kvec_\nvec}{0}{0}{}=(\transDiffusion k_\nvec^2+\mass)^{-1}$, whose factor $k_\nvec^{-2}$ cancels with 
$\kvec_\nvec\cdot\kvec_{\nvec}$ unless $\kvec_\nvec=\nullvec$, in which case the expression vanishes.
\Eref{chichitilde_diagrams_withPot_wPert_2D_line2_v3} is the lowest order correction in both $\extPot$ and $\drift_0$ to the density.

Of the remaining two diagrams, the leftmost in \Erefs{chichitilde_diagrams_withPot_wPert_2D_line3} and \eref{chichitilde_diagrams_withPot_wPert_2D_line4}, only the latter contributes once $t_0\to-\infty$ forces $\ell_0=0$ and $\int\dint{\varphi}$ forces $\ell=0$,
because the single tumbling vertex, \Eref{def_tumbleW}, requires the angular modes to be offset by $1$. The only diagram left to consider is thus
\begin{multline}\elabel{chichitilde_diagrams_withPot_wPert_2D_line4_v2}
D_d(\kvec_\nvec, \ell, \omega;\kvec_\pvec, \ell_0, \omega_0) \corresponds
\barepropS{\kvec_\nvec,\omega}{\ell}{}{}\!\!\!\!\velpert\!\!\!\!\barepropS{}{}{}{}\!\!\!\!\velpert\!\!\!\!\barepropS{}{}{}{}\!\!\!\!\pertprop{}{}\!\!\!\!\barepropS{}{}{\kvec_\pvec,\omega_0}{\ell_0}
\\
\corresponds
\barepropGwPert{\kvec_\nvec}{\omega}{\ell}{}
\left(
\sum_{\ell_1}
\tumbleW{\ell,\ell_1}{}{\kvec_\nvec}
\barepropGwPert{\kvec_\nvec}{\omega}{\ell_1}{}
\tumbleW{\ell_1,\ell_0}{}{\kvec_\nvec}
\right)
\barepropGwPert{\kvec_\nvec}{\omega}{\ell_0}{}
(-\kvec_\nvec\cdot\kvec_{\nvec+\pvec})
\extPot_{\nvec+\pvec}
\barepropGwPert{-\kvec_\pvec}{\omega}{\ell_0}{}\deltabar(\omega+\omega_0)
\ .
\end{multline}
After taking $t_0\to-\infty$ and $\int\dint{\varphi}$, the only two contributions from the sum are $\ell_1=-1,1$, so that
\begin{equation}\elabel{chichitilde_diagrams_withPot_wPert_2D_line4_v3}
\lim_{t_0\to-\infty} \int_0^{2\pi}\dint{\varphi} 
D_d(\rvec, \varphi, t;\rvec_0, \varphi_0, t_0)
=
\frac{1}{2\sysL^4}\sum_{\nvec}\exp{\imag\kvec_\nvec\cdot\rvec}
\drift_0^2 k_{\nvec}^4
\extPot_{\nvec}
(\transDiffusion k_\nvec^2+\mass)^{-2}(\transDiffusion k_\nvec^2+\rotDiffusion+\mass)^{-1}
\end{equation}
using
\begin{equation}
\tumbleW{0,\pm1}{}{\kvec}
=
- \frac{\drift_0}{2} [\imag k_x \pm k_y ]
\quad\text{ and }\quad
\tumbleW{\pm1,0}{}{\kvec}
=
\frac{\drift_0}{2} [-\imag k_x \pm k_y ]
\end{equation}
from \Eref{def_tumbleW},
so that $\tumbleW{0,1}{}{\kvec}\tumbleW{1,0}{}{\kvec}=\tumbleW{0,-1}{}{\kvec}\tumbleW{-1,0}{}{\kvec}=-\drift_0^2k^2/4$. After taking the limit $\mass\downarrow0$ in \Eref{chichitilde_diagrams_withPot_wPert_2D_line4_v3} the factor $k_\nvec^4$ cancels with the same factor in $(\transDiffusion k_\nvec^2+\mass)^{-2}$ whenever $\nvec\ne\nullvec$, so that finally
\begin{subequations}
\elabel{density2D_final}
\begin{align}
\density_0(\rvec) &= \sysL^{-2} + 
\lim_{\mass\downarrow0}
\lim_{t_0\to-\infty} \int_0^{2\pi}\dint{\varphi} 
D_b(\rvec, \varphi, t;\rvec_0, \varphi_0, t_0)
+
D_d(\rvec, \varphi, t;\rvec_0, \varphi_0, t_0)
\\
\elabel{2DABPdensity_final_wPert}
&=
\sysL^{-2}
-
\frac{1}{\sysL^4}\sum_{\nvec\ne\nullvec}\exp{\imag\kvec_\nvec\cdot\rvec}
\frac{\extPot_{\nvec}}{\transDiffusion}
\left\{
1-
\frac{\drift_0^2}{2\transDiffusion}
(\transDiffusion k_\nvec^2+\rotDiffusion)^{-1}
\right\}
+ \order{\drift_0^4}
+ \ldots
\end{align}
\end{subequations}
to be compared to the purely drift-diffusive \Eref{density_expanded_in_E_easy}
and to
\Erefs{2DABPdensity_final} or \eref{2DABPdensity_final_in_k}, 
derived using Mathieu functions, rather than the perturbative approach to self-propulsion here. By inspection of the two expressions, \Erefs{2DABPdensity_final} and \eref{2DABPdensity_final_wPert}, the two are equivalent provided
\begin{multline}
1-
\frac{\drift_0^2}{2\transDiffusion}
(\transDiffusion k_\nvec^2+\rotDiffusion)^{-1}
+ \order{\drift_0^4}
=
\transDiffusion
\Bigg\{
	\big(\transDiffusion + \drift_0^2/(2\rotDiffusion) + \order{q^4}/k_\nvec^2 \big)^{-1} 
	\\+
	\frac{k_\nvec^4\drift_0^2 }{2\rotDiffusion^{2}}
	\Big[
	\big(\transDiffusion k_\nvec^2 + \drift_0^2k_\nvec^2/(2\rotDiffusion) + \order{q^4} \big)^{-1}	
	-
	\big(\transDiffusion k_\nvec^2 + \rotDiffusion + \rotDiffusion \order{q^2}\big)^{-1}
	\Big]
\Bigg\}
\end{multline}
which can indeed be shown by expanding all terms on the right hand side to order $\drift_0^2$ only, absorbing all other terms into $\order{\drift_0^4}$, for example
\begin{equation}
\transDiffusion\big(\transDiffusion + \drift_0^2/(2\rotDiffusion) + \order{q^4}/k_\nvec^2 \big)^{-1} 
=
1-\frac{\drift_0^2}{2\transDiffusion\rotDiffusion}
+\order{\drift_0^4}\ .
\end{equation}

This concludes the discussion of ABPs in two dimensions. In the following sections, we will calculate MSD and density in an external potential for ABPs in three dimensions, using the above as a template.

\section{Active Brownian Particles in $3$ dimensions}\label{sec:ABP3D}
In three spatial dimensions, the diffusion of the director takes place on a curved manifold, which renders the Laplacian rather unwieldy, \Eref{def_nabla2Omega}. The spherical harmonics provide a suitable eigensystem for this operator, but the self-propulsion term spoils the diagonalisation. As in the two-dimensional case, \Sref{ABP2D}, there are two possible ways ahead: Either find suitable special functions or deal with the self-propulsion term perturbatively. In the following subsection, we focus on the former.
\subsection{"Three-dimensional Mathieu functions"}\label{sec:ABP3D_mathieu}
The harmonic part of the action is \Eref{def_action1} with \Erefs{def_driftVecOmega} and \eref{def_nabla2Omega} in \Eref{def_Liouville1}, more explicitly 
\begin{multline}\elabel{action_3DABP}
    \action_{01}[\chitilde,\chi]=\int\dTHREEint{x}
	\int_0^{\pi}\dint{\theta}
	\int_0^{2\pi}\dint{\varphi}\sin\theta\,
	\int\dint{t}
    \chitilde(\rvec,\theta,\varphi,t)\\
    \bigg(
    -\partial_t + \transDiffusion\nabla_\rvec^2
	- \drift_0
	(\sin\!\theta\cos\!\varphi\ \partial_x
	+\sin\!\theta\sin\!\varphi\ \partial_y
	+\cos\!\theta\ \partial_z)
	+ 
	\rotDiffusion \big(
	\frac{1}{}\partial_\theta\sin\!\theta\ \partial_\theta + 
\frac{1}{\sin^2 \!\theta}\partial_\varphi^2
\big)
	- \mass
    \bigg) 
    \chi(\rvec,\theta,\varphi,t)
\end{multline}
using the notation $\rvec=(x,y,z)^\transpose$.

The eigensystem that we will use now is modelled along the two-dimensional Mathieu functions, \Eref{orthoSystem_2DABP}. Specifically, it is
\begin{subequations}
\elabel{orthoSystem_3DABP}
\begin{align}
    \chi(\rvec,\theta,\varphi,t) &= \int\dintbar{\omega}\dTHREEintbar{k}\sum_{\ell=0}^\infty\sum_m
    \exp{-\imag\omega t}\exp{\imag\kvec\cdot\rvec}
    \Mathieu{3}{\ell}{m}{\theta-\tau(\kvec),\varphi-\sigma(\kvec)}{\qvec(\kvec)}
    \chi_\ell^m(\kvec,\omega)\\
    \chitilde(\rvec,\theta,\varphi,t) &= \int\dintbar{\omega}\dTHREEintbar{k}\sum_{\ell=0}^\infty\sum_m
    \exp{-\imag\omega t}\exp{\imag\kvec\cdot\rvec}
    \Mathieutilde{3}{\ell}{m}{\theta-\tau(-\kvec),\varphi-\sigma(-\kvec)}{\qvec(-\kvec)}
    \chitilde_\ell^m(\kvec,\omega)
\end{align}
\end{subequations}
with 
\begin{equation}\elabel{def_Mathieu3Tilde}
\Mathieutilde{3}{\ell}{m}{\theta,\varphi}{\qvec(\kvec)}=
\frac{1}{4\pi}
\Mathieu{3}{\ell}{m}{\theta,\varphi}{\qvec(\kvec)}
\end{equation}
orthonormal, similar to \Eref{orthonormality_Mathieu2D}
\begin{equation}\elabel{Mathieu3_orthonormality}
    	\deltabar(\kvec+\kvec')
    	\int_0^{\pi}\dint{\theta}
	\int_0^{2\pi}\dint{\varphi}\sin\theta\,
\Mathieu{3}{\ell}{m}{\theta-\tau(\kvec),\varphi-\sigma(\kvec)}{\qvec(\kvec)}
\Mathieutilde{3}{\ell'}{m'}{\theta-\tau(-\kvec),\varphi-\sigma(-\kvec)}{\qvec(-\kvec')}
=
\deltabar(\kvec+\kvec')
\delta_{\ell,\ell'}
\delta_{m,m'}
\end{equation}
and
$q(|\kvec|)$ of the two-dimensional case, \Eref{def_q}, replaced by 
\begin{equation}
\elabel{def_qvec}
\qvec(\kvec)=\imag\frac{\drift_0\kvec}{\rotDiffusion}
\end{equation}
and 
two instead of one arbitrary functions $\sigma(\kvec)$ and $\tau(\kvec)$ \textit{cf.} \Eref{orthonormality_Mathieu2D}.
The benefit of those, however, is rather limited compared to the two-dimensional case, where $\kvec\cdot\driftVec$ was rewritten as $k\drift_0\cos(\varphi-\alpha)$. We therefore choose $\sigma\equiv\tau\equiv0$ and demand 
\begin{equation}\elabel{Mathieu3_eigenEquation}
\bigg(
-\imag \frac{\drift_0}{\rotDiffusion}
	(k_x \sin\theta\cos\varphi
	+ k_y \sin\theta\sin\varphi
	+ k_z \cos\theta
	)
	+ 
	\frac{1}{\sin \theta}\partial_\theta\sin\theta\partial_\theta + 
\frac{1}{\sin^2 \theta}\partial_\varphi^2
\bigg)
\Mathieu{3}{\ell}{m}{\theta,\varphi}{\qvec}
	=
- \lambda_\ell^m(\qvec)
\Mathieu{3}{\ell}{m}{\theta,\varphi}{\qvec}
\ .
\end{equation}
These eigenfunctions are characterised in some detail in \Appref{app_3DMathieu}. For now, it suffices to know that they exist and that the eigenvalues $\lambda_\ell^m(\qvec)$ are indeed discrete and indexed in both $\ell$ and $m$.
Having relegated the details of the eigensystem to the appendix, the bare propagator can now be determined without much ado, 
\begin{subequations}
\begin{align}\elabel{3DABP_propagator}
    \ave{
    \chi_{\ell}^m(\kvec,\omega)
    \chitilde_{\ell_0}^{m_0}(\kvec_0,\omega_0)
    }_0
    =
    \frac{\delta_{\ell,\ell_0}\delta_{m,m_0} \deltabar(\kvec + \kvec_0)\deltabar(\omega + \omega_0)}{-\imag \omega + \transDiffusion\kvec^2 + \rotDiffusion \MathieuLambda{\ell}{m}{\qvec(\kvec)} + \mass}
    = 
    \delta_{\ell,\ell_0}\delta_{m,m_0} \deltabar(\kvec + \kvec_0)\deltabar(\omega + \omega_0)
    \barepropGMathieu{\kvec}{\omega}{\ell}{m}\\
    \corresponds
    \ABPpropX{\kvec,\omega}{\ell,m}{\kvec_0,\omega_0}{\ell_0,m_0}
    \ ,
\end{align}
\end{subequations}
and we proceed with the MSD.

\subsubsection{Mean squared displacement}\label{sec:3DABP_Mathieu_MSD}
The MSD derives from the propagator \Eref{3DABP_propagator} via a double derivative. As in the two-dimensional case, the derivation simplifies considerably with uniform initialisation. Further, because the eigenequation \eref{Mathieu3_eigenEquation} is particularly simple whenever $k_x=k_y=0$, we may write the MSD in terms of derivatives with respect to $k_z$ only,
\begin{multline}\elabel{MSD3Dv1}
    \spave{\rvec^2}(t) = -3\left.\partial_{k_z}^2\right|_{\kvec=\nullvec} 
	\int_0^\pi\dint{\theta}
	\int_0^{2\pi}\dint{\varphi}\sin\theta\,
	\frac{1}{4\pi}
	\int_0^\pi\dint{\theta_0}
	\int_0^{2\pi}\dint{\varphi_0}\sin\theta_0\\
    \times \sum_{\ell=0}^\infty\sum_m
\Mathieu{3}{\ell}{m}{\theta,\varphi}{\qvec(k_z\evec_z)}
\Mathieutilde{3}{\ell}{m}{\varphi_0,\theta_0}{\qvec(k_z\evec_z)}
    \exp{-(\transDiffusion k^2 + \rotDiffusion \lambda_\ell^m(\qvec(k_z\evec_z)))t}
	\ .
\end{multline}
Similar to \Eref{MathieuIntegral}, we use coefficients
\begin{equation}\elabel{integral_Mathieu3}
\int_0^\pi\dint{\theta}
	\int_0^{2\pi}\dint{\varphi}\sin\theta\,
\Mathieu{3}{\ell}{m}{\theta,\varphi}{q\evec_z}
= 4 \pi 
\MathieuANought{\ell}{0,0}{q\evec_z}
\delta_{m,0}
\end{equation}
as derived in more detail in \Eref{integral_Mathieu3_app}. These coefficients we need to determine at most up to order $q^2$ in order to express $\spave{\rvec^2}(t)$ in closed form. Using the notation $q=\imag\drift_0k_z/\rotDiffusion$ in 
$\qvec(k_z\evec_z)=q\evec_z$, \Eref{MSD3Dv1} produces with \Eref{integral_Mathieu3}
\begin{equation}
    \spave{\rvec^2}(t) = -3\left.\partial_{k_z}^2\right|_{\kvec=\nullvec} 
    \left\{
    \left(1-\frac{q^2}{24}\right)^2 \exp{-(\transDiffusion k_z^2 + \rotDiffusion\MathieuLambda{0}{0}{q\evec_z})t}
    +
    \left(\frac{q}{2\sqrt{3}}\right)^2 \exp{-(\transDiffusion k_z^2 + \rotDiffusion\MathieuLambda{1}{0}{q\evec_z}}
    \right\}
\end{equation}
from the expansion of 
$\MathieuANought{0}{0,0}{q\evec_z}$, \Eref{MathieuA3D_00}, 
and
$\MathieuANought{1}{0,0}{q\evec_z}$, \Eref{MathieuA3D_10}, 
with 
$\MathieuANought{\ell}{0,0}{q\evec_z}\in\order{q^2}$ for $\ell\ge2$, \Eref{MathieuA3D_ell}. 
The two eigenvalues $\MathieuLambda{0}{0}{q\evec_z}=-q^2/6+\ldots$ and $\MathieuLambda{1}{0}{q\evec_z}=2+\ldots$, \Erefs{MathieuLambda3D_00} and \eref{MathieuLambda3D_10}, need to be known only to leading order in $q$. Performing the derivative and re-arranging terms then produces the final result
\begin{equation}\elabel{3DABP_MSD}
    \spave{\rvec^2}(t) =
    6 \transDiffusion t + \frac{\drift_0^2}{2\rotDiffusion^2}
    \left(
    \exp{-2\rotDiffusion t} -1 + 2 \rotDiffusion t
    \right)
\end{equation}
structurally similar to the result in two dimensions, \Eref{MSD_2d}, and indeed identical to the MSD of Run-and-Tumble particles in three dimensions, Eq.~(49) of \cite{ZhangPruessner:2022}, if the tumble rate $\alpha$ is replaced by $2\rotDiffusion$ \cite{Cates:2013, Cates:2015}. The effective diffusion constant in three dimensions is
\begin{equation}\elabel{def_effDiffusion3D}
\effDiffusion{3D} = \transDiffusion + \frac{\drift_0^2}{6\rotDiffusion}
\ ,
\end{equation}
to be compared to \Eref{def_effDiffusion}.

\subsubsection{External Potential}\label{sec:3DABP_Mathieu_extPot}
Allowing for an external potential and confining the particles to a finite space means that the fields are now written as
\begin{subequations}
\elabel{orthoSystem_3DABP_finite}
\begin{align}
    \chi(\rvec,\theta,\varphi,t) &= \int\dintbar{\omega}\frac{1}{\sysL^3}\sum_{\nvec\in\Zset^3}\sum_{\ell=0}^\infty\sum_{m}
    \exp{-\imag\omega t}\exp{\imag\kvec\cdot\rvec}
    \Mathieu{3}{\ell}{m}{\theta,\varphi}{\qvec(\kvec)}
    \chi_\ell(\kvec,\omega)\\
    \chitilde(\rvec,\theta,\varphi,t) &= \int\dintbar{\omega}\frac{1}{\sysL^3}\sum_{\nvec\in\Zset^3}\sum_{\ell=0}^\infty\sum_{m}
    \exp{-\imag\omega t}\exp{\imag\kvec\cdot\rvec}
    \Mathieutilde{3}{\ell}{m}{\theta,\varphi}{\qvec(-\kvec)}
    \chitilde_\ell(\kvec,\omega)
\end{align}
\end{subequations}
similar to \Erefs{orthoSystem_2DABP_finite} and \eref{orthoSystem_3DABP}
and any expectation needs to be taken with the perturbative action 
\begin{equation}\elabel{action_3DABP_pert_k2}
\action_{P1}=\int\dintbar{\omega}
    \frac{1}{\sysL^{9}}\sum_{\nvec_1,\nvec_2,\nvec_3}
    \sysL^3\delta_{\nvec_1+\nvec_2+\nvec_3,\nullvec}
    \sum_{\ell_1,\ell_3}
    \sum_{m_1,m_3}
    (\kvec_{\nvec_1}\cdot\kvec_{\nvec_2}) 
    \chitilde_{\ell_1,m_1}(\kvec_{\nvec_1},-\omega)
    \extPot_{\nvec_2}
    \chi_{\ell_3,m_3}(\kvec_{\nvec_3},\omega)
	\MMprojection{3}{\ell_1,\ell_3}{m_1,m_3}{-\kvec_{\nvec_1},-\kvec_{\nvec_3}} \ ,
\end{equation}
similar to \Eref{action_2DABP_pert_k2}
with integral
\begin{equation}\elabel{def_MM3}
    \MMprojection{3}{\ell_1,\ell_3}{m_1,m_3}{-\kvec_{\nvec_1},-\kvec_{\nvec_3}}
= \int_0^\pi\dint{\theta}\int_0^{2\pi}\dint{\varphi}\sin\theta\,
    \Mathieu{3}{\ell_3}{m_3}{\theta,\varphi}{\qvec(\kvec_{\nvec_3})}
    \Mathieutilde{3}{\ell_1}{m_1}{\theta,\varphi}{\qvec(\kvec_{\nvec_1})} \ ,
\end{equation}
similar to \Eref{def_MM2}.

The resulting diagrammatics of the full propagator to leading order in the potential is 
\begin{equation}\elabel{density3DMathieu_diagrams}
    \ave{\chi(\kvec_\nvec,\ell,m,\omega)\chitilde(\kvec_\pvec,\ell_0,m_0,\omega_0)} \corresponds
        \ABPpropX{\kvec_\nvec,\omega}{\ell,m}{\kvec_\pvec,\omega_0}{\ell_0,m_0}
        +
        \ABPpropX{\kvec_\nvec,\omega}{\ell,m}{}{}\!\!\!\!\pertprop{}{}\!\!\!\!\ABPpropX{}{}{\kvec_\pvec,\omega_0}{\ell_0,m_0}
        +
        \ldots \ ,
\end{equation}
as in \Eref{density2DMathieu_diagrams} but
with the index $\ell$ replaced by the couple $\ell,m$.
The arguments that follow to extract the stationary density are similar to those in 
\Sref{2DABP_Mathieu_extPot}. Yet, to be able to draw on the results in \Appref{app_3DMathieu}, all $\kvec$-vectors featuring in the three-dimensional Mathieu functions need to be parallel to $\evec_z$. Because amputated diagrams draw all $\kvec$-dependence from spatial variation of the potential, we thus demand $\extPot(\rvec)=\extPot(z)$, \ie that the potential depends only on the $z$-component of $\rvec=(x,y,z)^\transpose$. Its Fourier transform can therefore be written in the form
\begin{equation}\elabel{extPot3D_z_only}
    \extPot_{\nvec}
    =\iiint 
    \limits_0^{\quad\sysL}
    \dTHREEint{r}\extPot(\rvec)\exp{-\imag\kvec_\nvec\cdot\rvec}
    =\sysL^2 \delta_{n_x,0}\delta_{n_y,0}\extPot_{n_z}
\end{equation}
where $\nvec=(n_x,n_y,n_z)^\transpose$. The notation of \Eref{extPot3D_z_only} is slightly ambigious,
as $\extPot_{n_z}$ on the right is a Fourier transform in the $z$-direction only, while $\extPot_{\nvec}$ on the left is a Fourier transform in all space, yet the alternatives seem to obfuscate even more.

Just like in 
\Sref{2DABP_Mathieu_extPot}, the stationary limit enforces $\kvec_\pvec=\nullvec$, as well as $\ell_0=0$, as otherwise $\MathieuLambda{\ell}{m}{\qvec(0)}\ne0$, \Eref{Mathieu_q0_lambda}, in the propagator \Eref{3DABP_propagator}.

Turning our attention to the projection $\MMprojection{3}{\ell_1,0}{m_1,m_3}{\kvec_\nvec,\kvec_\pvec}$ for $\pvec=\nullvec$, this vanishes whenever $m_3\ne0$ as $\Mathieu{3}{\ell=0}{m}{\theta,\varphi}{\qvec=\nullvec}=\sqrt{4\pi}\SphericalHarmonics{0}{m}{\theta,\varphi}=\delta_{m,0}$, \Eref{Mathieu3_as_SphericalHarmonics}, 
because the spherical harmonics that the three-dimensional Mathieu functions are written in with $\ell=0$ allow only for $m=0$, so that $\SphericalHarmonics{0}{m}{\theta,\varphi}=\delta_{m,0}/\sqrt{4\pi}$. It follows that
\begin{equation}\elabel{Pot3Dk_MathieuAmputated_MM2term}
    \MMprojection{3}{\ell,0}{m,m_0}{\kvec_\nvec,\nullvec}
= \delta_{m_0,0}\int_0^\pi\dint{\theta}\int_0^{2\pi}\dint{\varphi}\sin\theta\,
    \Mathieutilde{3}{\ell}{m}{\theta,\varphi}{\qvec(\kvec_{\nvec})}
= \delta_{m_0,0}
\delta_{m,0} 
\MathieuA{\ell,0}{0,0}{\qvec(\kvec_{\nvec})}
\ ,
\end{equation}
for all $\kvec_\nvec\parallel\evec_z$. \Eref{Pot3Dk_MathieuAmputated_MM2term} corresponds to \Eref{Pot2Dk_MathieuAmputated_MM2term}. This produces
\begin{multline}\elabel{limit_chichitilde_3DMathieu_step2}
    \lim_{t_0\to-\infty}\ave{\chi(\kvec_\nvec,\theta,\varphi,t)\chitilde(\kvec_\pvec,\theta_0,\varphi_0,t_0)} 
	=        
	\Mathieu{3}{0}{0}{\theta,\varphi}{\nullvec}
	\Mathieutilde{3}{0}{0}{\theta_0,\varphi_0}{\nullvec}
	\sysL^2\delta_{\nvec+\pvec,\nullvec}\delta_{\pvec,\nullvec}\\
        +
		\sum_{\ell=0}^\infty\sum_m
		(-\kvec_\nvec\cdot\kvec_{\nvec+\pvec})\extPot_{\nvec+\pvec}\barepropGMathieu{\kvec_\nvec}{0}{\ell}{m}\delta_{\pvec,\nullvec}
\delta_{m,0} 
\MathieuA{\ell,0}{0,0}{\qvec(\kvec_{\nvec})}
	\Mathieu{3}{\ell}{m}{\theta,\varphi}{\qvec(\kvec_\nvec)}
	\Mathieutilde{3}{0}{0}{\theta_0,\varphi_0}{\nullvec}
	+
        \ldots \ .
\end{multline}
similar to \Eref{limit_chichitilde_2DMathieu_step2}, with $\barepropGMathieu{\kvec_\nvec}{0}{\ell}{m}$ denoting the bare propagator \Eref{3DABP_propagator} and again assuming $\kvec_\nvec\parallel\evec_z$. This is guaranteed by \Eref{extPot3D_z_only}, which means that $\extPot_\nvec$ vanishes for all $\kvec_\nvec\nparallel\evec_z$, so that the second term in \Eref{limit_chichitilde_3DMathieu_step2} only ever contributes when $\kvec_\nvec\parallel\evec_z$. 

Since $\Mathieutilde{3}{0}{0}{\theta_0,\varphi_0}{\nullvec}=1/(4\pi)$, it cancels the $4\pi$ of the integral over the final angles $\theta$ and $\varphi$ of both terms on the right hand side of \Eref{limit_chichitilde_3DMathieu_step2}, the second one by \Eref{integral_Mathieu3},
\begin{multline}\elabel{limit_chichitilde_3DMathieu_step3}
    \int_0^\pi\dint{\theta}\int_0^{2\pi}\dint{\varphi}\sin\theta\,\lim_{t_0\to-\infty}\ave{\chi(\kvec_\nvec,\theta,\varphi,t)\chitilde(\kvec_\pvec,\theta_0,\varphi_0,t_0)} 
	\\=        
	\sysL^3\delta_{\nvec+\pvec,\nullvec}\delta_{\pvec,\nullvec}
        +
		\sum_{\ell=0}^\infty
		(-\kvec_\nvec\cdot\kvec_{\nvec+\pvec})\extPot_{\nvec+\pvec}\barepropGMathieu{\kvec_\nvec}{0}{\ell}{m}\delta_{\pvec,\nullvec}
\delta_{m,0} 
		\big(
		\MathieuA{\ell,0}{0,0}{\qvec(\kvec_\nvec)}
		\big)^2
+
        \ldots \ ,
\end{multline}
similar to \Eref{limit_chichitilde_2DMathieu_step3}. Expanding to order $q^2=-\drift_0^2 k_z^2/\rotDiffusion^2$, \Eref{def_qvec}, means to keep only $\ell=0,1$ in the sum in \Eref{limit_chichitilde_3DMathieu_step3}, as $\MathieuA{\ell,0}{0,0}{q\evec_z}\in\order{q^\ell}$, \Eref{MathieuA3D_ell}. This produces an expression similar to \Eref{lim_of_prop}, and using \Eref{MathieuA3D_00} and \eref{MathieuA3D_10} for $\MathieuA{\ell,0}{0,0}{}$, as well as \Erefs{MathieuLambda3D_00} and \eref{MathieuLambda3D_10} for $\MathieuLambda{\ell}{0}{}$, finally gives
\begin{multline}\elabel{3DABPdensity_final}
\density_0(\rvec)
= \sysL^{-3} - 
\frac{1}{\sysL^4}\sum_{n_z\ne0}\exp{\imag zk_{n_z}}
\extPot_{n_z}
\Bigg\{
	\big(\transDiffusion + \drift_0^2/(6\rotDiffusion) + \order{q^4}/k_{n_z}^2 \big)^{-1} 
	+
	\frac{k_{n_z}^4\drift_0^2 }{12\rotDiffusion^{2}}
	\Big[
	\big(\transDiffusion k_{n_z}^2 + \drift_0^2k_{n_z}^2/(6\rotDiffusion) + \order{q^4} \big)^{-1}\\
	-
	\big(\transDiffusion k_{n_z}^2 + 2 \rotDiffusion + \rotDiffusion \order{q^2}\big)^{-1}
	\Big]
\Bigg\}
		+ \order{q^4}+
        \ldots \ ,
\end{multline}
using $\rvec=(x,y,z)^\transpose$ for the position,
$k_{n_z}=2\pi n_z/\sysL$ for the mode, and \Eref{extPot3D_z_only}, which cancels $\sysL^2$. 
\Eref{3DABPdensity_final} corresponds to \Eref{2DABPdensity_final} in two dimension.
Expanding rather in small $k_{n_z}$, similar to \Eref{2DABPdensity_final_in_k},  produces
\begin{equation}\elabel{3DABPdensity_final_in_k}
\density_0(\rvec)
=
\sysL^{-3} - 
\frac{1}{\sysL^4}\sum_{n_z\ne0}\exp{\imag zk_{n_z}}
\frac{\extPot_{n_z}}{\effDiffusion{3D}}
\left\{
1+\frac{\drift_0^2k_{n_z}^2}{12\rotDiffusion^2}
\bigg(
1-\frac{11}{90}\frac{\drift_0^2}{\rotDiffusion\effDiffusion{3D}}
\bigg)
+\order{k_\nvec^4}
\right\}
+\order{\extPot^2}
\end{equation}
which has non-trivial corrections compared to the purely drift-diffusive \Eref{density_expanded_in_E_easy}, beyond setting 
in the latter $\driftVec_0=\nullvec$ and replacing the diffusion constant
$\transDiffusion$ by $\effDiffusion{3D}$ of \Eref{def_effDiffusion3D}. The difference in $\sysL^{-4}$ compared to $\sysL^{-2d}$ in \Eref{density_expanded_in_E_easy} is solely due to the definition of $\extPot_{n_z}$ in \Eref{extPot3D_z_only}. 

This concludes the calculation of properties of three-dimensional active Brownian particles on the basis of "three-dimensional Mathieu functions". In the next two sections, we will reproduce the results above on the basis of a perturbation theory in the drift.







\subsection{Perturbation in small $\drift_0$}\label{sec:ABP3D_pert}
Along the lines of \Sref{ABP2D_pert}, we return now to calculate the MSD and the stationary density while treating the self-propulsion as a perturbation over diffusion, so that we can use the eigensystem
\begin{subequations}
\elabel{orthoSystem_3DABP_wPert}
\begin{align}
    \chi(\rvec,\theta,\varphi,t) &= \int\dintbar{\omega}\dTHREEintbar{k}\sum_{\ell=0}^\infty\sum_{m=-\ell}^\ell
    \exp{-\imag\omega t}\exp{\imag\kvec\cdot\rvec}
	(4\pi)^{1/2} \SphericalHarmonics{\ell}{m}{\theta,\varphi} 
    \chi_\ell^m(\kvec,\omega)\\
    \chitilde(\rvec,\theta,\varphi,t) &= \int\dintbar{\omega}\dTHREEintbar{k}\sum_{\ell=0}^\infty\sum_{m=-\ell}^\ell
    \exp{-\imag\omega t}\exp{\imag\kvec\cdot\rvec}
(4\pi)^{-1/2}
\Big(\SphericalHarmonics{\ell}{m}{\theta,\varphi}\Big)^*
    \chitilde_\ell^m(\kvec,\omega) \ .
\end{align}
\end{subequations}
where 
\begin{equation}
    \SphericalHarmonics{\ell}{m}{\theta,\varphi}=\sqrt{\frac{(2\ell+1)}{4\pi}\frac{(\ell-m)!}{(\ell+m)!}}P_\ell^m(\cos\!\theta)\exp{\imag m \varphi}
\end{equation} 
denote the well-known spherical harmonics with $P_\ell^m$ denotes the standard associated Legendre polynomials which carry a factor $(-1)^m$ in their definition, so that, for example $\SphericalHarmonics{1}{1}{\theta,\varphi}=-\frac{1}{2}\sqrt{\frac{3}{2\pi}}\sin\!\theta\exp{\imag\varphi}$. Our definition of $Y_\ell^m$ conforms to \texttt{SphericalHarmonicsY[$\ell,m,\theta,\varphi$]} \cite{Mathematica:10.0.2.0}. The spherical harmonics $Y_\ell^m$ are  orthonormal 
\begin{equation}
    \int_0^\pi\dint{\theta}\int_0^{2\pi}\dint{\varphi}\sin\theta\, 
    \SphericalHarmonics{\ell}{m}{\theta,\varphi}
    \Big(\SphericalHarmonics{\ell'}{m'}{\theta,\varphi}\Big)^*
    =
    \delta_{\ell,\ell'}\delta_{m,m'}
\end{equation}
and obey the eigenvalue equation
\begin{equation}
    \nabla^2_\Omega \SphericalHarmonics{\ell}{m}{\theta,\varphi}
    =
    \left(
    \frac{1}{\sin\!\theta}\partial_\theta\sin\!\theta\ \partial_\theta + 
\frac{1}{\sin^2 \!\theta}\partial_\varphi^2
\right)
\SphericalHarmonics{\ell}{m}{\theta,\varphi}
=
-\ell(\ell+1)
\SphericalHarmonics{\ell}{m}{\theta,\varphi}\ .
\end{equation}
The 
harmonic part of the action \Eref{def_action02} now reads 
\begin{equation}\elabel{action_3DABP_wPert}
    \action_{02}[\chitilde,\chi]=
- \int\dint{\omega}\dTHREEintbar{k}
\sum_{\ell=0}^\infty\sum_{m=-\ell}^\ell
\chitilde_\ell^m(-\kvec,-\omega)
(-\imag\omega+\transDiffusion k^2+\rotDiffusion\ell(\ell+1)+\mass)
\chi_\ell^m(\kvec,\omega)
\end{equation}
so that the bare propagators are
\begin{equation}\elabel{def_barepropGwPert_3D}
    \ave{
    \chi_{\ell}^m(\kvec,\omega)
    \chitilde_{\ell_0}^{m_0}(\kvec_0,\omega_0)
    }_0
    =
    \frac{\delta_{\ell,\ell_0}\delta_{m,m_0} \deltabar(\kvec + \kvec_0)\deltabar(\omega + \omega_0)}{-\imag \omega + \transDiffusion\kvec^2 + \rotDiffusion \ell(\ell+1) + \mass}
    = 
    \delta_{\ell,\ell_0}\delta_{m,m_0} \deltabar(\kvec + \kvec_0)\deltabar(\omega + \omega_0)
    \barepropGwPert{\kvec}{\omega}{\ell}{m}\\
    \corresponds
    \barepropX{\kvec,\omega}{\ell,m}{\kvec_0,\omega_0}{\ell_0,m_0}
    \!\!\!\ ,
\end{equation}
differing from \Eref{def_barepropGwPert} only by the additional factor $\delta_{m,m_0}$ and by the eigenvalue of the spherical Laplacian being $\ell(\ell+1)$ rather than $\ell^2$.
In the absence of an external potential the perturbative part of the action reads
\begin{equation}
	    \elabel{def_actionP2_3DABP}
	\action_{P2}=
	\int\dTHREEintbar{k}\dTHREEint{k'}\dintbar{\omega}
	\sum_{\ell,\ell'=0}^\infty
	\sum_{m=-\ell}^\ell
	\sum_{m'=-\ell'}^{\ell'}
	\chi_\ell^m(\kvec,\omega)
    \chitilde_{\ell'}^{m'}(\kvec',-\omega)
	\deltabar(\kvec+\kvec')
\tumbleW{\ell',\ell}{m',m}{\kvec}
\end{equation}
similar to \Eref{def_actionP2_2DABP}
with tumbling vertex
\begin{equation}\elabel{def_tumbleW3D}
\tumbleW{\ell',\ell}{m',m}{\kvec}
=
- \drift_0
\int_0^\pi\dint\theta\int_0^{2\pi}\dint{\varphi}\sin\!\theta
    \Big(\SphericalHarmonics{\ell'}{m'}{\theta,\varphi}\Big)^*
(\imag k_x\sin\!\theta\cos\!\varphi
+\imag k_y\sin\!\theta\sin\varphi
+\imag k_z\cos\!\theta)
	\SphericalHarmonics{\ell}{m}{\theta,\varphi}
\end{equation}
In principle, any product like $\sin\theta\cos\varphi\SphericalHarmonics{\ell}{m}{\theta,\varphi}$ can be written in terms of a sum of spherical harmonics (see \Appref{app_CG_coeff}), 
\begin{subequations}
\elabel{pert_expansion_in_CG}
\begin{align}
    \sin\!\theta\cos\!\varphi Y_\ell^m(\theta,\varphi)&=\sum_{L,M}\sqrt{\frac{2\ell+1}{2(2L+1)}}c_{1,0,\ell,0}^{L,0}Y_L^M(\theta,\varphi)\bigg(-c_{1,1,\ell,m}^{L,M}+c_{1,-1,\ell,m}^{L,M}\bigg)\ ,\\
    \sin\!\theta\sin\!\varphi Y_\ell^m(\theta,\varphi)&=\imag\sum_{L,M}\sqrt{\frac{2\ell+1}{2(2L+1)}}c_{1,0,\ell,0}^{L,0}Y_L^M(\theta,\varphi)\bigg(c_{1,1,\ell,m}^{L,M}+c_{1,-1,\ell,m}^{L,M}\bigg)\ ,\\
    \cos\!\theta\SphericalHarmonics{\ell}{m}{\theta,\varphi}&=\sum_{L,M}\sqrt{\frac{2\ell+1}{2L+1}}c_{1,0,\ell,0}^{L,0}c_{1,0,\ell,m}^{L,M}Y_{L}^M(\theta,\varphi)\ ,
\end{align}
\end{subequations}
where $c_{\ell',m',\ell,m}^{L,M}$ are the Clebsch-Gordan coefficients, further detailed \cite{Racah:1942} in \Appref{app_CG_coeff}.


The coupling matrix $\tumbleW{\ell',\ell}{m',m}{\kvec}$  can therefore be written as 
\begin{equation}
\elabel{Coupling_matrix_W}
    \tumbleW{\ell',\ell}{m',m}{\kvec}=\tumbleW{\ell',\ell}{m',m}{k_x\evec_x}+\tumbleW{\ell',\ell}{m',m}{k_y\evec_y}+\tumbleW{\ell',\ell}{m',m}{k_z\evec_z} \ .
\end{equation}
the simplest among the three terms is
\begin{subequations}
\elabel{Coupling_matrix_Wz}
\begin{align}
\tumbleW{\ell',\ell}{m',m}{k_z\evec_z} &=
-\imag\drift_0 k_z \sum_{L,M}\sqrt{\frac{2\ell+1}{2L+1}}c_{1,0,\ell,0}^{L,0}c_{1,0,\ell,m}^{L,M}\delta_{m',M}\delta_{\ell',L}
\\
&=-\imag\drift_0k_z\delta_{m,m'}\left\{
    \sqrt{
\frac{(\ell+1+m)(\ell+1-m)}{(2\ell+1)(2\ell+3)}
}\delta_{\ell',\ell+1}
    +
    \sqrt{
\frac{(\ell+m)(\ell-m)}{(2\ell+1)(2\ell-1)}
}\delta_{\ell',\ell-1}
    \right\}\elabel{simple_3D_tumbleW} \ ,
\end{align}
\end{subequations}
with the other two stated in \Appref{app_CG_coeff}.
With this framework in place, we proceed to calculate the MSD and the density at stationarity.


%




\subsubsection{Mean squared displacement}\label{sec:3DABP_wPert_MSD}
We follow \Sref{2DABP_wPert_MSD} to calculate the MSD of active Brownian particles in three dimensions, which is 
\begin{equation}\elabel{MSDx3D}
\overline{\rvec^2}(t-t_0) = 3 \overline{z^2}(t-t_0) = - 3 \left.\partial_{k_z}^2\right|_{\kvec=0} \int_0^\pi\dint{\theta}\int_0^{2\pi}\dint{\varphi}\sin\theta\,
\frac{1}{4\pi}
\int_0^\pi\dint{\theta_0}\int_0^{2\pi}\dint{\varphi_0}\sin\theta_0\,
\int\dTHREEintbar{k_0}
\ave{
    \chi(\kvec,\theta,\varphi,t)
    \chitilde(\kvec_0,\theta_0,\varphi_0,t_0)
    }
\end{equation}
corresponding to \Eref{MSDx2D}.
With the help of the diagrammatics \Eref{2D_propagator_perturbative} and the reasoning before \Eref{tumble_derivative}, this can essentially be done by inspection. The derivative of the bare propagator will contribute $-3(-2\transDiffusion t)$ to the MSD. The only non-trivial contribution is due to $\barepropS{}{}{}{}\!\!\!\!\velpert\!\!\!\!\barepropS{}{}{}{}\!\!\!\!\velpert\!\!\!\!\barepropS{}{}{}{}$
, which has a very simple structure because the integration over $\theta$ and $\varphi$, forces $\ell=0$ and $m=0$, given that $\SphericalHarmonics{\ell}{m}{}$ are orthogonal and $\SphericalHarmonics{0}{0}{}=1/\sqrt{4\pi}$ and so integrals over $\SphericalHarmonics{\ell}{m}{}$ vanish unless $\ell=m=0$. The same reasoning applies to $\ell_0=m_0=0$. 
The summation that appears in the last term of \Eref{2D_propagator_perturbative} therefore becomes
\begin{equation}\elabel{WHW_3D}
    \sum_{\ell_1=0}^\infty\sum_{m_1=-\ell_1}^{\ell_1}
    \tumbleW{0,\ell_1}{0,m_1}{k_z\evec_z}
    \barepropGwPert{k_z\evec_z}{\omega}{\ell_1}{m_1}
    \tumbleW{\ell_1,0}{m_1,0}{k_z\evec_z}
    =
    -\drift_0^2k_z^2\frac{1}{3} 
    \frac{1}{-\imag\omega+\transDiffusion k_z^2 + 2 \rotDiffusion}
\end{equation}
because only $\ell_1=1$ contributes, resulting in a factor $\ell_1(\ell_1+1)=2$ in front of $\rotDiffusion$ in the propagator.
Differentiating twice with respect to $k_z$ and taking the inverse Fourier transform of the expression with the bare propagators attached,
\begin{equation}\elabel{3DABP_MSD_terms_wPert}
    -\frac{2}{3}\drift_0^2\lim_{\mass\downarrow0}
    \int\dintbar{\omega} \exp{-\imag\omega t}
    \frac{1}{(-\imag\omega+\mass)^2}
    \frac{1}{-\imag\omega+2\rotDiffusion+\mass}
    =
    \frac{2}{3}\drift_0^2\left\{
    \frac{-t}{2\rotDiffusion}
    +\frac{1}{4\rotDiffusion^2}
    -\frac{1}{4\rotDiffusion^2}\exp{-2\rotDiffusion t}
    \right\}
\end{equation}
finally reproduces \Eref{3DABP_MSD} exactly.

\subsubsection{External potential}\label{sec:3D_wPert_extPot}
Calculating the stationary density $\density_0(\rvec)$ in the presence of an external potential perturbatively in the self-propulsion and the potential follows the pattern outlined in \Sref{2DABP_wPert_extPot}. Because the volume is finite, the representation we choose for the fields is
\begin{subequations}
\elabel{orthoSystem_3DABP_wPert_extPot}
\begin{align}
    \chi(\rvec,\varphi,t) &= \int\dintbar{\omega}
    \frac{1}{\sysL^3}\sum_{\nvec\in\Zset^3}\sum_{\ell=0}^\infty\sum_{m=-\ell}^\ell
    \exp{-\imag\omega t}\exp{\imag\kvec_\nvec\cdot\rvec}
	(4\pi)^{1/2} \SphericalHarmonics{\ell}{m}{\theta,\varphi} 
    \chi_\ell^m(\kvec_\nvec,\omega)\\
    \chitilde(\rvec,\varphi,t) &= \int\dintbar{\omega}\frac{1}{\sysL^3}\sum_{\nvec\in\Zset^3}\sum_{\ell=0}^\infty\sum_{m=-\ell}^\ell
    \exp{-\imag\omega t}\exp{\imag\kvec_\nvec\cdot\rvec}
(4\pi)^{-1/2}
\Big(\SphericalHarmonics{\ell}{m}{\theta,\varphi}\Big)^*
    \chitilde_\ell^m(\kvec_\nvec,\omega) \ .
\end{align}
\end{subequations}
rather than \Eref{orthoSystem_3DABP_wPert}, with the sums over $\nvec\in\Zset^3$ replacing the integrals over $\kvec$.

Given the orthogonality of the spherical harmonics, the perturbative part of the action now reads
\begin{multline}
	    \elabel{def_actionP2_3DABP_wPert}
	\action_{P2}=
	\int \dintbar{\omega}
	\frac{1}{\sysL^6}\sum_{\nvec_1,\nvec_2}
	\sum_{\ell,\ell'=0}^\infty
	\sum_{m=-\ell}^\ell
	\sum_{m'=-\ell'}^{\ell'}
	\chi_\ell^m(\kvec_{\nvec_1},\omega)
    \chitilde_{\ell'}^{m'}(\kvec_{\nvec_2},-\omega)
	\delta_{\nvec_1+\nvec_2,\nullvec}
\tumbleW{\ell',\ell}{m',m}{\kvec_{\nvec_1}}\\
 	+
 	\frac{1}{\sysL^{9}}\sum_{\nvec_1,\nvec_2,\nvec_3}
     \sysL^3\delta_{\nvec_1+\nvec_2+\nvec_3,\nullvec}
     \sum_{\ell_1,\ell_3}\sum_{m=-\ell_3}^{\ell_3}\sum_{m'=-\ell_1}^{\ell_1}
     (\kvec_{\nvec_1}\cdot\kvec_{\nvec_2}) 
     \chitilde^{m'}_{\ell_1}(\kvec_{\nvec_1},-\omega)
     \extPot_{\nvec_2}
     \chi^m_{\ell_3}(\kvec_{\nvec_3},\omega)
\end{multline}
corresponding to \Eref{def_actionP2_2DABP_wPert_k} in two dimensions, but with the tumbling vertex $\tumbleW{\ell',\ell}{m',m}{\kvec}$ given by \Eref{def_tumbleW3D}.

In principle, $\extPot_{\nvec}$ is arbitrary, but in order to stay with the simple form of $\tumbleW{\ell,\ell'}{m,m'}{k_z\evec_z}$, \Eref{simple_3D_tumbleW}, we restrict $\extPot(\rvec)$ to vary only with $z$, making again use of the notation \Eref{extPot3D_z_only}. The sole contribution that needs our attention is due to the single term \Eref{chichitilde_diagrams_withPot_wPert_2D_line4_v2}, which draws on \Eref{WHW_3D}.

Using $\tumbleW{0,1}{0,0}{k_z\evec_z}\tumbleW{1,0}{0,0}{k_z\evec_z}=-\drift_0^2k_z^2/3$, the density can be read off \Eref{2DABPdensity_final_wPert}
\begin{equation}\elabel{3Ddensity_wPert}
\density_0(\rvec)
=
   \sysL^{-3}
-
\frac{1}{\sysL^4}\sum_{n_z\ne0}\exp{\imag k_{n_z}z }
\frac{\extPot_{n_z}}{\transDiffusion}
\left\{
1-
\frac{\drift_0^2}{3\transDiffusion}
(\transDiffusion k_{n_z}^2+2\rotDiffusion)^{-1}
\right\}
+ \order{\drift_0^4}
+ \ldots
\end{equation}
with the denominator of the fraction $\drift_0^2/(3\transDiffusion)$ changed from $\drift_0^2/(2\transDiffusion)$ in \Eref{2DABPdensity_final_wPert}, and the internal propagator changed to 
$(\transDiffusion k_{n_z}^2+2\rotDiffusion)^{-1}$ from
$(\transDiffusion k_{\nvec}^2+\rotDiffusion)^{-1}$. 
Comparing to 
\Eref{density_expanded_in_E_easy}, \Eref{3Ddensity_wPert} 
reveals corrections to order $\drift_0^2$ beyond replacing $\transDiffusion$ at $\driftVec_0=\nullvec$ in \Eref{density_expanded_in_E_easy} by $\effDiffusion{3D}$ \Eref{def_effDiffusion3D}.

\Eref{3Ddensity_wPert} correctly reproduces the result based on the expansion in "three-dimensional Mathieu functions", \Eref{3DABPdensity_final}, as
\begin{multline}
1-
\frac{\drift_0^2}{3\transDiffusion}
(\transDiffusion k_{n_z}^2+2\rotDiffusion)^{-1}
+\order{\drift_0^4}
=
    \transDiffusion
    \Bigg\{
	\big(\transDiffusion + \drift_0^2/(6\rotDiffusion) + \order{q^4}/k_{n_z}^2 \big)^{-1} 
	+
	\frac{k_{n_z}^4\drift_0^2 }{12\rotDiffusion^{2}}
	\Big[
	\big(\transDiffusion k_{n_z}^2 + \drift_0^2k_{n_z}^2/(6\rotDiffusion) + \order{q^4} \big)^{-1}\\
	-
	\big(\transDiffusion k_{n_z}^2 + 2 \rotDiffusion + \rotDiffusion \order{q^2}\big)^{-1}
	\Big]
\Bigg\}
\ .
\end{multline}

\section{Discussion and Conclusion}
There are two main perspectives on the present results, first about the concrete observables calculated and second about the formalism.
\paragraph{Observables:}
In this work we calculated the mean squared displacement and the stationary density of ABPs in two and three dimensions using two distinct methods. The MSD calculations were performed primarily for methodological purposes, although the full time-dependence may be a new result.\cite{Bechinger:2016,FrankeGruler:1990,Cates:2013,ZhangPruessner:2022}.

In two dimensions, we determined the exact MSD of ABPs using Mathieu functions with full time dependence in \Eref{MSD_2d} as well as using the perturbative approach leading to \Eref{MSDx2D_omega}. In three dimensions, similarly to the two dimensional case, we used "three-dimensional Mathieu functions" for computation of the exact full time-dependent MSD in \Eref{3DABP_MSD} and spherical harmonics in \Eref{3DABP_MSD_terms_wPert}. The effective diffusion constants for both, the two and three dimensional case in \Eref{def_effDiffusion} and \Eref{def_effDiffusion3D} respectively, are identical to the Run-and-Tumble motion case using appropriate conversion \cite{ZhangPruessner:2022,Cates:2013}.

Generalizing to $d>3$ dimensions using the perturbative approach and so-called hyperspherical harmonics \cite{Wen:1985,Askey:1983}, the expression for the MSD of ABPs is \cite{Askey:1983}
\begin{equation}
    \spave{\rvec^2}(t) =2dDt+\frac{2\drift_0^2}{(d-1)^2 D_r^2 }\bigg(\exp{-(d-1)D_r t }-1+(d-1) D_r t\bigg) \ .
\end{equation}
with the details to be found in \cite{ZZTHESIS}. The effective diffusion constant
\begin{equation}
    \effDiffusion{dD}=D+\frac{\drift_0^2}{d(d-1)D_r}
\quad\text{for}\quad
d\ge2
\end{equation}
is consistent with \cite{Cates:2013}.

The MSD was calculated irrespective of the final orientation of the ABP and integrated about the initial angle without loss of generality. This leads to a significant reduction in the number of terms in the final MSD expression as discussed after \Eref{MSDv1} in two dimensions. The same method was used in the three dimensional case with ``three-dimensional Mathieu functions" in \Sref{3DABP_Mathieu_MSD}. For the self-propulsion perturbative case, it is a matter of convenience to determine the mean squared displacement in only one spatial direction, which however requires uniform initialisation in order to avoid bias, as discussed in two dimensions after \Eref{orthoSystem_2DABP_wPert_finite}. The same approach was taken in three dimensions in \Sref{3DABP_wPert_MSD}.

The calculation of the stationary density in the presence of the external potential is, to our knowledge, a new result with the present generality. The stationary density in two dimensions was computed perturbatively in presence of small potentials in
\Erefs{2DABPdensity_final}, \eref{2DABPdensity_final_in_k} and \eref{density2D_final} to first order in the potential. These expressions differ only in that they are expanded in different parameters, i.e. the parameter $q$ of the Mathieu function in \eref{2DABPdensity_final}, the spatial momentum $\kvec_n$ in \eref{2DABPdensity_final_in_k} and $\kvec, \drift_0$ in \eref{density2D_final}.  
We further assumed a finite volume $\sysL^d$, to guarantee a stationary state in the absence of a potential, similar to drift-diffusive case discussed in \Sref{app-field-theory}. As a result of using Fourier sums, the potentials and the resulting observables are periodic. 
The two-dimensional MSD is in agreement with experimental results \cite{Howse:2007}.

In three dimensions, the stationary density calculation with the ``three dimensional Mathieu-functions" in the external potential was simplified by allowing the potential to vary in only one direction, \Eref{extPot3D_z_only}, \Appref{app_3DMathieu}. In two dimensions, this assumption was avoided by the re-orientation of the coordinate system in \Sref{ABP2D_mathieu}.
In the perturbative approach, \Sref{ABP3D_pert}, the above-mentioned assumption helped avoid having to compute general Clebsch-Gordan coefficients in \Eref{def_tumbleW3D}, but determined nevertheless in \Appref{app_CG_coeff}.

The stationary density in three dimensions in \Eref{3DABPdensity_final} is calculated to first order in the potential and expanded to second order in $\qvec$. The perturbative calculation in the self-propulsion results in \Eref{3Ddensity_wPert}. The restriction on the potential in three dimensions is more easily lifted in the perturbation theory using spherical harmonics, since the Clebsch-Gordan coefficients can be derived without much ado. On the other hand, characterising the "three-dimensional Mathieu functions" requires at the very least a careful analysis of the completeness of the eigensystem, more easily done for $\qvec\parallel\evec_z$.

Despite the restrictions on the potential, the perturbation theory covers a wide class of potentials. As is analysed in detail for drift-diffusive systems in \Sref{app-field-theory}, 
the resummed pertubation theory recovers known closed form expressions, including
a confining potential. The most interesting case of ``active sedimentation" in a gravitational potential in a finite vessel can be realised within the restrictions mentioned above by long-stretched, periodic ratchet in the $z$-direction with period $\sysL$, 
\begin{equation}\elabel{pot_example}
    \extPot(\rvec) = \mu m g \rvec\cdot\evec_z
\end{equation}
with particle mass $m$, gravitational acceleration $g$ and mobility $\mu$. With $\rvec$ confined to  $[0,\sysL)^d$ by periodicity,
$\rvec\cdot\evec_z=-\epsilon\sysL$ with $\epsilon\in[0,1)$ is mapped to $\extPot((1-\epsilon)\sysL\evec_z)$ by periodicity in \Eref{def_extPot_nvec}, so that the potential "energy" \Eref{pot_example} jumps to the maximum right after reaching the minimum for decreasing $t$, implementing a sharp potential wall at $z=0$, the bottom of the vessel.

The ``barometric formulas" in \Erefs{2DABPdensity_final_in_k} and \eref{3DABPdensity_final_in_k}
show the corrections to the density that are generically due to the activity. Those cannot possibly affect the $0$-mode, which is explicitly excluded from the summation.
To leading order the density of ABPs behaves like that of an equilibrium system with no self-propulsion, $\driftVec_0=\nullvec$ and the translational diffusion constant $\transDiffusion$, \Eref{density_expanded_in_E_easy}, replaced by the effective diffusion constant $\effDiffusion{}$, \Eref{def_effDiffusion} and \eref{def_effDiffusion3D}.
Terms beyond that are generically due to the non-equilibrium nature of active matter.
The lowest order correction is quadratic in $k_\nvec$, namely $\drift_0^2 k_\nvec^2/(2\rotDiffusion^2)$ in two dimensions and $\drift_0^2 k_\nvec^2/(12\rotDiffusion^2)$ in three dimensions. It might be interesting to investigate how this can be measured in slowly varying potentials where contributions in large $\nvec$ are suppressed because $\extPot_\nvec$ vanishes there.








\paragraph{Methodology}
We have introduced all ingredients to set up a field theory of ABPs, in particular: the action, the propagators, the projections of the special functions such as 
$\MMprojection{2}{\ell_1,\ell_3}{}{-\kvec_{\nvec_1},-\kvec_{\nvec_3}}$,
\Erefs{def_MM2} and \eref{def_MM3}), and the tumble vertices (such as
$\tumbleW{\ell',\ell}{}{\kvec}$,
\Erefs{def_tumbleW} and \eref{def_tumbleW3D}. As well as providing the concrete mathematical framework, our derivations above highlight the advantages and disadvantages of using special functions rather than a perturbation theory. We will briefly summarise these here.

Using \emph{special functions}, \ie the Mathieu functions,
\Sref{ABP2D_mathieu} and \Sref{ABP3D_mathieu},
has the benefit that the bare propagator already contains the self-propulsion. As long as space $\rvec$ and its boundary conditions can be implemented through a Fourier-transform, this is the most efficient way to deal with \emph{free} problems, wrapping the self-propulsion in the properties of certain special functions. The resulting bare propagator is the exact and complete characterisation of the free stochastic particle movement.
When the special functions are well-known, literature provides the relevant properties of functions and eigenvalues. When they are not well-known, they can be determined \emph{perturbatively} in (hyper-)spherical harmonics, effectively performing the perturbation theory that would otherwise be done explicitly at the level of diagrams. In return, the wealth of literature on Sturm-Liouville problems is readily applicable, as touched on in \Eref{simpler_Mathieu3}. 

We were able to avoid having to determine the ``three-dimensional Mathieu functions" in greater detail by using a preferred direction, which generally requires integration of final and/or initial orientation. Thanks to the arbitrary function $\sigma(\kvec)$
in \Eref{orthoSystem_2DABP}, this was avoided two dimensions, where we could use the well-known Mathieu functions. This may be related to the metric of a $d$-dimensional sphere to be curved only in $d>2$. In three dimensions, despite having \emph{two} arbitrary functions $\sigma(\kvec)$ and $\tau(\kvec)$ in \Eref{orthoSystem_3DABP}, an elegant simplification of the eigenproblem does not seem to be available.

Using a \emph{perturbation theory}, \ie expanding diagrams in $\drift_0$, 
\Sref{ABP2D_pert} and \Sref{ABP3D_pert},
means that even the free particle has to be treated explicitly perturbatively. 
Although the full propagator now contains infinitely many terms, this does 
necessarily translate to infinitely many terms for every observable, as seen, for example, in the fully time-dependent MSD calculated exactly using perturbation theory in two and three dimensions, \Sref{2DABP_wPert_MSD} and \Sref{3DABP_wPert_MSD} respectively. The big advantage of using a perturbation theory is that orthogonality is not spoiled by a lack of momentum conservation, as is caused by an external potential or interaction. Such a lack of momentum conservation is what necessitates the introduction of the projection $\MMprojection{2}{\ell_1,\ell_3}{}{-\kvec_{\nvec_1},-\kvec_{\nvec_3}}$ in the case of special functions, \Eref{action_2DABP_pert_k2}
and similarly \eref{action_3DABP_pert_k2},
but this does not happen when using a perturbation theory, \Erefs{def_actionP2_2DABP_wPert_k} and \eref{def_actionP2_3DABP_wPert}.

It is striking that the perturbation theory in weak potentials recovers the barometric formula for a binding potential (\Appref{app-field-theory}).
The calculation of such stationary features is facilitated by the amputation mechanism as discussed in \Sref{2DABP_Mathieu_extPot} and similarly used in \cite{ZhangPruessner:2022}.

\paragraph{Outlook:}
We are currently working on a field theory of \emph{interacting} ABPs
displaying features of the Vicsek Model \cite{Vicsek:1995, Ginelli:2016}. 
As the upper critical dimension there is $d_c=4$, we aim to generalise the spherical harmonics to higher dimensions. This would also allow the use of dimensional regularisation in a renormalised field theory. While some characteristics of such hyperspherical functions follow a systematic pattern, allowing the spatial dimension to vary continuously like $d=4-\epsilon$ represents a significant challenge.

\section*{Acknowledgements}
The authors would like to thank Wilson Poon for providing relevant material. We also thank Rosalba Garcia-Millan for her hint on Mathieu functions.

\bibliography{articles}
\bibliographystyle{unsrt}

\appendix
\section{Three-dimensional Mathieu functions}
\label{sec:app_3DMathieu}
In the following, we want to characterise the "three-dimensional Mathieu functions", which we call that only because the eigenfunctions of the equations below in two dimensions are the well-known Mathieu-functions \cite{ZIENER2012,abramowitzStegun}.

We will first state the defining equations and then set out to calculate the eigenfunctions. Firstly, we demand that $\Mathieu{3}{\ell}{m}{\theta,\varphi}{\qvec}$ are eigenfunctions,
\Eref{Mathieu3_eigenEquation}
\begin{equation}\elabel{AppMathieu3_eigenEquation}
    \Big(
- \qvec\cdot\left(
\begin{array}{c}
     \sin\!\theta\cos\!\varphi \\
     \sin\!\theta\sin\!\varphi \\
     \cos\!\theta
\end{array}
\right)
+
    \frac{1}{\sin\!\theta}\partial_\theta\sin\!\theta\ \partial_\theta
    +
    \frac{1}{\sin^2\!\theta}\partial_\varphi^2
    \Big)
    \Mathieu{3}{\ell}{m}{\theta,\varphi}{\qvec}
=
- \lambda_\ell^m(\qvec) \Mathieu{3}{\ell}{m}{\theta,\varphi}{\qvec} \ .
\end{equation}
Further, we demand normalisation,
\begin{equation}\elabel{AppMathieu3_orthonormality}
    \int_0^\pi\dint{\theta}\int_0^{2\pi}\dint{\varphi}\sin\!\theta\ 
    \Mathieu{3}{\ell}{m}{\theta,\varphi}{\qvec}
    \Mathieutilde{3}{\ell'}{m'}{\theta,\varphi}{\qvec}
    =
    \delta_{\ell,\ell'}\delta_{m,m'}
\end{equation}
with \Eref{def_Mathieu3Tilde} to maintain close correspondence to the two-dimensional case \Eref{def_Mathieu2Tilde},
\begin{equation}\elabel{Appdef_Mathieu3Tilde}
\Mathieutilde{3}{\ell}{m}{\theta,\varphi}{\qvec(\kvec)}=
\frac{1}{4\pi}
\Mathieu{3}{\ell}{m}{\theta,\varphi}{\qvec(\kvec)}
\end{equation}
and
orthogonality following from the self-adjointness of the differential operator \Eref{AppMathieu3_eigenEquation} under the scalar product \Eref{AppMathieu3_orthonormality}. With that, all of Sturm-Liouville theory \cite{Zettl:2005} is at our disposal.
We will index these eigenfunctions with $\ell=0,1,\ldots$ and $m\in\Zset$. 

For $\qvec=\nullvec$ vanishing, \Eref{AppMathieu3_eigenEquation} is the eigen-equation of the spherical harmonics \cite{abramowitzStegun}. We therefore choose to express the three-dimensional Mathieu functions in spherical harmonics $\SphericalHarmonics{j}{k}{\theta,\varphi}$, 
\begin{equation}\elabel{Mathieu3_in_SphericalHarmonics}
    \Mathieu{3}{\ell}{m}{\theta,\varphi}{\qvec}
    =
    \sqrt{4\pi}
    \sum_{j=0}^\infty\sum_{k=-j}^j
    \MathieuA{\ell,j}{m,k}{\qvec}
    \SphericalHarmonics{j}{k}{\theta,\varphi}
    =
    \sum_{j=0}^\infty\sum_{k=-j}^j
    \MathieuA{\ell,j}{m,k}{\qvec}
    \sqrt{\frac{(2j+1)(j-k)!}{(j+k)!}}
    \Legendre{j}{k}{\cos\!\theta}
    \exp{\imag k \varphi}
\end{equation}
with the factor of $\sqrt{4\pi}$ solely to ease notation and 
$\Legendre{\ell}{m}{z}$ denoting the associated Legendre polynomials.
For $\qvec=\nullvec$,
$\MathieuA{\ell,j}{m,k}{\qvec}$ is diagonal, so that
\begin{subequations}\elabel{Mathieu_q0}
\begin{align}
    \elabel{Mathieu_q0_alpha}
    \MathieuA{\ell,j}{m,k}{\qvec}
    &= \left(\delta_{\ell,j}\delta_{m,k} + \order{\qvec}\right)\\
    \elabel{Mathieu_q0_lambda}
    \MathieuLambda{\ell}{m}{\qvec} &= \ell(\ell+1) + \order{\qvec}
    \ ,
\end{align}
\end{subequations}
where we assume that the coefficients $\MathieuA{\ell,j}{m,k}{\qvec}$ and eigenvalues $\MathieuLambda{\ell}{m}{\qvec}$ can be expanded in small $\qvec$. Explicitly, for $\qvec=\nullvec$,
\begin{subequations}\elabel{Mathieu3_as_SphericalHarmonics}
\begin{align}
    \Mathieu{3}{\ell}{m}{\theta,\varphi}{\nullvec} &= \sqrt{4\pi} \SphericalHarmonics{\ell}{m}{\theta,\varphi}\\
        \Mathieutilde{3}{\ell}{m}{\theta,\varphi}{\nullvec} &= \frac{1}{\sqrt{4\pi}} \big(\SphericalHarmonics{\ell}{m}{\theta,\varphi}\big)^* \ ,
\end{align}
\end{subequations}
with the asterisk denoting the complex conjugate and indeed
\begin{equation}
    \int_0^\pi\dint{\theta}\int_0^{2\pi}\dint{\varphi}\sin\!\theta
    \Mathieu{3}{\ell}{m}{\theta,\varphi}{\nullvec} \Mathieutilde{3}{\ell'}{m'}{\theta,\varphi}{\nullvec}
    = \delta_{\ell,\ell'}\delta_{m,m'}  \ .
\end{equation}

We focus henceforth on those $\qvec$ that have no projection in the $x$ and $y$ directions, $\qvec=(0,0,q)^\transpose=q\evec_z$, which simplifies \Eref{AppMathieu3_eigenEquation} dramatically, rendering it similar to the expression in two dimensions, \Eref{Mathieu_eqn}. 
Only the $\theta$-dependence is affected by the presence of $\qvec\parallel\evec_z$, so that 
$\MathieuA{\ell,j}{m,k}{\qvec}\propto\delta_{m,k}$ is diagonal in $m,k$ for such $\qvec$,
resulting in a simple structure and indexing of the azimuthal dependence, \ie on $\varphi$,
\begin{equation}\elabel{simpler_Mathieu3}
    \Mathieu{3}{\ell}{m}{\theta,\varphi}{q\evec_z}
    =
    \exp{\imag m \varphi}
    \sum_{j=0}^\infty
    \MathieuA{\ell,j}{m,m}{q\evec_z}
    \sqrt{\frac{(2j+1)(j-m)!}{(j+m)!}}
    \Legendre{j}{m}{\cos\!\theta} \ .
\end{equation}
Because we will integrate $\Mathieu{3}{\ell}{m}{\theta,\varphi}{q\evec_z}$ over $\varphi$, we can further narrow down the scope of this section to $m=0$, which simplifies the square root in \Eref{simpler_Mathieu3} to $\sqrt{2j+1}$.

Because the Legendre polynomials obey
\begin{equation}\elabel{cos_times_Legendre}
    cos\theta P_\ell^m=\frac{\ell-m+1}{2\ell+1}P_{\ell+1}^m+\frac{\ell+m}{2\ell+1}P_{\ell-1}^m\ ,
\end{equation}
the eigen-equation \Eref{AppMathieu3_eigenEquation} for $m=0$ and $\qvec=q\evec_z$ with \Eref{Mathieu3_in_SphericalHarmonics} 
is simply
\begin{equation}
0=
\sum_{j=0}^\infty
\MathieuA{\ell,j}{0,0}{q\evec_z}
\sqrt{2j+1}
\Bigg(
q\frac{j+1}{2j+1}\Legendre{j+1}{0}{\cos\!\theta}
+
q\frac{j}{2j+1}\Legendre{j-1}{0}{\cos\!\theta}
+
\big(
j(j+1)-\MathieuLambda{\ell}{0}{q\evec_z}
\big)
\Legendre{j}{0}{\cos\!\theta} 
\Bigg)
\ .
\end{equation}
Projecting out the Legendre polynomials one by one, the coefficients 
$\MathieuA{\ell,j}{0,0}{q\evec_z}$ obey
\begin{equation}\elabel{Mathieu3_recurrence}
0=
    \alpha_{\ell,j-1} q \frac{j}{\sqrt{2j-1}}
    +
    \alpha_{\ell,j+1} q \frac{j+1}{\sqrt{2j+3}}
    +
    \alpha_{\ell,j}\sqrt{2j+1} \big(j(j+1)-\MathieuLambda{\ell}{0}{}\big)
\end{equation}
where we have used 
\begin{equation}\elabel{def_alpha}
\alpha_{\ell,j}=\NC_\ell^0 \MathieuA{\ell,j}{0,0}{q\evec_z} 
\ ,
\end{equation}
with some normalisation $\NC_\ell^0$ to be used below,
and $\MathieuLambda{\ell}{}{}=\MathieuLambda{\ell}{0}{q\evec_z}$
to ease notation.

\Eref{Mathieu3_recurrence} can of course be written in matrix form
\begin{equation}
        \begin{pmatrix}
    0 & \frac{q}{\sqrt{3}} & 0 & \\
    \frac{q}{\sqrt{3}} & 2 & \frac{2q}{\sqrt{15}} & \ldots \\
    0 & \frac{2q}{\sqrt{15}} & 6 & \\
      & \vdots & & \ddots 
    \end{pmatrix}
    \begin{pmatrix}
    \alpha_{\ell,0}\\
    \alpha_{\ell,1}\\
    \alpha_{\ell,2}\\
    \vdots
    \end{pmatrix}
    =
    \MathieuLambda{\ell}{0}{q\evec_z}
    \begin{pmatrix}
    \alpha_{\ell,0}\\
    \alpha_{\ell,1}\\
    \alpha_{\ell,2}\\
    \vdots
    \end{pmatrix}
\end{equation}
By demanding that this equation is to be solved to a certain order in $q$ and by \emph{assuming} $\alpha_{\ell,j}\in\order{q^{|\ell-j|}}$, only a finite number of matrix elements and vector elements need to be considered at a time, while also constraining the order of the eigenvalue. Using \Eref{Mathieu_q0} as the starting point, the first order equation for $\ell=0$ is
\begin{equation}\elabel{matrix_procedure0}
    \begin{pmatrix}
    0 & \frac{q}{\sqrt{3}} \\
    \frac{q}{\sqrt{3}} & 2
    \end{pmatrix}
    \begin{pmatrix}
    1\\
    u_1 q
    \end{pmatrix}
    =
    (0+vq) 
    \begin{pmatrix}
    1\\
    u_1 q
    \end{pmatrix}
+ \order{q^2} \ ,
\end{equation}
where we have labelled the unknown term of $\alpha_{0,1}$ by $u_1 q$ and the unknown term of $\MathieuLambda{0}{0}{q\evec_z}$ by $v q$.
The $0$ in the eigenvalue on the right hand side of \Eref{matrix_procedure0} is that of $\MathieuLambda{\ell}{0}{}$ at $\ell=0$, \Eref{Mathieu_q0_lambda}. Focussing solely on terms of order $q$ produces two equations, $0=vq$ and $q/\sqrt{3}+2u_1q=0$, determining both $v=0$ and $u_1=-1/(2\sqrt{3})$. With those in place, the next order equation is  
\begin{equation}\elabel{matrix_procedure1}
    \begin{pmatrix}
    0 & \frac{q}{\sqrt{3}} & 0 \\
    \frac{q}{\sqrt{3}} & 2 & \frac{2q}{\sqrt{15}} \\
    0 & \frac{2q}{\sqrt{15}} & 6
    \end{pmatrix}
    \begin{pmatrix}
    1\\
    -\frac{q}{2\sqrt{3}} + u_1 q^2\\
    u_2 q^2
    \end{pmatrix}
    =
    (0+vq^2) 
    \begin{pmatrix}
    1\\
    -\frac{q}{2\sqrt{3}} + u_1 q^2\\
    u_2 q^2
    \end{pmatrix}
+ \order{q^3} \ ,
\end{equation}
determining $u_1$, $u_2$ and $v$, now labelling terms of order $q^2$. 

The only additional difficulty for $\ell>0$ is that the number of rows and columns of the matrix to be considered going from one order to the next may increase by $2$ for the first $\ell$ orders. For example the first non-trivial order for $\ell=1$ is
\begin{equation}
    \begin{pmatrix}
    0 & \frac{q}{\sqrt{3}} & 0 \\
    \frac{q}{\sqrt{3}} & 2 & \frac{2q}{\sqrt{15}} \\
    0 & \frac{2q}{\sqrt{15}} & 6
    \end{pmatrix}
    \begin{pmatrix}
    u_0 q\\
    1\\
    u_2 q
    \end{pmatrix}
    =
    (2+vq) 
    \begin{pmatrix}
    u_0 q\\
    1\\
    u_2 q
    \end{pmatrix}
+ \order{q^2} \ .
\end{equation}

With this process, we have determined 
\begin{equation}\elabel{matrix_ell0_result}
    \Bigg[
    \begin{pmatrix}
    0 & \frac{q}{\sqrt{3}} & 0 & 0 & 0\\
    \frac{q}{\sqrt{3}} & 2 & \frac{2q}{\sqrt{15}} & 0 & 0\\
    0 & \frac{2q}{\sqrt{15}} & 6 & \frac{3q}{\sqrt{35}} & 0 \\
    0 & 0 & \frac{3q}{\sqrt{35}} & 12 & \frac{4q}{\sqrt{63}} \\
    0 & 0 & 0 & \frac{4q}{\sqrt{63}} & 20 \\
    \end{pmatrix}
    - \left(-\frac{q^2}{6}+\frac{11q^4}{1080}\right)\ident
    \Bigg]
    \begin{pmatrix}
    1\\
    -\frac{q}{2\sqrt{3}} + \frac{11q^3}{360\sqrt{3}}\\
    \frac{q^2}{18\sqrt{5}} - \frac{43q^4}{9072\sqrt{5}}\\
    -\frac{q^3}{360\sqrt{7}}\\
    \frac{q^4}{37800}
    \end{pmatrix}
    \in
  \order{q^5} 
\qquad\qquad\text{for }\ell=0
\ ,
\end{equation}
which can be improved further by demanding that the term of order $q^6$ in $\MathieuLambda{0}{0}{}$ is consistent with $\alpha_{0,2}=q^2/(18\sqrt{5})-43q^4/(9072\sqrt{5})$ in \Eref{matrix_ell0_result}, producing 
\begin{equation}\elabel{MathieuLambda3D_00}
    \MathieuLambda{0}{0}{q\evec_z}
    =-\frac{q^2}{6}+\frac{11q^4}{1080}-\frac{47q^6}{34020} + \order{q^7}
    \qquad \text{for }\ell=0 \ .
\end{equation}
For $\ell=1$, we found
\begin{equation}\elabel{matrix_ell1_result}
    \Bigg[
    \begin{pmatrix}
    0 & \frac{q}{\sqrt{3}} & 0 & 0 & 0\\
    \frac{q}{\sqrt{3}} & 2 & \frac{2q}{\sqrt{15}} & 0 & 0\\
    0 & \frac{2q}{\sqrt{15}} & 6 & \frac{3q}{\sqrt{35}} & 0 \\
    0 & 0 & \frac{3q}{\sqrt{35}} & 12 & \frac{4q}{\sqrt{63}} \\
    0 & 0 & 0 & \frac{4q}{\sqrt{63}} & 20 \\
    \end{pmatrix}
    - \left(2+\frac{q^2}{10}\right)\ident
    \Bigg]
    \begin{pmatrix}
    \frac{q}{2\sqrt{3}} - \frac{q^3}{40\sqrt{3}}\\
    1\\
    -\frac{q}{2\sqrt{15}} - \frac{11q^3}{700\sqrt{15}}\\
    \frac{3q^2}{100\sqrt{21}}\\
    -\frac{q^3}{3150\sqrt{3}}
    \end{pmatrix}
    \in
  \order{q^4} 
\qquad\qquad\text{for }\ell=1
\ ,
\end{equation}
\ie
\begin{equation}\elabel{MathieuLambda3D_10}
    \MathieuLambda{1}{0}{q\evec_z}
    = 2+\frac{q^2}{10} + \order{q^4}
    \qquad \text{for }\ell=1 \ .
\end{equation}
Matrix expressions such as \Erefs{matrix_ell0_result} and \eref{matrix_ell1_result} are easily verified with a computer algebra system, simply by determining whether the remainder has the expected order.




\Erefs{AppMathieu3_orthonormality} and \eref{Mathieu3_in_SphericalHarmonics} imply
\begin{equation}
    \sum_{j=0}^\infty 
    \sum_{k=-j}^j
    \left(\MathieuA{\ell,j}{m,k}{q\evec_z}\right)^2
= 1
\end{equation}
which determines the normalization $\mathcal{N}_\ell^0$ in \Eref{def_alpha} and with \Eref{matrix_ell0_result} this produces the desired coefficients for $\ell=0$,
\begin{subequations}\elabel{MathieuA3D_0}
\begin{align}
\elabel{MathieuA3D_00}
\MathieuA{0,0}{0,0}{q\evec_z} &=
1-\frac{q^2}{24}+\frac{383 q^4}{51840}+\order{q^5}\\
\MathieuA{0,1}{0,0}{q\evec_z} &=
-\frac{q}{2\sqrt{3}}+\frac{37 q^3}{720 \sqrt{3}}+\order{q^5}\\
\MathieuA{0,2}{0,0}{q\evec_z} &=
\frac{q^2}{18\sqrt{5}}-\frac{4 q^4}{567 \sqrt{5}}+\order{q^5}\\
\MathieuA{0,3}{0,0}{q\evec_z} &=
-\frac{q^3}{360\sqrt{7}} + \order{q^5}\\
\MathieuA{0,4}{0,0}{q\evec_z} &=
\frac{q^4}{37800}+\order{q^5}\ .
\end{align}
\end{subequations}
Similarly, we found for $\ell=1$ from \Eref{matrix_ell1_result},
\begin{subequations}
\begin{align}
\elabel{MathieuA3D_10}
\MathieuA{1,0}{0,0}{q\evec_z} &=
\frac{q}{2 \sqrt{3}}-\frac{q^3}{20\sqrt{3}}+\order{q^4}\\
\MathieuA{1,1}{0,0}{q\evec_z} &=
1-\frac{q^2}{20}+\order{q^4}\\
\MathieuA{1,2}{0,0}{q\evec_z} &=
-\frac{q}{2\sqrt{15}}+\frac{13 q^3}{1400 \sqrt{15}}+\order{q^4}\\
\MathieuA{1,3}{0,0}{q\evec_z} &=
\frac{1}{100}\sqrt{\frac{3}{7}} q^2+\order{q^4} \ .
\end{align}
\end{subequations}
For general $\ell$ we observe that in deed
\begin{equation}
    \elabel{MathieuA3D_ell}
    \MathieuA{\ell,0}{0,0}{q\evec_z}\in\order{q^\ell}
\end{equation}
by construction.

Finally, we calculate the integral of $\Mathieu{3}{\ell}{m}{\theta,\varphi}{\qvec(q\evec_z)}$ over $\varphi$ and $\theta$, \Eref{integral_Mathieu3}. Firstly, taking the integral over $\varphi$ of the definition \Eref{simpler_Mathieu3} enforces $m=0$,
\begin{equation}
\int_0^\pi\dint{\theta}
	\int_0^{2\pi}\dint{\varphi}\sin\!\theta
\Mathieu{3}{\ell}{m}{\theta,\varphi}{\qvec(\kvec)}
= 2 \pi \delta_{m,0}
\sum_{j=0}^\infty 
\MathieuA{\ell,j}{0,0}{\qvec(\kvec)}
\sqrt{2j+1} \int_0^\pi\dint{\theta}\sin\!\theta\, \Legendre{j}{0}{\cos\!\theta}
\ .
\end{equation}
Using the orthogonality of the Legendre polynomials,
\begin{equation}
    \int_0^\pi\dint{\theta}\sin\!\theta\, 
    \Legendre{j}{m}{\cos\!\theta}
    \Legendre{k}{m}{\cos\!\theta}
    =
    \frac{2(j+m)!}{(2j+1)(j-m)!}
    \delta_{j,k}
\end{equation}
and $\Legendre{0}{0}{}=1$, which means that the integral of $\Legendre{j}{0}{}$ is its projection against $\Legendre{0}{0}{}=1$, then gives the desired
\begin{equation}\elabel{integral_Mathieu3_app}
\int_0^\pi\dint{\theta}
	\int_0^{2\pi}\dint{\varphi}\sin\!\theta
\Mathieu{3}{\ell}{m}{\theta,\varphi}{\qvec(\kvec)}
= 4 \pi \delta_{m,0}
\MathieuA{\ell,0}{0,0}{\qvec(\kvec)}
\ .
\end{equation}

\section{Clebsch-Gordan coefficients}
\label{sec:app_CG_coeff}
A product of two spherical harmonics can be  expanded as \begin{equation}
\elabel{expand_spherical_harmonics}
    \SphericalHarmonics{\ell}{m}{\theta,\varphi}\SphericalHarmonics{\ell'}{m'}{\theta,\varphi}=\sum_{L,M}\sqrt{\frac{(2\ell+1)(2\ell'+1)}{4\pi (2L+1)}}c_{\ell',0,\ell,0}^{L,0}c_{\ell',m',\ell,m}^{L,M}\SphericalHarmonics{L}{M}{\theta,\varphi}\ ,
\end{equation}
where $c_{\ell',m',\ell,m}^{L,M}$ are the Clebsch-Gordan coefficients  \cite{Racah:1942}.
By rewriting trigonometric functions in \Eref{def_tumbleW3D} as spherical harmonics,
\begin{subequations}
\begin{align}
    \sin\!\theta\cos\varphi&=\sqrt{\frac{2\pi}{3}}\bigg(-\SphericalHarmonics{1}{1}{{\theta,\varphi}}+\SphericalHarmonics{1}{-1}{{\theta,\varphi}}\bigg)\\
    \sin\!\theta\sin\!\theta&=\imag\sqrt{\frac{2\pi}{3}}\bigg(\SphericalHarmonics{1}{1}{{\theta,\varphi}}+\SphericalHarmonics{1}{-1}{{\theta,\varphi}}\bigg)\\
    \cos\!\theta&=\sqrt{\frac{4\pi}{3}}\SphericalHarmonics{1}{0}{\theta,\varphi}\ ,
\end{align}
\end{subequations}
and using \Eref{expand_spherical_harmonics}, we arrive at \Eref{pert_expansion_in_CG}.

We list the Clebsch-Gordan coefficients of \Eref{pert_expansion_in_CG},
\begin{subequations}
\begin{align}
    c_{1,0,\ell,0}^{L,0}&=\delta_{L,\ell+1}\sqrt{\frac{1+\ell}{1+2\ell}}-\delta_{L,\ell-1}\sqrt{\frac{\ell}{1+2\ell}}\ ,\\
    c_{1,0,\ell,m}^{L,M}&=\delta_{m,M}\bigg(\delta_{L,\ell+1}\sqrt{\frac{(\ell+1-m)(\ell+1+m)}{(1+\ell)(1+2\ell)}}-\delta_{L,\ell}m\sqrt{\frac{1}{\ell(1+\ell)}}-\delta_{L,\ell-1}\sqrt{\frac{(\ell-m)(\ell+m)}{\ell(2\ell+1)}}\bigg)\ ,\\
    c_{1,1,\ell,m}^{L,M}&=
    \frac{\delta_{m+1,M}}{\sqrt{2}}\bigg(
    \delta_{L,\ell+1}
    \sqrt{\frac{(1+\ell+m)(2+\ell+m)}{(1+\ell)(1+2\ell)}}
    +\delta_{L,\ell-1}
    \sqrt{\frac{(\ell-1-m)(\ell-m)}{\ell(1+2\ell)}}
    +\delta_{L,\ell}
    \sqrt{\frac{(\ell-m)(1+\ell+m)}{\ell(1+\ell)}}
    \bigg)
     \ ,\\
     c_{1,-1,\ell,m}^{L,M}&=\frac{\delta_{m-1,M}}{\sqrt{2}}\bigg(\delta_{L,\ell+1}\sqrt{\frac{(1+\ell-m)(2+\ell-m)}{(1+\ell)(1+2\ell)}}
     +\delta_{L,\ell-1}\sqrt{\frac{(\ell+m-1)(\ell+m)}{\ell(1+2\ell)}}
-\delta_{L,\ell}\sqrt{\frac{(1+\ell-m)(\ell+m)}{\ell(1+\ell)}}    
     \bigg)\ .
\end{align}
\end{subequations}
By substituting  the above Clebsch-Gordan coefficients into \Eref{pert_expansion_in_CG}, the first two tumble vertices $W_{\ell',\ell}^{m',m}$ on the right hand side of \Eref{Coupling_matrix_W} are
\begin{subequations}
\begin{align}
    \tumbleW{\ell',\ell}{m',m}{k_x\evec_x}&=
-\imag\drift_0 k_x \sum_{L,M}\sqrt{\frac{2\ell+1}{2(2L+1)}}c_{1,0,\ell,0}^{L,0}\bigg(-c_{1,1,\ell,m}^{L,M}+c_{1,-1,\ell,m}^{L,M}\bigg)\delta_{m',M}\delta_{\ell',L}\\
&=-\imag \drift_0 k_x\bigg\{
\frac{\delta_{m',m+1}}{2}\bigg(-\delta_{\ell',\ell+1}\sqrt{\frac{(1+\ell+m)(2+\ell+m)}{(2\ell+3)(2\ell+1)}}+\delta_{\ell',\ell-1}\sqrt{\frac{(\ell-1-m)(\ell-m)}{(2\ell-1)(2\ell+1)}}\bigg)\\
&\quad\quad+\frac{\delta_{m',m-1}}{2}\bigg(\delta_{\ell',\ell+1}\sqrt{\frac{(1+\ell-m)(2+\ell-m)}{(2\ell+3)(2\ell+1)}}-\delta_{\ell',\ell-1}\sqrt{\frac{(\ell+m)(\ell+m-1)}{(2\ell+1)(2\ell-1)}}\bigg)
\bigg\}\ ,
\\
\tumbleW{\ell',\ell}{m',m}{k_y\evec_y}&=
\drift_0 k_y \sum_{L,M}\sqrt{\frac{2\ell+1}{2(2L+1)}}c_{1,0,\ell,0}^{L,0}\bigg(c_{1,1,\ell,m}^{L,M}+c_{1,-1,\ell,m}^{L,M}\bigg)\delta_{m',M}\delta_{\ell',L}\\
&= \drift_0 k_y\bigg\{
\frac{\delta_{m',m+1}}{2}\bigg(\delta_{\ell',\ell+1}\sqrt{\frac{(1+\ell+m)(2+\ell+m)}{(2\ell+3)(2\ell+1)}}-\delta_{\ell',\ell-1}\sqrt{\frac{(\ell-1-m)(\ell-m)}{(2\ell-1)(2\ell+1)}}\bigg)\\
&\quad\quad+
\frac{\delta_{m',m-1}}{2}\bigg(\delta_{\ell',\ell+1}\sqrt{\frac{(1+\ell-m)(2+\ell-m)}{(2\ell+3)(2\ell+1)}}-\delta_{\ell',\ell-1}\sqrt{\frac{(\ell+m)(\ell+m-1)}{(2\ell+1)(2\ell-1)}}\bigg)
\bigg\}
\end{align}
\end{subequations}
which together with the third term, $\tumbleW{\ell',\ell}{m',m}{k_z\evec_z}$ in \Eref{Coupling_matrix_Wz} complete the right hand side of \Eref{Coupling_matrix_W}.

\section{Field Theory of the Drift Diffusion in an Arbitrary Potential}\label{sec:app-field-theory}

In the following we derive the field theory for a drift-diffusion
particle in a finite volume with potential $\extPot(\rvec)$ and
determine the stationary particle density $\density_0(\rvec)$. 
Our motivation is two-fold: Firstly, we want to reproduce the barometric formula $\density_0(\rvec)\propto\exp{-\extPot(\rvec)/\transDiffusion}$ in the absence of drift, which is the equilibrium situation of pure diffusion, where the diffusion constant $\transDiffusion$ plays the role of the temperature. The barometric formula is Boltzmann's factor, which we want to see reproduced by the field theory as a matter of consistency, \Sref{barometric_formula}. The derivation will provide us with a template for the more complicated case of a drift. 
Secondly, we want to determine precisely those corrections due to a constant drift with velocity $\driftVec_0$. We will compare our field-theoretic, perturbative result with the classical calculation in one dimension in \Sref{field_theory_produces_1D_drift}.

We study this phenomenon in a \emph{finite} volume with linear extent $\sysL$, so that the system develops into a stationary state even in the absence of an external potential, which will be treated perturbatively. In a finite volume, the Fourier series representation of any reasonable (finite, continuous, more generally fulfilling the Dirichlet conditions) potential exists. In an infinite volume, we would have to demand that the potential vanishes at large distances, which may result in the stationary density no longer being normalisable even in the presence of the potential.

\newcommand{\indicator}[1]{\mathcal{I}(#1)}
\newcommand{\Hbaro}{B}
\newcommand{\Eft}{E}
\newcommand{\Hdrift}{F}
\newcommand{\HdriftSub}{f}

To simplify the calculations, we further demand the domain to be periodic.
The annihilation field $\chi(\rvec,t)$ and the
Doi-shifted creation field $\chitilde(\rvec,t)$ at position $\rvec$ and time $t$ can then be written as
\begin{subequations}
\begin{align}
\chitilde(\rvec,t) &= \frac{1}{\sysL^d}\sum_{\nvec\in\Zset^d}\int_{-\infty}^\infty\dintbar{\omega} \exp{-\imag\omega t} \exp{\imag \kvec_{\nvec}\cdot\rvec} \chitilde(\kvec_\nvec,\omega)\\
\chi(\rvec,t) &= \frac{1}{\sysL^d}\sum_{\nvec\in\Zset^d}\int_{-\infty}^\infty\dintbar{\omega} \exp{-\imag\omega t} \exp{\imag \kvec_{\nvec}\cdot\rvec} \chi(\kvec_\nvec,\omega)
    \elabel{inverseFT}
\end{align}
\end{subequations}
with the notation in keeping with the rest of the present work. The finite volume we have chosen to be a cube with linear extension $\sysL$, easily generalisable to a cuboid or similar. The Fourier modes are $\kvec_\nvec=2\pi\nvec/\sysL$ with $\nvec=(n_1,n_2,\ldots,n_d)^\transpose$ a $d$-dimensional integer in $\Zset^d$. We further use the notation  $\dbar=\dint/\!(2\pi)$ and later, correspondingly, $\deltabar(\omega)=2\pi\delta(\omega)$ to cancel that factor of $2\pi$.

The time-independent potential $\extPot(\rvec)$  follows the  convention \Eref{def_extPot_nvec}, restated here for convenience in arbitrary dimensions
\begin{subequations}
\begin{align}
    \extPot(\rvec)&=\frac{1}{\sysL^d}\sum_{\nvec\in\Zset^d} \extPot_\nvec\exp{\imag\kvec_\nvec\cdot\rvec}\\
    \extPot_\nvec&=
    \int_0^{\sysL}\ldots\int_0^{\sysL}
    \ddint{r}\extPot(\rvec)\exp{-\imag\kvec_\nvec\cdot\rvec}
    \ .
\end{align}
\end{subequations}

\subsection{Action}
The Fokker-Planck equation of the fully time-dependent density $\density(\rvec,t)$ of particles at position $\rvec$ and time $t$, subject to diffusion with constant $\transDiffusion$ and drift with velocity $\driftVec$, in a potential $\extPot(\rvec)$ can be written as 
\begin{equation}\elabel{FPE_app}
\partial_t\density(\rvec,t) = 
    \transDiffusion\nabla_\rvec^2 
    \density(\rvec,t)
    - \nabla_\rvec\cdot\big( \density(\rvec,t) (\driftVec-\nabla\extPot(\rvec))\big)
\end{equation}
removing all ambiguity about the action of the gradient operators by stating that they act on all objects to their right. 
The action of particles with diffusion constant $\transDiffusion$ and moving ballistically with velocity $\driftVec_0$ is \cite{ZhangPruessner:2022}
\begin{equation}\elabel{harmonic_action}
    \action_0[\chitilde,\chi]=
    -\int\dintbar{\omega}\frac{1}{\sysL^d}\sum_{\nvec\in\Zset^d}
    \chitilde(-\kvec_\nvec,-\omega)
    \chi(\kvec_\nvec,\omega)
    \big(
    -\imag\omega+\transDiffusion k_\nvec^2 + \imag\driftVec_0\cdot\kvec_\nvec+\mass
    \big)
\end{equation}
with $k_\nvec=|\kvec_\nvec|$ and $\mass\downarrow0$ a mass to regularise the infrared and establish causality. An additional external potential results in an additional space-dependent drift term and thus in the action
\begin{subequations}
\begin{align}\elabel{pert_action}
	\action_{P}&=\int\ddint{r}\int\dint{t}
    \chitilde(\rvec,t)
	\nabla_\rvec\cdot\big(\nabla_\rvec\extPot(\rvec)\big)
    \chi(\rvec,t)\\
    \elabel{pert_action_k}
    &=\int\dintbar{\omega}
    \frac{1}{\sysL^{3d}}\sum_{\nvec_1,\nvec_2,\nvec_3}
    \kvec_{\nvec_1}\cdot\kvec_{\nvec_2}\,
    \chitilde(\kvec_{\nvec_1},-\omega)
    \extPot_{\nvec_2}
    \chi(\kvec_{\nvec_3},\omega)
    \sysL^d\delta_{\nvec_1+\nvec_2+\nvec_3,\nullvec}
\end{align}
\end{subequations}\eg \Erefs{def_actionP1} or \eref{action_2DABP_pert_X}
to be treated perturbatively, \Eref{perturbative_action_use}.

The bare propagator follows immediately from \Eref{harmonic_action},
\begin{equation}\elabel{bareProp_app_baro}
    \ave{\chi(\kvec_\nvec,\omega)\chitilde(\kvec_\pvec,\omega_0)}_0=\frac{\deltabar(\omega+\omega_0)\sysL^d\delta_{\nvec+\pvec,\nullvec}}
    {-\imag\omega+\transDiffusion k_\nvec^2+\imag\driftVec_0\cdot\kvec+\mass}
    =\deltabar(\omega+\omega_0)\sysL^d\delta_{\nvec+\pvec,\nullvec}
    \barepropG{\kvec_\nvec}{\omega}
    \corresponds
    \bareprop{\nvec,\omega}{\pvec,\omega'}
\end{equation}
and the perturbative part of the action provides a single vertex with a ``bauble"
\begin{equation}\elabel{pertVertex_app_baro}
    \kvec_{\nvec_1}\cdot\kvec_{-(\nvec_1+\nvec_3)}
    \extPot_{-(\nvec_1+\nvec_3)}
    \corresponds \singlepertpropPot{\nvec_1}{\nvec_3}{-(\nvec_1+\nvec_3)}
\end{equation}
to be summed over.

\subsection{Propagator}
With the help of \Erefs{bareProp_app_baro} and \eref{pertVertex_app_baro}
the full propagator can be written perturbatively,
\begin{align}\elabel{full_propagator_attempt1}
    \ave{\chi(\kvec_\nvec,\omega)\chitilde(\kvec_\pvec,\omega_0)}
    &\corresponds
    \bareprop{\nvec,\omega}{\pvec,\omega'}
    +
    \bareprop{\nvec,\omega}{}\!\!\!\!\pertprop{}{}\!\!\!\!\bareprop{}{\pvec,\omega_0}
    +
    \bareprop{\nvec,\omega}{}\!\!\!\!\pertprop{}{}\!\!\!\!\!\!\!\!\!\!\!\!\!\!\!\!\!\bareprop{}{}\!\!\!\!\pertprop{}{}\!\!\!\!\bareprop{}{\pvec,\omega_0}
    + \ldots\\
    &=\deltabar(\omega+\omega_0)
    \Big\{
    \sysL^d\delta_{\nvec+\pvec,0}\barepropG{\kvec_\nvec}{\omega}
    +
    \barepropG{\kvec_\nvec}{\omega}
    \kvec_{-\nvec}\cdot\kvec_{\nvec+\pvec}
    \extPot_{\nvec+\pvec}
    \barepropG{-\kvec_\pvec}{\omega}
    + \ldots
    \Big\} \ .
\end{align}
More systematically, the full propagator may be expanded as
\begin{equation}\elabel{prop_expansion}
    \ave{\chi(\kvec_\nvec,\omega)\chitilde(-\kvec_\pvec,\omega_0)} =
    \deltabar(\omega+\omega_0)
    \sum_{j=0}^\infty
    \Eft_j(\kvec_\nvec,\kvec_\pvec,\omega)
\end{equation}
where we changed the sign of $\kvec_\pvec$ in the creator field compared to \Eref{full_propagator_attempt1} to improve notation below.
The expansion terms $\Eft_j$ are defined as follows
\begin{subequations}
\begin{align}
\elabel{def_E0}
\Eft_0(\kvec_\nvec,\kvec_\pvec,\omega)&=
\sysL^d\delta_{\nvec-\pvec,0}\barepropG{\kvec_\nvec}{\omega}
\\&\nonumber
\corresponds\bareprop{\nvec,\omega}{-\pvec,\omega_0},\\
\elabel{def_E1}
\Eft_1(\kvec_\nvec,\kvec_\pvec,\omega)&=
-\barepropG{\kvec_\nvec}{\omega}
\kvec_\nvec\cdot\kvec_{\nvec-\pvec}
\extPot_{\nvec-\pvec}
\barepropG{\kvec_\pvec}{\omega}
\\&\nonumber
\corresponds\bareprop{\nvec,\omega}{}\!\!\!\!\pertprop{}{}\!\!\!\!\bareprop{}{-\pvec,\omega_0},\\
\Eft_2(\kvec_\nvec,\kvec_\pvec,\omega)&=
\barepropG{\kvec_\nvec}{\omega}
\frac{1}{\sysL^d}\sum_{\qvec\in\Zset^d}
\big(\kvec_\nvec\cdot\kvec_{\nvec-\qvec}\big)\extPot_{\nvec-\qvec}\barepropG{\kvec_\qvec}{\omega} \,
\big(\kvec_\qvec\cdot\kvec_{\qvec-\pvec}\big)\extPot_{\qvec-\pvec}\barepropG{\kvec_\pvec}{\omega} \,
\nonumber
\\&=-\barepropG{\kvec_\nvec}{\omega}
\frac{1}{\sysL^d}\sum_{\qvec\in\Zset^d}
\big(\kvec_\nvec\cdot\kvec_{\nvec-\qvec}\big)\extPot_{\nvec-\qvec}
\Eft_1(\kvec_\qvec,\kvec_\pvec,\omega)
\\&\corresponds\bareprop{\nvec,\omega}{}\!\!\!\!\pertprop{}{}\!\!\!\!\!\!\!\!\!\!\!\!\!\!\!\!\!\bareprop{\qquad\sum \qvec}{}\!\!\!\!\pertprop{}{}\!\!\!\!\bareprop{}{-\pvec,\omega_0}\nonumber\\
&\nonumber=
\frac{1}{\sysL^d}\sum_{\qvec\in\Zset^d}
\bareprop{\nvec,\omega}{}\!\!\!\!\pertprop{}{\qvec}\Eft_1(\kvec_\qvec,\kvec_\pvec,\omega)\elabel{second_order_one},
\end{align}
\end{subequations}
with the general recurrence
\begin{equation}\elabel{Ej_recurrence}
    \Eft_{j+1}(\kvec_\nvec,\kvec_\pvec,\omega)
=-\barepropG{\kvec_\nvec}{\omega}
\frac{1}{\sysL^d}\sum_{\qvec\in\Zset^d}
\big(\kvec_\nvec\cdot\kvec_{\nvec-\qvec}\big)\extPot_{\nvec-\qvec}
\Eft_j(\kvec_\qvec,\kvec_\pvec,\omega)
\end{equation}
amounting to a convolution to be analysed further below.


\subsection{Stationary state}
The density $\density_0(\rvec)$ in the stationary or steady state is found by taking of the propagator the limit $t_0\to-\infty$ of the time of the initial deposition,
\begin{equation}\elabel{def_density_app}
    \density_0(\rvec)=\lim_{t_0\to-\infty}
    \ave{\chi(\rvec,t)\chitilde(\rvec_0,t_0)}
\end{equation}
expected to be independent of $\rvec_0$ and $t$, provided the system is ergodic and possesses a (unique) stationary state.
Provided all internal legs are dashed, \ie they carry a factor $\kvec$,
the diagrammatic mechanics of this limit amounts to an amputation of the incoming leg in any composite diagram. This is most easily understood by taking the inverse Fourier transform from $\omega,\omega_0$ to $t,t_0$, while leaving the spatial dependence in $\kvec$, starting from the propagators that have the form \Eref{bareProp_app_baro}. The inverse Fourier transform will result in exponentials of the form $\exp{-\imag t_0 s}$, where $s$ is the location of the pole of the propagators, here $s=\imag (\transDiffusion k_\nvec^2+\imag\driftVec_0\cdot\kvec_\nvec+\mass)$ with $\Im{(s)}\geq0$. Only if the imaginary part of $s$ vanishes, the exponential will not vanish as $t_0\to-\infty$. This happens only when $\nvec=\nullvec$ and in the limit of $\mass\downarrow0$. The resulting residue is thus determined by the remainder of the diagram evaluated at the pole $\omega=0$ and with the incoming leg replaced by $\delta_{\pvec,\nullvec}$, so that an inverse Fourier transform in $\kvec_\pvec$ to $\rvec_0$ results in a factor $1/\sysL^{d}$. As for the individual propagator \Eref{def_E0} in \Eref{prop_expansion}, it contributes indeed only this background, $1/\sysL^d$. Writing therefore $\Eft_i$ with only one arguments as the desired limit, 
\begin{align}\elabel{def_Ei_stat}
\Eft_i(\kvec_\nvec)&=\lim_{t_0\to-\infty}\lim_{\mass\downarrow0}
\int\dintbar{\omega_0}
\exp{-\imag\omega_0(t_0-t)}
\frac{1}{\sysL^d}\sum_{\pvec\in\Zset}
\exp{\imag\kvec_\pvec\cdot\rvec_0}
\Eft_i(\kvec_\nvec,\kvec_\pvec,-\omega_0)\\
\nonumber
&\corresponds\bareprop{\nvec}{}\!\!\!\!\!\!\!\!\underbrace{\pertprop{}{}\!\!\!\!\bareprop{
}{}\!\!\!\!\pertprop{}{}\dots\bareprop{}{}\!\!\!\!\pertpropZHIJIAO{}{}}_{\text{$i$ baubles }}
\end{align}
allows the stationary density to be written in the form
\begin{equation}\elabel{density_expanded_in_E}
    \density_0(\rvec)=
    \frac{1}{\sysL^d}\sum_{\nvec\in\Zset}\exp{\imag\kvec_\nvec\cdot\rvec}
    \sum_{j=0}^\infty
    \Eft_j(\kvec_\nvec)
\end{equation}
via \Erefs{prop_expansion} and \eref{def_density_app}
and gives
\begin{equation}\elabel{E0_stat}
\Eft_0(\kvec_\nvec)=\sysL^d\delta_{\nvec,\nullvec}\frac{1}{\sysL^d}
= \delta_{\nvec,\nullvec}
\end{equation}
from \Eref{def_E0} and
\begin{equation}\elabel{E1_stat}
\Eft_1(\kvec_\nvec)=-\barepropG{\kvec_\nvec}{0}
\kvec_\nvec\cdot\kvec_{\nvec}\indicator{\nvec\ne\nullvec}
\extPot_{\nvec}
\frac{1}{\sysL^d}
= - 
\frac{1}{\sysL^d}
\frac{k_\nvec^2\extPot_{\nvec}\indicator{\nvec\ne\nullvec}}{\transDiffusion k_\nvec^2 + \imag\driftVec_0\cdot\kvec_\nvec}
\ ,
\end{equation}
using \Eref{bareProp_app_baro} and
where
\begin{equation}\elabel{def_indicator}
    \indicator{\nvec\ne\nullvec}
    =1-\delta_{\nvec,\nullvec}
\end{equation}
accounts for the subtlety that the right hand side of \Eref{def_E1} strictly vanishes for $\nvec=\nullvec$, because of the $\kvec_\nvec$ prefactor and all other terms being finite as $\mass\ne0$. However, in \Eref{def_Ei_stat} the limit $\mass\downarrow0$ has to be taken so that  $t_0\to-\infty$ produces a finite result, resulting in
\begin{equation}
    \lim_{\mass\downarrow0}
    \frac{\kvec_\nvec}{\kvec_\nvec^2+\mass}
    = \frac{\kvec_\nvec}{\kvec_\nvec^2} \indicator{\nvec\ne\nullvec}=
\begin{cases}
\nullvec & \text{for $\kvec=\nullvec$}\\
\kvec_\nvec/\kvec_\nvec^2 & \text{otherwise}\ ,
\end{cases}
\end{equation}
using the convention $0\infty=0$. The indicator function \Eref{def_indicator} needs to be in place for every prefactor $\kvec$ that also features in the denominator of a propagator $\barepropG{\kvec}{0}$ so that the limits $\mass\downarrow0$ and $t\rightarrow\infty$ can safely be taken.

For easier comparison, we state the expansion of the density to first order in the potential for drift-diffusion particles in a finite volume determined explicitly so far,
\begin{equation}\elabel{density_expanded_in_E_easy}
        \density_0(\rvec)=
    \frac{1}{\sysL^d} - \frac{1}{\sysL^{2d}}
    \sum_{\substack{\nvec\in\Zset\\ \nvec\ne\nullvec}}
    \exp{\imag\kvec_\nvec\cdot\rvec}
\frac{k_\nvec^2 \extPot_\nvec}{\transDiffusion k_\nvec^2 + \imag\driftVec_0\cdot\kvec_\nvec}
+ \ldots
\end{equation}
using \Erefs{density_expanded_in_E}, \eref{E0_stat} and \eref{E1_stat}.

With the indicator function \Eref{def_indicator} suppressing unwanted singularities,
the recurrence \Eref{Ej_recurrence} remains essentially unchanged at staionarity, as only the outgoing propagator needs to be adjusted to $\omega=0$,
\begin{subequations}
\elabel{E_recurrence_final}
\begin{align}
        \Eft_{j+1}(\kvec_\nvec)
&=-\barepropG{\kvec_\nvec}{0}
\frac{1}{\sysL^d}\sum_{\qvec\in\Zset^d}
\indicator{\nvec\ne\nullvec}
\big(\kvec_\nvec\cdot\kvec_{\nvec-\qvec}\big)\extPot_{\nvec-\qvec}
\Eft_j(\kvec_\qvec) \\
&=-\frac{\indicator{\nvec\ne\nullvec}}{\transDiffusion k_\nvec^2+\imag\driftVec_0\cdot\kvec_\nvec}
\frac{1}{\sysL^d}\sum_{\qvec\in\Zset^d}
\big(\kvec_\nvec\cdot\kvec_{\nvec-\qvec}\big)\extPot_{\nvec-\qvec}
\Eft_j(\kvec_\qvec) 
\\
&\corresponds\bareprop{\nvec}{}\!\!\!\!\!\!\!\!\underbrace{\pertprop{}{}\!\!\!\!\!\!\!\!\!\bareprop{\sum_{\qvec}
}{}\!\!\!\!\pertprop{}{}\dots\bareprop{}{}\!\!\!\!\pertpropZHIJIAO{}{}}_{\text{$i+1$ baubles }}
\end{align}
\end{subequations}
using \Eref{bareProp_app_baro}. In one dimension we can further simplify,
\begin{equation}\elabel{E_recurrence_final_1D}
        \Eft_{j+1}(k_n)
=-
\frac{\indicator{n\ne0}}{\transDiffusion k_n + \imag\drift_0}
\frac{1}{\sysL}\sum_{q\in\Zset}
\extPot_{n-q}k_{n-q}
\Eft_j(k_q) \ .
\end{equation}

From \Eref{E_recurrence_final} it can be gleaned that in real-space $\Eft_j(\rvec)$ obeys the recurrence
\begin{equation}\elabel{E_j_ODE}
    \big(\transDiffusion\nabla_\rvec^2 - \driftVec\cdot\nabla_\rvec\big) \Eft_{j+1}(\rvec) =
    - \nabla_\rvec \big(\Eft_j(\rvec) \nabla_\rvec\extPot(\rvec)\big) \ ,
\end{equation}
where we have refrained from introducing a separate symbol for the inverse Fourier transform $\Eft_j(\rvec)$ of $\Eft_j(\kvec)$.
It is instructive to demonstrate that this implies that \Eref{density_expanded_in_E} 
therefore solves the Fokker-Planck \Eref{FPE_app} at stationarity, \ie we want to show that using \Eref{E_j_ODE} in \Eref{density_expanded_in_E} provides a solution of \Eref{FPE_app} with $\partial_t \rho=0$,
\begin{equation}\elabel{FPE_stationary_app}
0 = 
    \transDiffusion\nabla_\rvec^2 
    \density_0(\rvec)
    - \nabla_\rvec\cdot\big( \density_0(\rvec) (\driftVec-\nabla\extPot(\rvec))\big)
\ .
\end{equation}
To see this, we sum both sides of \Eref{E_j_ODE} over $j=0\ldots$, which on the left-hand side produces $\density_0$ of \Eref{density_expanded_in_E}, including the vanishing term 
\begin{equation}
    \big(\transDiffusion\nabla_\rvec^2 - \driftVec\cdot\nabla_\rvec\big) \Eft_{0}(\rvec) =
    0
\end{equation}
as $\Eft_0(\rvec)=\sysL^{-d}$, \Eref{E0_stat}, so that \Eref{E_j_ODE} implies
\begin{equation}\elabel{E_j_solves_stat_FPE}
    \big(\transDiffusion\nabla_\rvec^2 - \driftVec\cdot\nabla_\rvec\big) 
    \sum_{j=0}^\infty \Eft_j(\rvec)
%
    =
    - \nabla_\rvec \big(
    \sum_{j=0}^\infty \Eft_j(\rvec)
    \nabla_\rvec\extPot(\rvec)\big) \ ,
\end{equation}
and thus \Eref{FPE_stationary_app}, with the $\Eft_j$ in \Eref{density_expanded_in_E} determined by \Eref{E_recurrence_final}.

\Eref{E_recurrence_final} is the central result of the present section.
By means of \Eref{density_expanded_in_E}, it allows the calculation of
the stationary density starting from $\Eft_0$, \Eref{E0_stat}. Without
drift, the density should produce the Boltzmann factor, which we will
recover in the next section. While this result is easy to obtain from a
differential equation approach in real-space $\rvec$, it is very
instructive to perform this calculation in $\kvec$-space. With drift, on
the other hand, we can compare the density $\rho_0$ resulting from 
\Erefs{E_recurrence_final}  to the known stationary density
in one dimension \cite{Pavliotis:2014}, which is done in \Sref{field_theory_produces_1D_drift}.



\subsection{Barometric formula}
\label{sec:barometric_formula}
In the absence of a drift, $\driftVec=\nullvec$, the density will evolve into the Boltzmann
factor
\begin{equation}\elabel{Boltzmann_result}
	\density_0(\rvec)=\frac{1}{\NC}
	\exp{-\extPot(\rvec)/\transDiffusion}
\end{equation}
with normalisation constant 
\begin{equation}\elabel{Boltzmann_result_normalisation}
	\NC=\int\ddint{r}\exp{-\extPot(\rvec)/\transDiffusion} \ .
\end{equation}
\Eref{Boltzmann_result} is the stationary solution of the Fokker-Planck \Eref{FPE_app} without drift,
\begin{equation}\elabel{FPE_stationary_no_drift}
	0=\transDiffusion\nabla_\rvec^2\density_0(\rvec) 
	+ \nabla_\rvec
	\big(\density_0(\rvec)\nabla_\rvec\extPot(\rvec)\big)
\ ,
\end{equation}
which demands that the probability current is spatially uniform
\cite{Risken:1989}. 
Provided the potential is bounded from above and from below and with
boundary conditions given, the solution is unique 
 by \Eref{Boltzmann_result}, including the normalisation constant given by
\Eref{Boltzmann_result_normalisation}.

In the following, we show the equivalence of \Eref{Boltzmann_result} and 
the field-theoretic result \Erefs{density_expanded_in_E},  \eref{E0_stat}
and \eref{E_recurrence_final}, by expanding the density $\density_0$ from \Eref{Boltzmann_result} in terms of some $\Hbaro_j$, to be characterised next.

\subsubsection{Properties of $\Hbaro_j(\kvec_\nvec)$}
We expand
the exponential in \Eref{Boltzmann_result} order by order in the potential
\begin{equation}
\density_0(\rvec)=\NC^{-1} \sum_{j=0}^\infty 
\frac{1}{j!}\left(
-\frac{\extPot(\rvec)}{\transDiffusion}
\right)^j
\end{equation}
and introduce Fourier-modes
\begin{equation}\elabel{H_j_modes}
\Hbaro_j(\kvec_\nvec)
=
\int_0^\sysL\ldots\int_0^\sysL\ddint{r}
\exp{-\imag\kvec_\nvec\cdot\rvec}
\frac{1}{j!}\left(
-\frac{\extPot(\rvec)}{\transDiffusion}
\right)^j
\end{equation}
so that
\begin{equation}\elabel{expansion_density_in_H_modes}
    \density_0(\rvec)=\NC^{-1} \sum_{j=0}^\infty 
    \frac{1}{\sysL^d}\sum_{\nvec\in\Zset} \exp{\imag\kvec_\nvec\cdot\rvec}
    \Hbaro_j(\kvec_\nvec) \ ,
\end{equation}
similar to \Eref{density_expanded_in_E}.
The normalisation \Eref{Boltzmann_result_normalisation} can therefore be written as a sum over the $\nullvec$-modes of $\Hbaro_j$ in \Eref{H_j_modes}
\begin{equation}
\elabel{norm_in_FT_space}
    \NC = \sum_{j=0}^\infty \Hbaro_j(\nullvec) \ .
\end{equation}
That $\Eft_j(\rvec)$ as used in \Eref{density_expanded_in_E} and defined in \Eref{def_Ei_stat} solves \Eref{FPE_stationary_no_drift} has been established in \Eref{E_j_solves_stat_FPE} via the recurrence \Eref{E_recurrence_final}. By construction, the $\Eft_j$, in \Eref{density_expanded_in_E} provide an expansion of the density order by order in the potential $\extPot$. 
The $\Hbaro_j$, \Eref{expansion_density_in_H_modes}, on the other hand, are an expansion in orders of the potential of only the exponential, lacking, however, the normalisation. 
Both $\Eft_j$ and $\Hbaro_j$ are, by construction, of order $j$ in the potential $\extPot$.
While generally $\Eft_j(\kvec_\nvec) \ne \Hbaro_j(\kvec_\nvec)/\NC$, we will show in the following that
\begin{equation}\elabel{density_with_expanded_denominator} 
\density_0(\kvec_\nvec)=
    \sum_{j=0}^\infty 
    \Eft_j(\kvec_\nvec) = \frac{\sum_{j=0}^\infty  \Hbaro_j(\kvec_\nvec)}{\Hbaro_0(\nullvec)+\Hbaro_1(\nullvec)+\ldots} \ ,
\end{equation}
taking the Fourier transform on both sides of \Erefs{density_expanded_in_E} and \eref{expansion_density_in_H_modes}, and using \Eref{norm_in_FT_space}. 
Equivalently, we show that 
$\Hbaro_j(\kvec_\nvec)$
is 
the sum of the terms of $j$th order in $\extPot$ of $\NC\sum_j\Eft_j(\kvec_\nvec)$,
\begin{equation}\elabel{barometric_formula_aim}
    \sum_{i=0}^j \Hbaro_i(\nullvec)\Eft_{j-i}(\kvec_\nvec)
    = \Hbaro_j(\kvec_\nvec) \ .
\end{equation}
We will thus show, beyond \Eref{E_j_solves_stat_FPE}, that the $\Eft_j$ are correctly normalised, providing us with an expansion of the density $\density_0$ order by order in the potential.
Showing this, will further provide us with a useful, non-trivial algebraic identity.


To proceed, we firstly confirm \Eref{barometric_formula_aim} for $j=0,1$ by determining $\Hbaro_j(\kvec)$ directly from \Eref{H_j_modes}
\begin{subequations}
\begin{align}
\elabel{H_0}
    \Hbaro_0(\kvec_\nvec)&=\sysL^d \delta_{\nvec,\nullvec}\\
\elabel{H_1}
    \Hbaro_1(\kvec_\nvec)&=-\frac{\extPot_\nvec}{\transDiffusion}
\end{align}
\end{subequations}
so that \Eref{E0_stat} indeed confirms
\begin{equation}\elabel{barometric_formula_aim_H0}
    \Hbaro_0(\nullvec) \Eft_0(\kvec_\nvec) = 
    \sysL^d \delta_{\nvec,\nullvec} = \Hbaro_0(\kvec_\nvec)
\end{equation}
and \Erefs{E0_stat} and \eref{E1_stat} with 
$\driftVec=\nullvec$ confirm
\begin{equation}\elabel{barometric_formula_aim_H1}
    \Hbaro_0(\nullvec) \Eft_1(\kvec_\nvec)
    + 
    \Hbaro_1(\nullvec) \Eft_0(\kvec_\nvec)
    = 
    - \sysL^d 
    \frac{\extPot_\nvec\indicator{\nvec\ne\nullvec}}{\transDiffusion}\frac{1}{\sysL^d}
    -
    \frac{\extPot_\nullvec}{\transDiffusion}\delta_{\nvec,\nullvec}
    =
    - \frac{\extPot_\nvec}{\transDiffusion}
    \big(
        \indicator{\nvec\ne\nullvec} + \delta_{\nvec,\nullvec}
    \big)
= \Hbaro_1(\kvec_\nvec)
\end{equation}
as $\extPot_\nullvec\delta_{\nvec,\nullvec}=\extPot_\nvec\delta_{\nvec,\nullvec}$ and 
\begin{equation}\elabel{indicator_id}
1=\indicator{\nvec\ne\nullvec} + \delta_{\nvec,\nullvec}\ .
\end{equation}

To characterise $\Hbaro_j$ more generally, we 
introduce the recursion formula
\begin{equation}\elabel{H_j_recursion}
    \Hbaro_{j+1}(\kvec_\nvec) = - \frac{1}{(j+1)\transDiffusion}\frac{1}{\sysL^d}
    \sum_{\qvec\in\Zset^d}\extPot_{\nvec-\qvec}\Hbaro_j(\kvec_\qvec)
\end{equation}
by elementary calculations on the basis of \Eref{H_j_modes} and similar to \Eref{E_recurrence_final}. Starting from \Eref{H_0}, it can be used to produce for $j\ge2$
\begin{equation}\elabel{H_j_expansion_start}
    \Hbaro_j(\kvec_{\nvec}) = \frac{
    \sysL^d}{j!}\left(\frac{-1}{\sysL^d\transDiffusion}\right)^j
    \sum_{\nvec_{j-1},\ldots,\nvec_1}
    \underbrace{
    \extPot_{\nvec-\nvec_{j-1}}
    \extPot_{\nvec_{j-1}-\nvec_{j-2}}
    \ldots
    \extPot_{\nvec_{2}-\nvec_{1}}
    \extPot_{\nvec_1}
    }_{\text{$j$ factors of $\extPot$}} \ .
\end{equation}
This expression
can be simplified further by substituting dummy variables according to $\pvec_i=\nvec_i-\nvec_{i-1}$ for $i=j-1,j-2,\ldots,1$ with the convention $\nvec_0=\nullvec$. As for 
\begin{equation}\elabel{p_j}
    \pvec_j=\nvec-\nvec_{j-1}=\nvec-\pvec_{j-1}-\nvec_{j-2}=\ldots=\nvec-\pvec_{j-1}-\pvec_{j-2}-\ldots-\pvec_1
\ ,
\end{equation} 
this can be allowed to run in the sum by enforcing \Eref{p_j} via a Kronecker $\delta$-function, so that eventually
\begin{equation}\elabel{H_j_easy_form}
    \Hbaro_j(\kvec_{\nvec}) = \frac{
    \sysL^d}{j!}\left(\frac{-1}{\sysL^d\transDiffusion}\right)^j
    \sum_{\pvec_j,\ldots,\pvec_1}
    \extPot_{\pvec_j}
    \extPot_{\pvec_{j-1}}
    \ldots
    \extPot_{\pvec_1}
    \delta_{\pvec_1+\ldots+\pvec_j,\nvec} \ .
\end{equation}
Even when calculated with a convolution in mind, \Eref{H_j_easy_form} is valid for $j=1$ by direct comparison with \Eref{H_1} and, by comparison with \Eref{H_0}, even $j=0$ with the convention that the final $\delta_{\pvec_1+\ldots+\pvec_j,\nvec}$ degenerates into $\delta_{\nullvec,\nvec}$ for $j=0$.

In the form \Eref{H_j_easy_form}, we can show easily an important identity, based on the observation
\begin{equation}
    \sum_{\pvec_j,\ldots,\pvec_1}
    \kvec_{\pvec_j}
    \extPot_{\pvec_j}
    \extPot_{\pvec_{j-1}}
    \ldots
    \extPot_{\pvec_1}
    \delta_{\pvec_1+\ldots\pvec_j,\nvec}
=
    \sum_{\pvec_j,\ldots,\pvec_1}
    \kvec_{\pvec_i}
    \extPot_{\pvec_j}
    \extPot_{\pvec_{j-1}}
    \ldots
    \extPot_{\pvec_1}
    \delta_{\pvec_1+\ldots\pvec_j,\nvec}
\end{equation}
for all $i\in\{1,2,\ldots,j\}$, which labels the index of $\kvec_{\pvec_i}$ on the right-hand side as the $\pvec_i$ are all equivalent dummy variables. Summing this sum for $i=1,2,\ldots,j$ therefore gives $j$ times the same result, while the sum of the $\kvec_{\pvec_i}$ gives $\kvec_{\pvec_1}+\ldots+\kvec_{\pvec_j}=\kvec_{\nvec}$ inside the sum, according to the Kronecker $\delta$-function. In conclusion for $j\ge1$
\begin{equation}
    j \sum_{\pvec_j,\ldots,\pvec_1}
    \kvec_{\pvec_j}
    \extPot_{\pvec_j}
    \extPot_{\pvec_{j-1}}
    \ldots
    \extPot_{\pvec_1}
    \delta_{\pvec_1+\ldots\pvec_j,\nvec} =
    \kvec_{\nvec}
    \sum_{\pvec_j,\ldots,\pvec_1}
    \extPot_{\pvec_j}
    \extPot_{\pvec_{j-1}}
    \ldots
    \extPot_{\pvec_1}
    \delta_{\pvec_1+\ldots\pvec_j,\nvec} \ , 
\end{equation}
and thus
\begin{equation}\elabel{final_trick_expression}
    \sum_{\pvec_j,\ldots,\pvec_1}
    \indicator{\nvec\ne\nullvec}
    \frac{\kvec_{\nvec}\cdot\kvec_{\pvec_j}}{k_\nvec^2}
    \extPot_{\pvec_j}
    \extPot_{\pvec_{j-1}}
    \ldots
    \extPot_{\pvec_1}
    \delta_{\pvec_1+\ldots\pvec_j,\nvec} =
    \frac{1}{j}
    \sum_{\pvec_j,\ldots,\pvec_1}
    \extPot_{\pvec_j}
    \extPot_{\pvec_{j-1}}
    \ldots
    \extPot_{\pvec_1}
    \delta_{\pvec_1+\ldots\pvec_j,\nvec} \ ,
\end{equation}
where the indicator function on the left avoids the divergence of $\kvec_{\nvec}\cdot\kvec_{\pvec_j}/k_\nvec^2$.
Using \Eref{final_trick_expression} in \Eref{H_j_easy_form} gives
\begin{subequations}
\begin{align}
    \indicator{\nvec\ne\nullvec} \Hbaro_j(\kvec_{\nvec}) &= 
    \indicator{\nvec\ne\nullvec} \frac{
    \sysL^d}{(j-1)!}\left(\frac{-1}{\sysL^d\transDiffusion}\right)^j
    \sum_{\pvec_j,\ldots,\pvec_1}
    \frac{\kvec_{\nvec}\cdot\kvec_{\pvec_j}}{k_\nvec^2}
    \extPot_{\pvec_j}
    \extPot_{\pvec_{j-1}}
    \ldots
    \extPot_{\pvec_1}
    \delta_{\pvec_1+\ldots\pvec_j,\nvec}\\
&=
    \indicator{\nvec\ne\nullvec} 
    \frac{
    \sysL^d}{(j-1)!}\left(\frac{-1}{\sysL^d\transDiffusion}\right)^j
    \sum_{\nvec_{j-1},\ldots,\nvec_1}
    \frac{\kvec_{\nvec}\cdot\kvec_{\nvec-\nvec_{j-1}}}{k_\nvec^2}
    \extPot_{\nvec-\nvec_{j-1}}
    \extPot_{\nvec_{j-1}-\nvec_{j-2}}
    \ldots
    \extPot_{\nvec_2-\nvec_1}
    \extPot_{\nvec_1}\\
    \elabel{magical_id}
&= - \frac{\indicator{\nvec\ne\nullvec}}{\transDiffusion}\frac{1}{\sysL^d}
    \sum_{\nvec_{j-1}\in\Zset^d}
    \frac{\kvec_{\nvec}\cdot\kvec_{\nvec-\nvec_{j-1}}}{k_\nvec^2}
    \extPot_{\nvec-\nvec_{j-1}}
    \Hbaro_{j-1}(\kvec_{\nvec_{j-1}})
\end{align}
\end{subequations}
where the second equality is a matter of labelling the indices again in $\nvec$ and $\nvec_{1\ldots j-1}$ using \Eref{p_j} in reverse, and thus reverting from the indexing in \Eref{H_j_easy_form} to that in \Eref{H_j_expansion_start}, which does not carry a Kronecker $\delta$-function. 
In this form the summation can again be written recursively, similarly to \Eref{H_j_recursion}, which is done after the third equality, by absorbing the summation over $\nvec_{j-2},\ldots,\nvec_1$ into $\Hbaro_{j-1}(\kvec_{\nvec_{j-1}})$.
\Eref{magical_id} is the key outcome of the present section.

\subsubsection{Relationship between $\Eft_j(\kvec_\nvec)$ and $\Hbaro_j(\kvec_\nvec)$}
\label{sec:Relationship_Eft_Hbaro}
We proceed with confirming 
\Eref{barometric_formula_aim} 
by induction, using \Eref{magical_id} in the course. To access this identity, we firstly rewrite \Eref{barometric_formula_aim} using \Eref{indicator_id}, so that for $j\ge0$
\begin{equation}\elabel{barometric_formula_aim2}
    \Hbaro_{j+1}(\kvec_\nvec) =
    \Hbaro_{j+1}(\kvec_\nvec) (\indicator{\nvec\ne\nullvec} + \delta_{\nvec,\nullvec}) =
 - \frac{\indicator{\nvec\ne\nullvec}}{\transDiffusion}\frac{1}{\sysL^d}
    \sum_{\nvec_{j}\in\Zset^d}
    \frac{\kvec_{\nvec}\cdot\kvec_{\nvec-\nvec_{j}}}{k_\nvec^2}
    \extPot_{\nvec-\nvec_{j}}
    \Hbaro_{j}(\kvec_{\nvec_{j}})
    + \Hbaro_{j+1}(\kvec_\nvec) \delta_{\nvec,\nullvec} \ .
\end{equation}
Of the two terms on the right hand side, the first can be rewritten using \Eref{barometric_formula_aim}, which is \emph{assumed} to hold for $j$. This allows us to replace $\Hbaro_{j}(\kvec_{\nvec_{j}})$ by a sum.
In the second term we replace $\delta_{\nvec,\nullvec}$ by $\Eft_0(\kvec_n)$ according to \Eref{E0_stat}, resulting in
\begin{equation}\elabel{barometric_formula_aim3}
    \Hbaro_{j+1}(\kvec_\nvec) =
 - \frac{\indicator{\nvec\ne\nullvec}}{\transDiffusion}\frac{1}{\sysL^d}
    \sum_{\nvec_{j}\in\Zset^d}
    \frac{\kvec_{\nvec}\cdot\kvec_{\nvec-\nvec_{j}}}{k_\nvec^2}
    \extPot_{\nvec-\nvec_{j}}
 \sum_{i=0}^j \Hbaro_i(\nullvec)\Eft_{j-i}(\kvec_{\nvec_j})
+ \Hbaro_{j+1}(\nullvec) \Eft_0(\kvec_n)
\ .
\end{equation}
The summation over $\nvec_{j}\in\Zset^d$ can now be performed on $\Eft_{j-i}$ using \Eref{E_recurrence_final} at $\driftVec=\nullvec$, \begin{equation}
    \Hbaro_{j+1}(\kvec_\nvec) =
 \sum_{i=0}^j \Hbaro_i(\nullvec)\Eft_{j-i+1}(\kvec_{\nvec})
+ \Hbaro_{j+1}(\nullvec) \Eft_0(\kvec_n)
=
 \sum_{i=0}^{j+1} \Hbaro_i(\nullvec)\Eft_{j+1-i}(\kvec_{\nvec})
 \quad\text{ for }\ j\ge0
\ ,
\end{equation}
completing our demonstration that \Eref{barometric_formula_aim} holds for $j+1$ if it holds for $j$. With the induction basis \Erefs{barometric_formula_aim_H0} or \eref{barometric_formula_aim_H1} this concludes our demonstration that the 
diagrams \Eref{def_Ei_stat} or equivalently
\eref{E_recurrence_final}
produce a normalised expansion of the density $\density_0$
\eref{density_expanded_in_E} identical to the Boltzmann factor, order by order in the external potential. The field-theoretic approach thus determines the stationary density without the need of an \latin{a posteriori} normalisation. And while the field theory is an expansion necessarily in a sufficiently weak potential and necessarily in a finite volume, no such restriction applies to the resulting Boltzmann factor. 


In the next section we will similarly demonstrate the correspondence between classical and field-theoretic stationary density in the presence of a drift $\drift$.

\subsection{Density in one dimension with drift}\label{sec:field_theory_produces_1D_drift}
In this section, we will demonstrate that the expansion \Eref{density_expanded_in_E} in terms of diagrams \Eref{def_Ei_stat} produces the correct, normalised, stationary result in the presence of a potential and a drift in one dimension. The restriction to one dimension is because we are not aware of a closed-form expression beyond.

Following the pattern in \Sref{barometric_formula}, we want to show that the diagrammatic expansion \Eref{density_expanded_in_E} order by order in the potential reproduces a known expression once this is written out order by order in potential. The difficulty here is similar to that in \Sref{barometric_formula}, namely that it is fairly easy to expand the known expression for the stationary density \emph{without} the normalisation. Working in Fourier-space, we will once again write both the expansion and the normalisation in modes. 

The first step is to characterise the expansion terms $\Hdrift$ and to bring them in a form that lends itself more naturally to a comparison to the field theory. This characterisation involves some rather involved manipulations. 
In a second step, we will then demonstrate that the resulting expansion agrees with the diagrammatic one.

The classical stationary solution of the Fokker-Planck \Eref{FPE_app} 
of a drift-diffusion particle 
in one spatial dimension on a periodic domain and subject to a periodic potential defined so that 
$\extPot(x+\sysL)=\extPot(x)$ for $x\in [0,\sysL)$ is \cite{Risken:1989,Pavliotis:2014},
\begin{equation}
\density_0(x)=
\NC^{-1}
Z_-(x)\int_x^{x+\sysL}\!\!\!\!\dint{y}Z_+(y)\elabel{DriftDiffusionSolution}
\quad\text{ with }\quad
Z_\pm(x)=\Exp{\pm\frac{\extPot(x)-\drift_0x}{\transDiffusion}}
\ ,
\end{equation}
similar to \Eref{Boltzmann_result} defined with an explicit normalisation,
\begin{equation}\elabel{normalisation_density_expansion_drift_diffusion}
    \NC=\int_0^\sysL\!\!\dint{x} 
    Z_-(x)\int_x^{x+\sysL}\!\!\!\!\dint{y}Z_+(y) \ .
\end{equation}
The density 
\begin{equation}\elabel{density_expansion_drift_diffusion}
    \density_0(x)
    =\NC^{-1}\sum_{j=0}^\infty\frac{1}{\sysL}\sum_{n=-\infty}^\infty 
    \exp{\imag k_nx} \Hdrift_j(k_n)
\end{equation}
can be written systematically in orders of the potential up to the normalisation to be determined below.
If $\Hdrift_j(k_n)$ is the term of order $j$ in $\extPot$, expanding $Z_\pm(x)=\exp{\mp\drift_0x/\transDiffusion}\sum_n(\pm\extPot(x)/\transDiffusion)^n/n!$ in \Eref{DriftDiffusionSolution}, immediately gives
\begin{equation}\elabel{def_HdriftSub}
    \Hdrift_j(k_n)=\sum_{i=0}^j\frac{(-1)^i}{i!}\frac{1}{(j-i)!}\HdriftSub_{j,i}(k_n)
\quad\text{ with }\quad
\HdriftSub_{j,i}(k_n)=
    \int_0^L\dint{x} \exp{-\imag k_nx}
    \left(
    \frac{\extPot(x)}{\transDiffusion}
    \right)^i
    \exp{\drift_0x/\transDiffusion}
    \int_x^{x+\sysL}\dint{y}
    \left(
    \frac{\extPot(y)}{\transDiffusion}
    \right)^{j-i}
    \exp{-\drift_0y/\transDiffusion}
\ .
\end{equation}
An easy route to perform the Fourier transformation on the right  is to split it into two and perform separate Fourier transformations of $(-\extPot(x)/\transDiffusion)^i$ and of the second term, $\int_x^{x+L}\!\! \dint y\dots$ tying both together again by a convolution afterwards. While the first term produces essentially \Eref{H_j_modes}, the second term involving the integral is more complicated. Writing however
\begin{equation}
    \exp{-\imag k_nx + \drift_0x/\transDiffusion} = 
    \frac{1}{-\imag k_n + \drift_0/\transDiffusion}
    \ddx \exp{-\imag k_nx + \drift_0x/\transDiffusion} \ ,
\end{equation}
which is well-defined as $-\imag k_n + \drift_0/\transDiffusion\ne0$ for all $n$ by virtue of $k_n$ and $\drift_0$ being real and $\drift_0\ne0$, allows the second integral to be removed by an integration by parts. Surface terms then drop out as $\extPot(x)$ is periodic. Following at first the pattern of \Eref{H_j_expansion_start} the resulting Fourier series reads
\begin{multline}
    \HdriftSub_{j,i}(k_n) = \frac{1}{\sysL}\sum_{n_0=-\infty}^\infty
    \left(\frac{1}{\sysL\transDiffusion}\right)^i
    \sysL \sum_{n_1,\ldots,n_{i-1}}
    \extPot_{n_0-n_1}\ldots\extPot_{n_{i-2}-n_{i-1}}\extPot_{n_{i-1}}\\
\times
    \frac{1-\exp{-\drift_0\sysL/\transDiffusion}}{-\imag k_{n-n_0} + \drift_0/\transDiffusion}
    \left(\frac{1}{\sysL\transDiffusion}\right)^{j-i}
    \sysL \sum_{m_1,\ldots,m_{j-i-1}}
    \extPot_{n-n_0-m_1}\ldots\extPot_{m_{j-i-2}-n_{j-i-1}}\extPot_{n_{j-i-1}}
\ .
\end{multline}
This expression has some shortcomings that are mostly down to notation, which can be seen in the first sum by setting $i=0$ or in the second sum by setting $i=j$. In both cases, the sums are ill-defined, with the first sum running over indeces $n_1,\ldots,n_{-1}$ for $i=0$ and similarly the last sum for $i=j$. Choosing $i=1$ or $i=j-1$ causes similar problems. However, \Eref{def_HdriftSub} allows for all of these choices. The problem is resolved by changing the indeces similar to \Eref{p_j}, first in the individual Fourier-transforms and then after tying them together in a convolution 
\begin{equation}\elabel{HdriftSub_improved}
    \HdriftSub_{j,i}(k_n)=\frac{1}{(\sysL\transDiffusion)^j}
\sum_{p_1,\ldots,p_j}\extPot_{p_1}\ldots\extPot_{p_j}
\sysL\delta_{p_1+\ldots+p_j,n}
\frac{1-\exp{-\drift_0\sysL/\transDiffusion}}{-\imag k_{p_1+\ldots+p_{j-i}} + \drift_0/\transDiffusion}
\end{equation}
with the convention of $k_{p_1+\ldots+p_{j-i}}=0$ for $i=j$, as can be verified by direct evaluation of the corresponding Fourier transform.
Similarly, we demand $\delta_{p_1+\ldots+p_j,n}=\delta_{0,n}$ for $j=0$, so that $\HdriftSub_{0,0}(k_n)=\sysL\delta_{0,n} (1-\exp{-\drift_0\sysL/\transDiffusion})\transDiffusion/\drift_0$,  confirmed by direct evaluation of \Eref{def_HdriftSub}.
We state in passing $\Hdrift_j$ for $j=0,1$ by direct evaluation of \Erefs{def_HdriftSub} and \eref{HdriftSub_improved},
\begin{subequations}\elabel{Hdrift_0and1}
\begin{align}\elabel{Hdrift_0}
    \Hdrift_0(k_n)&=
    \sysL\delta_{0,n} \frac{1-\exp{-\drift_0\sysL/\transDiffusion}}{\drift_0/\transDiffusion}\\
    \elabel{Hdrift_1}
    \Hdrift_1(k_n)&=
    \frac{1}{\transDiffusion}
    \extPot_n
    \left\{
    \frac{1-\exp{-\drift_0\sysL/\transDiffusion}}{-\imag k_n+\drift_0/\transDiffusion}
    -
    \frac{1-\exp{-\drift_0\sysL/\transDiffusion}}{\drift_0/\transDiffusion}
    \right\}
    \\
    &=\frac{\imag k_n\extPot_n}{\drift_0}
    \frac{1}{-\imag k_n+\drift_0/\transDiffusion}
    \ .
\end{align}
\end{subequations}
We proceed by deriving a recurrence formula for $\Hdrift_j$.

\subsubsection{Properties of $\Hdrift_j(k_n)$}
\Eref{HdriftSub_improved} is to be used to compile $\Hdrift_j(k_n)$ according to \Eref{def_HdriftSub}, 
\begin{equation}
    \elabel{Hdrift_so_far}
    \Hdrift_j(k_n)=
    \sum_{i=0}^j\frac{(-1)^i}{i!}\frac{1}{(j-i)!}
    \frac{1}{(\sysL\transDiffusion)^j}
\sum_{p_1,\ldots,p_j}\extPot_{p_1}\ldots\extPot_{p_j}
\sysL\delta_{p_1+\ldots+p_j,n}
\frac{1-\exp{-\drift_0\sysL/\transDiffusion}}{-\imag k_{p_1+\ldots+p_{j-i}} + \drift_0/\transDiffusion}
\end{equation}
which we rewrite for $j\ge1$ as
\begin{equation}
        \Hdrift_j(k_n)=\sum_{i=1}^j\frac{(-1)^i}{i!}\frac{1}{(j-i)!}\HdriftSub_{j,i}(k_n)
-
\sum_{i=1}^j
{j \choose i}
\frac{(-1)^i}{j!}\HdriftSub_{j,0}(k_n)
\end{equation}
where the second term is the term for $i=0$ which has been removed from the first sum and multiplied by a clever unity for $j\ge1$,
\begin{equation}
    1=-\sum_{i=1}^j
{j \choose i} (-1)^i
\end{equation}
as $(1-1)^j=0$ for $j\ge1$, so that
\begin{equation}
    \Hdrift_j(k_n)=\sum_{i=1}^j\frac{(-1)^i}{i!}\frac{1}{(j-i)!}
    \frac{1}{(\sysL\transDiffusion)^j}
\sum_{p_1,\ldots,p_j}\extPot_{p_1}\ldots\extPot_{p_j}
\sysL\delta_{p_1+\ldots+p_j,n}
\left[
\frac{1-\exp{-\drift_0\sysL/\transDiffusion}}{-\imag k_{p_1+\ldots+p_{j-i}} + \drift_0/\transDiffusion}
-
\frac{1-\exp{-\drift_0\sysL/\transDiffusion}}{-\imag k_{p_1+\ldots+p_{j}} + \drift_0/\transDiffusion}
\right]   \ ,
\end{equation}
using ${j \choose i}/j!=1/(i!(j-i)!)$.
The square bracket can be simplified further as $k_{p_1+\ldots+p_{j}}=k_n$ by the Kronecker $\delta$-function and 
\begin{equation}
\elabel{app_k_sums}
    k_n-k_{p_1+\ldots+p_{j-i}}=k_{p_{j-i+1}+\ldots+p_j}
=k_{p_{j-i+1}} + k_{p_{j-i+2}} + \ldots + k_{p_{j}}
\end{equation}
so that 
\begin{multline}\elabel{Hdrift_j_simplified_long_sum}
    \Hdrift_j(k_n)=
    \sum_{i=1}^j\frac{(-1)^i}{i!}\frac{1}{(j-i)!}
    \frac{1}{(\sysL\transDiffusion)^j}\\
    \times
\sum_{p_1,\ldots,p_j}\extPot_{p_1}\ldots\extPot_{p_j}
\sysL\delta_{p_1+\ldots+p_j,n}
\left[
\frac{(1-\exp{-\drift_0\sysL/\transDiffusion})(-\imag (k_{p_{j-i+1}}+k_{p_{j-i+2}}+\dots+k_{p_j}))}
{
(-\imag k_{p_1+\ldots+p_{j-i}} + \drift_0/\transDiffusion)
(-\imag k_n + \drift_0/\transDiffusion)
}
\right]   \ .
\end{multline}
The last term in the numerator of the square bracket, $(-\imag (k_{p_{j-i+1}}+k_{p_{j-i+2}}+\dots+k_{p_j}))$, which is $(-\imag)$  times \Eref{app_k_sums} can be split into $i$ terms, resulting in $i$ sums over $p_1,\dots,p_j$ for each $i$. But every sum is in fact equal, as the $p_1,\ldots,p_j$ are dummy variables, so that the sum involving, say $p_{j-i+1}$, in the numerator can be turned into the sum involving $p_j$, simply by swapping the dummy variables $p_{j-i+1}$ and $p_j$ and then noticing that all other terms, such as $\extPot_{p_1}\ldots\extPot_{p_j}$ or $k_{p_1+\ldots+p_{j-i}}$ are invariant under such permutations. In fact, the indeces of the dummy variables divide naturally into two sets, namely $\{1,\ldots,j-i\}$ as used in $k_{p_1+\ldots+p_{j-i}}$ in one of the factors in the denominator, at the complement, $\{j-i+1,\ldots,j\}=\{1,\ldots,j\}\setminus\{1,\ldots,j-i\}$.
In other words for $j\ge1$
\begin{subequations}
\begin{align}\elabel{Hdrift_j_simplified_i}
    \Hdrift_j(k_n)&=
    \sum_{i=1}^j\frac{(-1)^i}{i!}\frac{1}{(j-i)!}
    \frac{1}{(\sysL\transDiffusion)^j}
\sum_{p_1,\ldots,p_j}\extPot_{p_1}\ldots\extPot_{p_j}
\sysL\delta_{p_1+\ldots+p_j,n}
\left[
\frac{i(1-\exp{-\drift_0\sysL/\transDiffusion})(-\imag k_{p_j})}
{
(-\imag k_{p_1+\ldots+p_{j-i}} + \drift_0/\transDiffusion)
(-\imag k_n + \drift_0/\transDiffusion)
}
\right]   \\
\elabel{Hdrift_j_simplified_take_out_p_j}
&=    
    \frac{1}{\sysL\transDiffusion}
\sum_{p_j}
\frac{\imag k_{p_j}}{(-\imag k_n + \drift_0/\transDiffusion)}
\extPot_{p_j}
\sum_{i=1}^j\frac{(-1)^{i-1}}{(i-1)!}\frac{1}{(j-i)!}
    \frac{1}{(\sysL\transDiffusion)^{j-1}}
    \\ \nonumber &\qquad\times
\sum_{p_1,\ldots,p_{j-1}}\extPot_{p_1}\ldots\extPot_{p_{j-1}}
\sysL\delta_{p_1+\ldots+p_{j-1},n-p_j}
\left[
\frac{(1-\exp{-\drift_0\sysL/\transDiffusion})}
{(-\imag k_{p_1+\ldots+p_{j-i}} + \drift_0/\transDiffusion)}
\right]   \\
\elabel{Hdrift_j_simplified_shift_i}
&=    
    \frac{1}{\sysL\transDiffusion}
\sum_{p_j}
\frac{\imag k_{p_j}}{(-\imag k_n + \drift_0/\transDiffusion)}
\extPot_{p_j}
\sum_{i=0}^{j-1}\frac{(-1)^i}{i!}\frac{1}{(j-1-i)!}
    \frac{1}{(\sysL\transDiffusion)^{j-1}}
    \\ \nonumber &\qquad\times
\sum_{p_1,\ldots,p_{j-1}}\extPot_{p_1}\ldots\extPot_{p_{j-1}}
\sysL\delta_{p_1+\ldots+p_{j-1},n-p_j}
\left[
\frac{(1-\exp{-\drift_0\sysL/\transDiffusion})}
{(-\imag k_{p_1+\ldots+p_{j-1-i}} + \drift_0/\transDiffusion)}
\right]   \\
& =\elabel{Hdrift_recursion}
    \frac{1}{\sysL\transDiffusion}
\sum_{p_j}
\frac{\imag k_{p_j}}{(-\imag k_n + \drift_0/\transDiffusion)}
\extPot_{p_j}
\Hdrift_{j-1}(k_n-k_{p_j})
\ ,
\end{align}
\end{subequations}
where we have used the equivalence of the $i$ terms in \Eref{Hdrift_j_simplified_long_sum} to write \Eref{Hdrift_j_simplified_i} which carries a factor of $i$ in the numerator in the square bracket, taken the summation over $p_j$ and other terms outside the rest of the expression in \Eref{Hdrift_j_simplified_take_out_p_j}, shifted $i$ by unity in \Eref{Hdrift_j_simplified_shift_i}, to finally re-express the sum as a convolution of a pre-factor with $\Hdrift_{j-1}(k_n)$ using \Eref{Hdrift_so_far}. \Eref{Hdrift_recursion} is the central result of thise section, providing a simple recurrence formula for $\Hdrift_j(k_n)$ for all $j\ge1$. We will use \Eref{Hdrift_recursion} in the following section to demonstrate the equivalence of the classical result \Eref{density_expansion_drift_diffusion} and the field theoretic result \Eref{density_expanded_in_E} with \Eref{def_Ei_stat} in the presence of drift in one dimension.

\subsubsection{Relationship between $\Eft_j(k_n)$ and $\Hdrift_j(k_n)$} 
Similar to \Sref{Relationship_Eft_Hbaro}, we will now demonstrate that the density $\density_0$ expanded in $\Hdrift_j(k_n)$ according to \Eref{density_expansion_drift_diffusion} with recurrence \Eref{Hdrift_recursion} and with normalisation \Eref{normalisation_density_expansion_drift_diffusion}, is equal to that expanded in $\Eft_j(k_n)$ according to \Erefs{density_expanded_in_E} and, in one dimension, \eref{E_recurrence_final_1D}. We will do so by demonstrating that the $j$th order  in $\extPot$ of $\NC\density_0$ with $\density_0$ according to \Eref{density_expanded_in_E} equals $\Hdrift_j(k_n)$. Expressing $\NC$ in terms of $\Hdrift_j(k_n)$ from \Erefs{normalisation_density_expansion_drift_diffusion} and \eref{density_expansion_drift_diffusion}, this amounts to the task of showing
\begin{equation}\elabel{Hdrift_from_Eft}
\sum_{i=0}^j \Hdrift_i(0)\Eft_{j-i}(k_n)
    = \Hdrift_j(k_n) \ .
\end{equation}
similar to \Eref{barometric_formula_aim}, but now allowing for $\drift_0\ne0$ and restricting ourselves to one dimension. \Eref{Hdrift_from_Eft} clearly holds for $j=0$, as
\begin{equation}
    \elabel{Hdrift_induction_base_case0}
\Hdrift_0(0)\Eft_{0}(k_n)=
\sysL \frac{1-\exp{-\drift_0\sysL/\transDiffusion}}{\drift_0/\transDiffusion} \delta_{n,0}
=
\Hdrift_0(k_n)
\end{equation}
using \Erefs{E0_stat} and \eref{Hdrift_0}. It also holds for the more complicated $j=1$, \Eref{Hdrift_1},
\begin{equation}\elabel{Hdrift_induction_base_case1}
    \Hdrift_1(0)\Eft_{0}(k_n) + 
    \Hdrift_0(0)\Eft_{1}(k_n)=
    \left(\sysL \frac{1-\exp{-\drift_0\sysL/\transDiffusion}}{\drift_0/\transDiffusion}\right)
    \left(-
    \frac{1}{\sysL} \frac{k_n\extPot_n\indicator{n\ne0}}
    {\transDiffusion k_n + \imag \drift_0 }
    \right)
    =
    \frac{\imag k_n \extPot_n}{\drift_0}
    \frac{1-\exp{-\drift_0\sysL/\transDiffusion}}
    {-\imag k_n + \drift_0/\transDiffusion}
\end{equation}
using \Erefs{E1_stat} and \eref{Hdrift_1} at $k_n=0$, the former in one dimension and the latter producing 
$\Hdrift_1(0)=0$. In the last equality of \Eref{Hdrift_induction_base_case1} we have dropped the indicator function as $k_n\indicator{n\ne0}=k_n$, as the denominator has no zero at $k_n=0$ given that $\drift_0\ne0$.


As in \Sref{Relationship_Eft_Hbaro}, we assume \Eref{Hdrift_from_Eft} to hold for $j$ and use \Erefs{E_recurrence_final_1D} and \eref{Hdrift_recursion} to show 
 in the following that \Eref{Hdrift_from_Eft} holds for $j+1$. Writing therefore for $j\ge0$
\begin{subequations}
\begin{align}
    \Hdrift_{j+1}(k_n)&=(\indicator{n\ne 0}+\delta_{n,0})\Hdrift_{j+1}(k_n)\\
    &=
    \frac{\indicator{n\ne0}}{\sysL\transDiffusion}
\sum_{q}
\frac{\imag k_{n-q}}{(-\imag k_n + \drift_0/\transDiffusion)}
\extPot_{n-q}
\Hdrift_{j}(k_q)
+
\Eft_{0}(k_n)
\Hdrift_{j+1}(0)
\end{align}
\end{subequations}
as in \Eref{barometric_formula_aim2} using \Eref{indicator_id} in the first line and in the second line using \Eref{Hdrift_recursion} with $p_j$ replaced by $n-q$ to rewrite $\Hdrift_{j+1}(k_n)$ and $\delta_{\nvec,\nullvec}$ replaced by $\Eft_{0}(k_n)$ according to \Eref{E0_stat}. The $\Hdrift_{j}(k_q)$ can now be replaced by \Eref{Hdrift_from_Eft},
\begin{equation}
   \Hdrift_{j+1}(k_n) 
   =
    \frac{\indicator{n\ne0}}{\sysL\transDiffusion}
\sum_{q}
\frac{\imag k_{n-q}}{(-\imag k_n + \drift_0/\transDiffusion)}
\extPot_{n-q}
\sum_{i=0}^j \Hdrift_i(0)\Eft_{j-i}(k_q)
+
\Eft_{0}(k_n)
\Hdrift_{j+1}(0)
\end{equation}
and the convolution performed on $\Eft_{j}(k_n)$, \Eref{E_recurrence_final_1D}, 
which gives
\begin{equation}\elabel{Hdrift_induction_step}
   \Hdrift_{j+1}(k_n) 
   =
\sum_{i=0}^j \Hdrift_i(0)
\Eft_{j-i+1}(k_n)
+
\Eft_{0}(k_n)
\Hdrift_{j+1}(0)
=
\sum_{i=0}^{j+1} \Hdrift_i(0)
\Eft_{j+1-i}(k_n)
\end{equation}
confirming \Eref{Hdrift_from_Eft} for $j+1$, provided it holds for $j$. With \Eref{Hdrift_induction_step} as the induction step and \Erefs{Hdrift_induction_base_case0} or \eref{Hdrift_induction_base_case1} as the base case, this confirms \Eref{density_expanded_in_E} with \eref{E_recurrence_final_1D} in the case of drift diffusion in one dimension.

\subsection{Summary}
In the present appendix, we have introduced a general formula to determine the stationary density $\density_0(\xvec)$ \Eref{density_expanded_in_E} digrammatically, recursively and perturbatively in the potential, \Eref{E_recurrence_final}. We have validated this expression for the case of pure diffusion producing the barometric formula, \Eref{Boltzmann_result}, in \Sref{barometric_formula} and for the case of drift-diffusion in one dimension, \Eref{DriftDiffusionSolution}, in \Sref{field_theory_produces_1D_drift}. In both cases, we have used induction to show the equivalence of the density derived from theory and the density derived from classical considerations. The key challenge is to compare the field theoretic terms to a certain order in the potential to the corresponding terms in the classical result, which is straight-forwardly expanded in powers of the potential only up to a normalisation, which needs to be expanded itself, \Erefs{density_with_expanded_denominator}, \eref{barometric_formula_aim} and \eref{Hdrift_from_Eft}.  

In order to treat the potential perturbatively, we had to allow for vanishing potential, which allows for a stationary state only if the system is finite. However, the resulting perturbation theory reproducing the Boltzmann factor in the case of no drift means that it applies equally for \emph{confining} potentials. This applies similarly to the case with drift, as the thermodynamic limit $\sysL\to\infty$ may be taken after resumming the perturbative result into the integral expression \Eref{DriftDiffusionSolution}.

\end{document}